\documentstyle[aps,pre,array,multicol,epsfig,eqsecnum]{revtex}      

\begin{document}
\draft

\title{The coherent scattering function in the reptation model: \\ 
       analysis beyond asymptotic limits}      
\author{Lothar Sch\"afer$^1$, Ute Ebert$^2$, and  Artur Baumg\"artner$^3$}
\address{$^1$ Universit\"at GH Essen, Universit\"atsstr.\ 5, 45117 Essen, 
        Germany}
\address{$^2$ Centrum voor Wiskunde en Informatica, 
        P.O.\ Box 94079, 1090 GB Amsterdam, The Netherlands} 
\address{$^3$ Institut f\"ur Festk\"orperforschung,
        Forschungszentrum J\"ulich, 52425 J\"ulich, Germany}
\date{\today}
\maketitle

\begin{abstract}
We calculate the coherent dynamical scattering function 
$S_{c}(q,t;N)$ of a flexible chain of length $N$, 
diffusing through an ordered background of topological obstacles. 
As an instructive generalization, we also calculate the
scattering function $S_{c}(q,t;M,N)$ for the central piece of length 
$M\le N$ of the chain.
Using the full reptation model, we treat global creep, tube length 
fluctuations, and internal relaxation within a consistent and unified 
approach. Our theory concentrates on the universal aspects of 
reptational motion, and our results in all details show excellent agreement 
with our simulations of the Evans-Edwards model, provided we allow 
for a phenomenological prefactor which accounts for non-universal 
effects of the micro-structure of the Monte Carlo chain, 
present for short times.
Previous approaches to the coherent structure function can be
analyzed as special limits of our theory.
First, the effects of internal relaxation 
can be isolated by studying the limit $N \rightarrow \infty$, $M$ fixed. 
The results do not support the model of a `Rouse chain in a tube'. 
We trace this back to the non-equilibrium initial conditions of 
the latter model. 
Second, in the limit of long chains $(M = N \rightarrow \infty)$ 
and times large compared  to the internal relaxation time 
$(t/N^{2} \rightarrow \infty)$, our theory reproduces the results 
of the primitive chain model. This limiting form applies only 
to extremely long chains, and for chain lengths accessible in practice,
effects of, e.g., tube length fluctuations are not negligible. 
\end{abstract}


\begin{multicols}{2}

\section{Introduction}

The equilibrium dynamics of a dense polymer system, i.e., a melt, 
a solution of high concentration, or a free chain moving through a gel, 
is an important topic of polymer physics. It has been investigated 
for many years, but still is not fully understood. The problem is 
quite complex, even if we concentrate on the motion of a single chain. 
Clearly, its motion is strongly hindered by the surrounding chains, 
which the chain considered cannot cross. This has led to the idea \cite{Z1} 
that the motion of the chain is confined to a tube roughly defined 
by its instantaneous configuration. Thus the tube is assumed to have 
a random walk configuration which changes only by the motion of the 
chain ends. The ends can retract into the tube which thus is effectively 
shortened, and they can creep out of the original tube, thus creating 
a new tube segment in some random direction. The interior parts 
of the tube are assumed to be fixed in space until they are reached 
by the diffusive motion of the chain ends. This concept of a tube is 
one basic ingredient of the `reptation' \cite{Z1,Z2} scenario, 
which certainly is valid provided the obstacles confining 
the chain motion form a rigid, time independent network. 
In a realistic system the surrounding chains are mobile, 
which sheds some doubt on the postulated existence of a 
well defined tube. Indeed, there exist other approaches \cite{Z3,Z4},
more in line with standard many body theory, which are not based 
on the tube concept.

Most work on the reptation model concentrates  on asymptotic results 
expected to hold for long chains in special time regions (see Sect.\ II.A). 
In comparison to experiments or simulations, these results often fail 
on the quantitative level \cite{Z5}, and partly other theories seem 
to be more satisfactory \cite{Z3,Z4}. Thus some work \cite{Z6} has been 
invested to incorporate additional physical effects like relaxation 
of the surrounding or specific interaction effects into asymptotic 
reptation theory. However, the evaluation of the pure reptation model 
outside asymptotic limits has found only little attention. 
In recent work \cite{Z70,Z7}, we presented such a calculation for 
the motion of individual segments of the chain. We found that 
asymptotic results, which for the quantities 
considered take the form of specific power laws, apply only to 
surprisingly long chains. Large time intervals are covered 
by crossover regions. Our crossover functions compare very well 
to simulations \cite{Z8} of the pure reptation model, i.e., to the motion 
of a flexible chain through a fixed regular lattice of impenetrable 
obstacles. Furthermore, also results of simulations of melts look 
quite similar to our analytical results for short chains. 
This is consistent with the observation that the tube diameter, 
i.e., the average distance among effective obstacles extracted 
from the simulations, is fairly large. Since in the pure reptation model, 
the tube width is of the order of the effective segment size of 
the reptating chain, this implies that to map melt dynamics on 
the reptation model, we have to consider a coarse grained chain 
of effective segment number $N/N_{e}$. Here $N$ is the chain length 
(polymerization index) 
of the physical macromolecule, whereas the `entanglement length' $N_{e}$ 
is the length of a subchain which shows a coil radius of the order 
of the obstacle spacing. In recent simulations \cite{Z9} of melts, 
a value $N/N_{e} \approx 14$ was reached, far below the value 
$N/N_{e} \agt 50$ needed according to our theory to clearly identify 
asymptotic power laws. Other recent simulations \cite{Z10} reach 
a value $N/N_{e} \approx 300$, but for this chain length they cover 
only times short compared to the characteristic time scales of 
the reptation model. Still, the onset of a first power law regime is seen, 
again consistent with our theory. Thus, concerning the motion of 
individual chain segments within a melt, there at present seems 
to be no need to invoke other mechanisms than pure reptation.

The motion of specific segments is  easily accessible only 
to computer experiments. Physical experiments often measure 
dynamic scattering functions. Asymptotic results of the reptation model 
for the coherent scattering function have been worked out 
previously \cite{Z11,Z12,Z2}, but our analysis of segment motion 
suggests that an evaluation of the scattering function outside 
asymptotic limits is needed. This is the topic of the present work. 
We use the same analytical reptation model as in our previous 
analysis \cite{Z7}. We also measured the coherent structure function 
in Monte Carlo simulations, again using the same implementation 
of the model as previously \cite{Z8}. This allows for a comparison 
among theory and data, where all parameters are fixed by our previous work. 
Some results of the simulations will be presented here, but a detailed 
comparison of our simulation results to the present and previous theories 
will be presented in a separate, less technical paper \cite{Z13}.           

In the next section, we briefly review the basic features of the 
reptation model and recall previous results for the coherent structure
function. In Sect.\ III, we introduce our analytical model and outline 
the structure of our approach. In Sect.\ IV, we consider those 
contributions to the coherent structure function in which the initial tube 
is not yet completely destroyed by the stochastic motion. 
A rigorous analysis is possible as long as end effects 
can be neglected. These end effects, known as `tube renewal' and 
`tube length fluctuations', can be treated only in some approximation. 
We here generalize an approach which in our previous work gave good results 
for the segment motion. In Sect.\ V, we compare our rigorous results 
for the motion within the initial tube  
to those of the model of a `Rouse chain in a coiled tube' \cite{Z11}. 
Pronounced differences are found and their origin is clarified. 
In Sect.\ VI, we derive an integral equation which takes complete 
tube destruction into account. For long chains and times large compared 
to the internal relaxation time 
of the chain, we recover the results of the `primitive chain' model 
\cite{Z12,Z2}, as shown in Sect.\ VII. Typical numerical results 
of our theory are discussed in Sect.\ VIII. It is found that tube length 
fluctuations, which have been neglected in previous calculations 
of the coherent scattering function, in fact determine the scattering 
up to times larger than the Rouse time. In Sect.\ VIII, we also present 
some results of our simulations, which compare favorably with our theory. 
Finally, Sect.\ IX contains a summary and conclusions. 
The full evaluation of the reptation model leads to quite involved
expressions, and some part of the analysis is summarized in appendices.   


\section{Review of the reptation scenario and of previous results 
for the coherent structure function}

\subsection{Basic dynamics and time scales}

As mentioned in the introduction, the reptation model assumes 
the existence of a tube defined by the instantaneous configuration 
of the chain together with the surrounding obstacles. The chain cannot 
leave the tube sideways since it would have to fold into a double-stranded 
conformation which costs too much entropy. Those parts of the chain, 
which lie stretched in the tube, essentially cannot move. In the interior 
of the tube only little wiggles of `spared length' are mobile, as illustrated 
in Fig.~1 for the special case of a lattice model. These wiggles carry 
out Brownian motion along the chain.
If a wiggle reaches a chain end, it may decay and prolong the tube 
by its spared length. Chain ends also may produce new wiggles 
which then diffuse 
into the interior of the tube. This shortens the tube by the spared length 
of the newly created wiggle. In the long run, this random motion 
of the chain ends leads to a complete destruction of the initial tube. 

This very simple dynamical model involves several time scales. 
It needs a microscopic time $T_{0}$ until the segment motion feels 
the existence of the constraints. $T_{0}$ generally is identified 
with the Rouse time of a chain of length equal to the entanglement length 
$N_{e}$:
\begin{equation}
T_{0} \sim N_{e}^{2} \: \: \:.
\end{equation}
$T_{0}$ is relevant for the short-time dynamics of melts, 
where the tube diameter typically is found to be quite large \cite{Z9,Z10}: 
$N_{e} \sim 10-40$. 
For reptation, $T_{0}$ defines the elementary time step, since this theory
does not deal with the unconstrained motion on scale of the tube diameter. 
A second scale $T_{2}$ is the time a wiggle needs to diffuse 
over the whole chain. Since in the coarse grained description, 
the wiggle has to diffuse a distance of $N/N_{e}$ steps, one finds 
\begin{equation}
T_{2} \sim T_{0} \left( \frac{N}{N_{e}}\right)^{2} \sim N^{2}~.
\end{equation}
$T_{2}$ thus is of the order of the Rouse time of the whole chain. 
Finally, the reptation time $T_{3}$ is needed to destruct 
the initial tube completely. Reptation theory \cite{Z1} predicts 
\begin{equation}
T_{3} \sim T_{0}  \left( \frac{N}{N_{e}}\right)^{3}
\end{equation}
as limiting result for long chains. Asymptotically power laws 
as function of $t$ and $N$ are predicted to hold for the segment motion 
or the motion of the center of mass in the time windows 
$~T_{0} \ll t \ll T_{2}~$, $~T_{2} \ll t \ll T_{3}~$, and $~T_{3} \ll t~$. 
As a typical result we quote the mean squared spatial displacement 
of some bead of the chain:
\begin{eqnarray*}
\left\langle~\overline{({\bf r}_{j}(t) - {\bf r}_{j}(0))^{2}}~\right\rangle 
\sim \left\{
\begin{array}{ll}
t^{1/4}            & \mbox{for } T_{0} \ll t \ll T_{2}\\
(t/N)^{1/2} \qquad & \mbox{for } T_{2} \ll t \ll T_{3}\\
t/N^{2}            & \mbox{for } T_{3} \ll t~.
\end{array}
\right.
\end{eqnarray*}
Here ${\bf r}_{j}(t)$ gives the spatial position of bead $j$ at time $t$. 
The bar indicates the (dynamic) average over the stochastic motion and 
the pointed brackets denote the (static) average over all tube configurations.

\subsection{Previous results for the coherent structure function}

We consider a chain of $N+1$ beads ($N$ segments), labelled by 
$j = 0,1,\ldots,N$. The coherent structure function is defined as
\begin{equation}
S_{c}(q,t,N) = \sum^{N}_{j,k=0} \left\langle~\overline{  
e^{\textstyle i {\bf q} ({\bf r}_{j}(t) - {\bf r}_{k}(0))}}~\right\rangle 
\: \: \:.
\end{equation}
By definition $S_{c}(q,t,N)$ refers to a single chain. It can be measured
by appropriately labelling a few chains in the system. Reptation results 
for $S_{c}(q,t,N)$ previously have been derived by Doi and Edwards 
\cite{Z12} and by de Gennes \cite{Z11}.   

Doi and Edwards have evaluated a simplified version of the reptation model,
where the internal motion of the chain is neglected. 
The physical chain is replaced by a `primitive chain', 
which only can slide along the tube so that all segments 
experience the same curvilinear displacement $\Delta \xi(t)$. This model 
therefore reduces the dynamics to diffusive motion of the single stochastic 
variable $\Delta \xi$. For the coherent structure function, it yields 
the result (see Ref.\ \cite{Z2}, chapter 6.3.4)
\begin{eqnarray}
\frac{S_{c}(q,t,N)}{S_{c}(q,0,N)} &=& 
\bar{S}_{DE} \left(q^{2} R_{g}^{2},\frac{t}{\tau_{d}}\right)~,
\\
\bar{S}_{DE}(Q,\tau) &=& \frac{Q}{D(Q)}\: \sum^{\infty}_{p=1} 
\frac{\sin^2\alpha_{p}~~
e^{\textstyle - \frac{4}{\pi^{2}} \alpha^{2}_{p} \tau}}
{\alpha^{2}_{p}~ (Q^{2}/4 + Q/2 + \alpha^{2}_{p})}~,
\end{eqnarray}
where $R_{g}^{2}$ is the radius of gyration and 
$\tau_{d} \sim T_{3} \sim N^{3}$. $D(Q)$ is the Debye function:
\begin{equation}
D(Q) = \frac{2}{Q^{2}}\:(e^{-Q} - 1 + Q)~.
\end{equation}
The $\alpha_{p} = \alpha_{p}(Q)$ are the positive solutions of
\begin{equation}
\alpha_{p}\: \tan\: \alpha_{p} = \frac{Q}{2} \: \: \:.
\end{equation}
Neglecting all internal motions, the result can be applied only
for $t \gg T_{2}$, i.e., in a time regime where the internal degrees 
of freedom are equilibrated. In the limit of large wave numbers 
$Q = q^{2} R_{g}^{2} \gg 1$, the result reduces to
\begin{equation}
\bar{S}_{DE}(Q,\tau) = \frac{8}{\pi^{2}}\:
\sum^{\infty}_{p=1}\:(2 p - 1)^{-2} \exp \left[- (2 p-1)^{2} \tau\right]~.
\end{equation}
This is the scattering from that part of the primitive chain which at time 
$t$ still resides in the initial tube \cite{Z1,Z11}. 

The limit $q^{2} R_{g}^{2} \gg 1$ has also been considered by de Gennes. 
Taking the internal relaxation of the chain into account, his result 
\cite{Z11} for the normalized coherent scattering function 
is a sum of two terms: 
\begin{eqnarray}
\bar{S}_{dG} (q,t,N) &=& \Big(1\!-\!B_{dG}(q)\Big)\;\bar{S}^{(\ell)}(q,t) 
+ B_{dG}(q)\;\bar{S}^{(c)}(t,N)
\nonumber\\
&& \\
&& B_{dG}(q) = 1 - \frac{N_{e}}{6 N}\:q^{2} R_{g}^{2}~.
\end{eqnarray}
The `creep term' $\bar{S}^{(c)}(t,N)$ is given by Eq.\ (2.9) 
and thus describes the large time behavior $t \gg T_{2}$. 
It tends to $1$ for $t/T_{3} \rightarrow 0$. The `local term' 
$\bar{S}^{(\ell)}(q,t)$ is taken from an approximate evaluation 
of the internal relaxation of an infinitely long one-dimensional 
Rouse chain, folded into the three-dimensional random walk configuration 
of a tube of $N/N_{e}$ segments. The result reads 
\begin{equation}
\bar{S}^{(\ell)}(q,t) = e^{t_{1}}\;\mbox{erfc} \sqrt{t_{1}} \: \: \:
\end{equation}
where 
\begin{equation}
t_{1} = \frac{3}{\pi^{2}} \frac{N}{N_{e}} (q^{2} R_{g}^{2})^{2} 
\frac{t}{\tau_{d}} = \frac{t}{T_{q}} \: \: \:.
\end{equation}
This introduces an additional $q$-dependent time scale
\begin{equation}
T_{q} = \frac{\pi^{2}}{3} \frac{N_{e}}{N} 
\frac{\tau_{d}}{(q^{2} R_{g}^{2})^{2}} \: \: \:,
\end{equation}
which in view of $R_{g}^{2} \sim N$, and $\tau_{d} \sim N^{3}$ is independent 
of $N$. $T_{q}$ governs the relaxation of segment density fluctuations 
on scale $q^{-1}$. In view of $q^{2} R_{g}^{2} \gg 1$, $T_{q}$ is much
smaller than $\tau_{d}$, and for times $t\alt T_q$, the creep term
is constant, $\bar S^{(c)}(t,N)\approx\bar S^{(c)}(0,N)=1$. 
On top of this plateau, $\bar{S}^{(\ell)}(q,t)$ yields a peak 
rapidly decreasing in time. Note that 
$\bar{S}^{(\ell)}(q,t)$ for $t_1\gg1$ behaves as 
$\bar{S}^{(\ell)}(q,t) = (\pi t_{1})^{-1/2}$. 
The amplitude of the peak is determined  by $B_{dG}$, 
which only depends on $q^{2}$ and $N_{e}$.

Both these approaches neglect end effects like tube length fluctuations, 
which are governed by the time scale $T_{2}$. The approximations 
involved greatly simplify the analysis but are no essential part 
of the reptation model. In the sequel, we present an analysis 
of the full model, accounting for the internal degrees of freedom 
and the finite chain length. Since all the dynamics is driven by 
the diffusion of the spared length as the only stochastic process, 
this yields a unified description of local relaxation, global creep 
and tube length fluctuations. We will find that tube length fluctuations, 
in particular, have an important influence for intermediate times 
and chain lengths. Internal relaxation, however, is of much less 
influence than the results referred to above suggest.  


\section{Formulation of the full reptation model} 

\subsection{Microscopic dynamics}

We here recall the essential features of our model. A more detailed 
discussion can be found in Ref.\ \cite{Z7}. 
The chain is modeled as a random walk of $N$ steps of fixed length 
$|{\bf r}_{j} - {\bf r}_{j-1}| = \ell_{0}$, $j = 1,\ldots,N$.
The motion is due to the diffusion of 
wiggles of spared length $\ell_{s}$. These are represented by particles 
hopping along the chain from bead to bead, with hopping probability 
$p$ per time step. The particles do not interact, and a given particle 
sees the others just as a part of the chain. If a particle passes 
a bead $j$, it tracks it along by a distance of the spared length 
$\ell_{s}$, which is taken to be the same for all particles. 
The end beads $j = 0,N$ of the chain are coupled to large reservoirs, 
which absorb and emit particles at such a rate that the equilibrium 
density $\rho_{0}$ of particles on the chain is maintained on average. 
Keeping track of the change of the particle number in these reservoirs, 
we control the motion of the chain ends: creation or decay of a wiggle 
at a chain end implies emission or absorption of the corresponding 
particle by the reservoir.  

For the motion of beads in the interior of the tube, the essential 
stochastic variable of the model is the number $n(j,t)$ of particles 
which passed over bead $j$ within time interval $[0,t]$. 
\begin{equation}
n(j,t) = n_{+}(j,t) - n_{-}(j,t)
\end{equation}
Here $n_{\pm}(j,t)$ is the number of particles that came from 
the `left' $(j' < j)$ or from the `right' $(j' > j)$, respectively. 
Consider, for instance, the motion of segment $j$ for a time interval 
in which it stays in the original tube. Its displacement in the tube 
is given by $\ell_{s} n(j,t)$, and since the tube has a random walk 
configuration, its spatial displacement is given by
\begin{equation}
\left\langle ~\overline{(r_{j}(t) - r_{j}(0))^{2}}~\right\rangle 
= \ell_{s}\; \ell_{0}\; \overline{|n (j,t)|}~.
\end{equation}
Since the underlying stochastic process is single particle hopping, 
the distribution function of $n(j,t)$ is easily calculated, 
with the result (Ref.\ \cite{Z7}, Eq.\ (3.22))  
\begin{equation}
{\cal P}_{1} (n;j,t) = \overline{\delta_{n,n(j,t)}} 
= e^ {\textstyle -\overline{n^{2}(j,t)}}\; 
I_{n}\left( ~\overline{n^{2}(j,t)}~\right),
\end{equation}
where $I_{n}(z)$ is the modified Bessel function of the first kind. 
The second moment $\overline{n^{2}(j,t)}$ is found as [see Ref.\ \cite{Z7}, 
Eqs.\ (3.24), (3.12), (3.9)]
\begin{eqnarray}
\overline{n^{2}(j,t)} &=& 2 \rho_{0} A_{1} (j,t)
\\
A_{1} (j,t) &=& \frac{p t}{N} + \frac{1}{2N} 
\sum^{N-1}_{\kappa=1}\:(1 - \alpha^{t}_{\kappa}) \;
\frac{\cos^{2} \left(\frac{\pi \kappa}{N} (j + \frac{1}{2})\right)}
{\sin^{2} \left(\frac{\pi \kappa}{2 N}\right)}
\\
\alpha_{\kappa} &=& 1 - 4 p \sin^{2} \frac{\pi \kappa}{2 N}.
\end{eqnarray}
Some useful properties of $A_{1}(j,t)$ are collected in \cite{Z7}, 
Appendix A. We also will need the first moment $\overline{|n (j,t)|}$, 
which from Ref.\ \cite{Z7}, Eqs.\ (3.26), (3.27), is found as 
\begin{eqnarray} 
\overline{|n (j,t)|} &=& \frac{2}{\sqrt{\pi}}\:
\left(\rho_{0} A_{1}(j,t)\right)^{1/2} \;
\left[ 1 - F_{1} (4 \rho_{0} A_{1} (j,t))\right]
\\
F_{1}(z) &=& \frac{1}{2 \sqrt{\pi}}\:\int_{0}^{z}\:d x\:x^{-3/2} e^{-x} 
\left(\left(1 - \frac{x}{z}\right)^{-1/2} - 1 \right) 
\nonumber\\
&&-\: \frac{1}{2 \sqrt{\pi}}\:\Gamma \left(- \frac{1}{2},z\right),
\end{eqnarray}
where $\Gamma (y,z)$ is the incomplete $\Gamma$-function. 
  
Except for microscopic times $t \alt 2/p$, $\alpha^{t}_{\kappa}$ 
can be approximated as  
\begin{equation}
\alpha^{t}_{\kappa} \approx 
\exp \left[ - 4 p t\:\sin^{2}\:\frac{\pi \kappa}{2 N}\right],
\end{equation}
so that the theory involves time only in the combination
\begin{equation}
\hat{t} = p t \: \: \:.
\end{equation}
In evaluating the theory, we will use $\hat{t}$ as time variable. 
For $\overline{n^{2}(j,t)} \agt 100$, which for $N \agt 100$ 
implies $\hat{t} \agt 10^{4}, {\cal P}_{1}(n;j,t)$ is well represented 
by a simple Gaussian
\begin{equation}
{\cal P}_{1}(n;j,t) \approx 
\left(2 \pi \;\overline{n^{2}(j,t)}\right)^{-1/2} 
\;\exp \left(- \frac{n^{2}}{~2\; \overline{n^{2}(j,t)}~}\right).
\end{equation} 
 
Knowledge of ${\cal P}_{1}(n;j,t)$ is sufficient as long 
as we consider motion inside the initial tube. End effects introduce 
a more complicated quantity. Within time interval $[0,t]$, the tube 
from the end $j = 0$ is destructed up to bead $j_{<}$, where
$j_<$ is defined as
\begin{equation}
j_{<} = \bar{\ell}_{s}\: n_{\mbox{\scriptsize{max}}}(0,t),
\end{equation}
\begin{equation}
n_{\mbox{{\scriptsize max}}}(0,t) = 
\max_{s \in [0,t]} \left[ - n (0,s)\right].
\end{equation}
Here
\begin{eqnarray*}
n (0,s) = m_{0} (s) - m_{0} (0),
\end{eqnarray*}
where $m_{0}(s)$ is the occupation number at time $s$ of the reservoir 
at chain end $0$. Thus $n_{\mbox{{\scriptsize max}}}(0,t)$
is the maximal negative fluctuation of the occupation number 
of reservoir $j = 0$ in the time interval $[0,t]$. In (3.12)
we also introduced
\begin{equation}
\bar{\ell}_{s} = \ell_{s}/\ell_{0} \: \: \:,
\end{equation}
measuring all lengths in units of the segment size $\ell_{0}$. 
Similarly, from the other end tube destruction within time $t$ 
proceeds to bead
\begin{equation}
j_{>} = N - \bar{\ell}_{s}\: n_{\mbox{{\scriptsize max}}}(N,t) \: \: \:,
\end{equation}
with $n_{\mbox{{\scriptsize max}}}(N,t)$ being the maximal negative 
fluctuation of the occupation number $m_{N}(s)$ of the reservoir 
at chain end $N$. The stochastic processes $m_{0}(s)$ or $m_{N}(s)$ 
are not Markovian, since a particle emitted by a reservoir can be 
reabsorbed by the same reservoir later. This induces a correlation 
which dies out only if the particle has time to reach the other reservoir, 
i.e., on time scale $T_{2}$. For such a correlated process, 
the distribution and the moments of 
$n_{\mbox{{\scriptsize max}}}$ cannot be calculated rigorously, 
even though arbitrary moments of $n(0,s)$, involving any number 
of time variables $s$, can be evaluated (see \cite{Z7}, Sect.\ III). 
As soon as tube renewal comes into play, we therefore have to resort 
to some approximation.  

Some important quantity entering our theory is the average 
$\overline{n_{\mbox{{\scriptsize max}}}(0,t)}$. It, for instance, 
yields the motion of the end-segment via the relation [Ref.\ \cite{Z7}, 
Eq.\ (2.12)]
\begin{eqnarray*}
\left\langle ~\overline{({\bf r}_{0}(t) - {\bf r}_{0}(0))^{2}}~\right\rangle 
= 2 \bar{\ell}_{S}\; \ell_{0}^{2} \;
\overline{n_{\mbox{{\scriptsize max}}}(0,t)} \: \: \:.
\end{eqnarray*}  
We use the expression (Ref.\ \cite{Z7}, Eq.\ (5.1))
\begin{equation}
\overline{n_{\mbox{{\scriptsize max}}}(0,t)} = 
\sum^{t}_{s=1}\:\frac{\;\overline{|n (0,s)|}\;}{2 s}\: \: \:,
\end{equation}
which is correct for a Markov process. Using in Eq.\ (3.16) 
the exact moments $\overline{|n (0,s)|}$ (Eq.\ (3.7)), 
we in essence approximate the correlated process by a sequence 
of Markov processes which for each time step $s$ yield the correct 
instantaneous value of $\overline{|n (0,s)|}$. This `mean hopping rate' 
approximation, which was discussed in more detail in \cite{Z7}, 
gives good results for larger times. For microscopic times, 
it underestimates $\overline{n_{\mbox{{\scriptsize max}}}(0,t)}$ 
by about a factor of 2, but with increasing time it approaches 
the full result for $\overline{n_{\mbox{{\scriptsize max}}}}$ 
as found in simulations. For $t \approx T_{2}$, the deviation 
for the motion of the end segment, which is most sensitive 
to our approximation, is of the order of 10 \% only 
(see Fig.~9 of Ref.\ \cite{Z8}).

\subsection{Outline of our calculation of the coherent structure function} 
            
The basic quantity to be considered, is the scattering from a pair of beads
\begin{equation}
S (q,t;j,k,N) = \left\langle~ 
\overline{e^{\textstyle i{\bf q}\left({\bf r}_{j}(t) - {\bf r}_{k}(0)\right)}}
~\right\rangle~,
\end{equation}
which must be summed over the beads to find the coherent structure 
function $S_{c}(q,t;N)$. To get some information on the contribution 
of the different parts of the chain, we consider a slight generalization 
in which we sum only over the $M+1$ central beads
\begin{equation}
S_{c} (q,t;M,N) = \sum^{\frac{N+M}{2}}_{j,k = \frac{N-M}{2}} S(q,t;j,k,N)~.
\end{equation}
Clearly, the coherent structure function of the full chain is
\begin{equation}
S_{c}(q,t;N) = S_{c} (q,t;N,N) \: \: \:.
\end{equation}

To calculate $S (q,t;j,k,N)$, we first consider the contribution 
$S^{(T)} (q,t;j,k,N)$ which results from those stochastic motions 
for which a part of the initial tube still exists at time $t$. 
(The upper index $(T)$ stands for `tube'.)~ We then can set up 
an integral equation for $S (q,t;j,k,N)$, in which $S^{(T)} (q,t;j,k,N)$ 
shows up as inhomogeneity (see Sect.\ VI). Furthermore, 
for $t \ll T_{3}$, contributions where the tube is destroyed completely, 
are negligible, and $S^{(T)}$ coincides with $S$. 

$S^{(T)} (q,t;j,k,N)$ incorporates the effects of internal relaxation 
and tube length fluctuations, and its calculation is the most 
tedious part of our analysis. We here need to simultaneously control 
the motion of segment $j$ and of the chain ends. More specifically, 
we will need the distribution function
\begin{equation}
{\cal P}^{(T)}_{\mbox{{\scriptsize max}},j} (n_{m},n_{j};t) 
= \overline{\Theta (j_{<} - j_{>})\; 
\delta_{n_m,n_{\mbox{\scriptsize max}}(0,t)}\; \delta_{n_{j},n(j,t)}}~,
\end{equation}
i.e., the simultaneous distribution of $n(j,t)$ and 
$n_{\mbox{{\scriptsize max}}}(0,t)$ under the constraint that a part 
of the initial tube still exists. Again the correlated nature 
of the stochastic motion of the chain ends prevents a rigorous evaluation 
of ${\cal P}^{(T)}_{\mbox{{\scriptsize max}},j}$, and we use 
random walk theory to construct an approximate functional form. 
The result depends on $n_{m}$, $n_{j}$ and $t$ only through the 
rescaled variables 
$n_{m}/\:\overline{n_{m}(t)}$, and $n_{j}/\:\overline{(n^{2}_{j}(t))}^{1/2}$, 
and in the spirit of our mean hopping rate approximation, 
we in these variables replace the normalizing factors 
$\overline{n_{m}(t)}$ and $\overline{(n^{2}_{j}(t))}^{1/2}$ 
of the random walk by their counterparts for the proper correlated process. 
In essence, this again amounts to replacing the correlated stochastic 
motion of the chain ends by a whole sequence of uncorrelated 
random walks, parametrized by an effective hopping rate $p'$. 
This hopping rate is adjusted such that the random walk which replaces 
the correlated process for final time $t$, at {\em that time} yields 
the correct moments $\overline{n_{m}(t)} 
= \overline{n_{\mbox{\scriptsize max}}(0,t)}$ and $\overline{n_{j}^{2}(t)} 
= \overline{(n (j,t))^{2}}$. (It in fact yields the correct Gaussian 
distribution of the single variable $n(j,t)$.)~ 
As discussed in Ref.\ \cite{Z7}, Sect.\ V.B, $p'$ changes from a value 
$\rho_{0} p$ at microscopic times to $\rho_{0} p/N$ for $t \gg T_{2}$. 
Since $\rho_{0} p$ governs the short time motion of a segment whereas 
$\rho_{0} p/N$ is the mobility of the primitive chain, the mean hopping 
rate approximation smoothly interpolates between these more rigorously 
accessible limits. This will be discussed again in Sect.\ IV.C, 
after we presented the details of our approach.

Our theory involves three important time (and segment index) 
dependent parameter-functions:  
\begin{equation}
c = c (t) = \sqrt{\frac{\pi}{2}} \;\bar{\ell}_{s} \;
\overline{n_{\mbox{{\scriptsize max}}}(0,t)}
\end{equation}
measures the extent of tube destruction and thus accounts 
for tube length fluctuations. 
For $\overline{n_{\mbox{{\scriptsize max}}}(0,t)}$, we use 
the approximation (3.16). It turns out that the time dependence of 
$c(t)$ which very slowly tends to its asymptotic limit  $c (t) 
\stackrel{t \rightarrow \infty}{\longrightarrow} \mbox{const}\:t^{1/2}$ 
(cf.\ Eq.\ (7.2)), is responsible for the well known crossover behavior 
of the reptation time: $T_{3} \sim N^{z_{\rm eff}}$, 
where $z_{\rm eff}$ slowly approaches its asymptotic value 
$z_{\rm eff} \rightarrow 3$ from above. (A detailed discussion 
of the reptation time will be given in a separate paper.)

A second function, $a (j,t)$, measures the coupling of the motion 
of an interior segment $j$ to the motion of a chain end. 
Initially, this coupling vanishes, but it increases with time 
due to particles created at a chain end and traveling over segment $j$. 
If this coupling is fully developed, all segments approximately have 
moved the same distance in the tube and the primitive chain model 
results. The precise definition of $a(j,t)$ is given in Eq.\ (4.17).

Finally, it should be noted that the effective mobility of a segment 
for $t\ll T_2$ depends on its position in the chain, an effect
already present for free Rouse type motion. This is taken into account 
by the function $b(j,t)$, which is defined in Eq.\ (4.36).

Having described the main ideas of our approach, we now turn 
to the details. We first construct and analyze the tube conserving 
contribution $S_c^{(T)} (q,t;M,N)$ to the structure function. 


\section{Tube conserving contribution to the structure function}

In this and the next section, we consider the contribution 
of those stochastic processes, which do not destroy the initial 
tube completely, i.e., we insist on the inequality   
\begin{equation}
j_{>} - j_{<} \geq 0 \: \: \:,
\end{equation}
since due to the definitions (3.12), (3.15), 
the tube has been destroyed up to segment $j_{<}$ from chain end $0$
or $j_{>}$ from chain end $N$, 
respectively. We first construct a formally exact expression 
for the corresponding contribution $S^{(T)} (q,t;j,k,N)$.
Its summation over indices $j$ and $k$ as in (3.18) and (3.19)
yields the tube conserving contribution to the coherent structure
function.

\subsection{Exact expression for $S^{(T)} (q,t;j,k,N)$}

Depending on the relation among $j$, $k$, $j_{<}$, and $j_{>}$, 
we have to distinguish several cases. We use the notation
\begin{equation}
j(t) = j + \bar{\ell}_{s}\; n(j,t) \: \: \:,
\end{equation}
and we illustrate the analysis with two typical cases shown in Fig.~2. 
Fig.~2~a gives a schematic sketch of a situation, in which the inequalities
\begin{eqnarray*}
j(t) \leq k~,~~ j_{<} \leq k~,~~j_{>} \geq j(t)   \: \: \:,
\end{eqnarray*}
hold. That means that tube renewal from chain end $0$ has not passed 
over the original position of segment $k$, and segment $j$ at time $t$ 
is not found in the part of the new tube created from chain end $N$. 
Furthermore the new position of segment $j$, if measured along the tube, 
is closer to the new position of chain end zero than the original position 
of segment $k$.
The relative ordering of $k$ and $j_{>}$, or of $j(t)$ and $j_{<}$ is 
unimportant. As is clear from Fig.~2~a, the path connecting $j(t)$ and $k$ 
has $k-j(t)$ steps, and since the chain configuration is a random walk, 
we find
\begin{eqnarray}
\left\langle e^{\textstyle i {\bf q} 
                ({\bf r}_{j}(t) - {\bf r}_{k}(0))}\right\rangle 
&=& e^{\textstyle - \bar{q}^{2} (k - j(t))},
\\
\mbox{where }~~~~~~~\bar{q}^{2} &=& \frac{q^{2} \ell_{0}^{2}}{6}.
\end{eqnarray}
As a result, the contribution of such configurations to 
$S^{(T)}(q,t;j,k,N)$ reads 
\begin{eqnarray*}
\overline{\Theta(k\!-\!j(t))\;\Theta(k\!-\!j_{<})\;\Theta (j_{>}\!-\!j(t))\; 
\Theta(j_{>}\!-\!j_{<})\;\;e^{- \bar{q}^{2} (k-j(t))}},
\end{eqnarray*}
where the discrete $\Theta$-function is defined as 
\begin{eqnarray*}
\Theta (n) = \left\{ 
\begin{array}{l@{\quad;\quad}l}
1 ~\: & n \equiv 0,1,2,\ldots\\
0 ~\: & n = -1,-2,\ldots\\
\end{array}
\right.
\end{eqnarray*}

Now consider a typical case of other type, shown in Fig.~2~b. 
It is defined by the inequalities
\begin{eqnarray*}
j (t) \leq k~,~~j_{<} > k~,
\end{eqnarray*}
and differs from the previous one in that tube renewal from chain end $0$ 
has passed over the original position of segment $k$. The thus created 
part of the new tube necessarily contains the new position of segment $j$, 
and the random walk connecting $j(t)$ to $k$ has 
$(j_{<} - k) + (j_{<} - j(t))$ steps. We thus find the contribution
\begin{eqnarray*}
\overline{\Theta(k\!-\!j(t))\;\Theta(j_{<}\!-\!k\!-\!1)\; 
\Theta(j_{>}\!-\!j_{<}) 
\;\;e^{- \bar{q}^{2}(2 j_{<} - k - j(t))}}~.
\end{eqnarray*}
The other cases compatible with $j_{>} \geq j_{<}$ are given by the relations 
$~[~j (t) \leq k,~ j_{>} < j (t)~]~$, $~[~j (t) > k,~ j_{<} \leq j (t),~ 
j_{>} \geq k~]~$, $~[~j (t) > k,~ j_{<} > j (t)~]~$, and
$~[~j (t) > k,~ j_{>} < k~]~$. 
Proceeding as above, we after some manipulations with the 
$\Theta$-functions arrive at the result
\begin{eqnarray}
S^{(T)}(q,t;j,k,N) &=& {\cal S}_{1} (q,t;j,k,N) + {\cal S}_{2} (q,t;j,k,N) 
\nonumber\\
&&+\; {\cal S}_{3} (q,t;j,k,N)~,
\end{eqnarray}
where 
\begin{equation}
{\cal S}_{1} (q,t;j,k,N) = \overline{e^{\textstyle - \bar{q}^{2}|k-j(t)|}}
\end{equation}
is the contribution ignoring $j_{<}$, $j_{>}$, thus ignoring all end effects. 
${\cal S}_{2}$ corrects ${\cal S}_{1}$ for the constraint $j_{>} \geq j_{<}$ 
\begin{equation}
{\cal S}_{2} (q,t;j,k,N) = \overline{
\left[ \Theta \left(j_{>} - j_{<}\right)-1\right] \;
e^{\textstyle - \bar{q}^{2}|k-j(t)|}}~,
\end{equation}
and ${\cal S}_{3}$ takes the newly created parts of the tube into account 
\end{multicols}
\begin{eqnarray}
{\cal S}_{3} (q,t;j,k,N)
&=& \overline{\Theta(k\!-\!j(t))\; \Theta(j_{<}\!-\!k-1) \;
\Theta(j_{>}\!-\!j_{<})\;
\left[e^{\textstyle - \bar{q}^{2}(2 j_{<}\!-\!k\!-\!j(t))} 
- e^{\textstyle - \bar{q}^{2}(k\!-\!j(t))}\right]}
\nonumber \\
& & ~+\; \overline{\Theta(j(t)\!-\!k\!-\!1)\; \Theta(j_{<}\!-\!j(t)\!-\!1)\; 
\Theta(j_{>}\!-\!j_{<})\;
\left[e^{\textstyle - \bar{q}^{2}(2 j_{<}\!-\!k\!-\!j(t))} 
- e^{\textstyle - \bar{q}^{2}(j(t)\!-\!k)}\right]}
\nonumber \\
& & ~+\; \overline{\Theta(k\!-\!j(t)) \;\Theta(j(t)\!-\!j_{>}-1)\; 
\Theta(j_{>}\!-\!j_{<})\;
\left[e^{\textstyle - \bar{q}^{2}(k\!+\!j(t)\!-\!2 j_{>})} 
- e^{\textstyle - \bar{q}^{2}(k\!-\!j(t))}\right]}
\nonumber \\
& & ~+\; \overline{\Theta(j(t)\!-\!k\!-\!1)\; \Theta(k\!-\!j_{>}\!-\!1)\; 
\Theta(j_{>}\!-\!j_{<})\; 
\left[e^{\textstyle - \bar{q}^{2}(k\!+\!j(t)\!-\!2 j_{>})} 
-e^{\textstyle - \bar{q}^{2}(j(t)\!-\!k)}\right]}~.
\end{eqnarray}
\begin{multicols}{2}
These expressions are formally exact, but, as pointed out in the 
previous section, to evaluate ${\cal S}_{2}$ and ${\cal S}_{3}$, 
we have to construct an approximation for the simultaneous distribution 
of $n(j,t)$, $j_{<}$, and $j_{>}$. ${\cal S}_{1}$ could be evaluated 
with the exactly known distribution of $n(j,t)$ (Eq.\ (3.3)). 
However, being interested in the universal features of the model, 
which only show up for larger times, we use the Gaussian approximation 
(3.11) and ignore the discreteness of the elementary hopping process. 
We also will take the chain as continuous, in the evaluation replacing 
segment summations by integrals. A priori these simplifications might 
influence  the short time behavior, but in practice they are found 
to have no measurable effects. For a check, we numerically have compared 
the continuous model to a fully discrete evaluation. For the properly 
normalized coherent structure function $S_{c}(q,t;M,N)/S_{c}(q,0;M,N)$, 
the difference for all times, including the microscopic range, is found 
to be of the order $10^{-3}$ and thus negligible. 

\subsection{The contribution ${\cal S}_{1}(q,t;j,k,N)$}

Combining equations (4.2) and (4.6) with the definition (3.3)
of ${\cal P}_1$, we find
\begin{equation}
{\cal S}_{1}(q,t;j,k,N) = \sum^{+ \infty}_{n=-\infty}
e^{\textstyle - \bar{q}^{2} |k - j - \bar{\ell}_s n|} \;
{\cal P}_{1} (n;j,t)~.
\end{equation}
With the Gaussian approximation (3.11) for ${\cal P}_{1}(n,j,t)$ 
and with $n$ taken continuous, this expression is easily evaluated to yield 
\begin{eqnarray}
\lefteqn{{\cal S}_{1}(q,t;j,k,N) }\\
&=& \frac{1}{2}\:e^{Q^{2}} \Big[ e^{2 \Delta Q}\:
\mbox{erfc}\: (Q + \Delta)
+ e^{- 2 \Delta Q}\:\mbox{erfc}\: (Q - \Delta)\Big]~,\nonumber
\end{eqnarray}
where 
\begin{eqnarray}
Q &=& \bar{q}^{2} \bar{\ell}_s\:\sqrt{\rho_{0} A_{1}(j,t)}  
\nonumber \\
\Delta &=& \frac{k-j}{2 \bar{\ell}_s \sqrt{\rho_{0} A_{1}(j,t)}} \: \: \:.
\end{eqnarray}
Note that Eqs.\ (3.2) and (3.7) imply
\begin{eqnarray*}
\sqrt{\rho_{0} A_{1}(j,t)} \sim 
\left\langle~ \overline{(r_{j}(t) - r_{j}(0))^{2}}~\right\rangle \: \: \:,
\end{eqnarray*}
so that in the result (4.10), $q$ and $k-j$ are measured relative 
to the mean displacement of segment $j$. Even though the Gaussian 
approximation from its derivation holds only for $\rho_{0} A_{1} \gg 1$, 
it for $\rho_{0} A_{1} \rightarrow 0$, i.e., $t \rightarrow 0$, 
reproduces the exact static behavior of our model:
\begin{equation}
{\cal S}_{1}(q,0;j,k,N) = e^{\textstyle - \bar{q}^{2} |k-j|} \: \: \:.
\end{equation}
As will be discussed in Sect.\ V.E, this is an important requirement 
for any theory of the dynamic scattering functions.

\subsection{Distribution function for ${\cal S}_{3}(q,t;j,k,N)$}

In view of the symmetry of the chain under reflection 
$j \rightarrow N -j$, the last two terms in Eq.\ (4.8) for ${\cal S}_{3}$, 
when summed over $j$ and $k$, yield contributions identical 
to the first two terms. We therefore can restrict the analysis 
to the first terms, which involve the distribution function referred 
to in Sect.\ III.B (Eq.\ (3.20)): 
\begin{eqnarray*}
{\cal P}^{(T)}_{\mbox{{\scriptsize max}},j} (n_{m},n_{j};t) = 
\overline{\Theta (j_{>} - j_{<})\; 
\delta_{n_{m},n_{\mbox{{\scriptsize max}}}(0,t)} \;\delta_{n_{j},n(j,t)}}~.
\end{eqnarray*}
For instance, in terms of this distribution function, the contribution 
to the coherent scattering function $S_{c}(q,t,N)$ of the first term 
in Eq.\ (4.8) reads
\begin{eqnarray*}
\lefteqn{
\sum_{j,k} \sum_{n_{m},n_{j}} \Theta (k\!-\!j\!-\!\bar{\ell}_s n_{j}) \;
\Theta (\bar{\ell}_s n_{m}\!-\!k\!-\!1)\;
          {\cal P}^{(T)}_{{\scriptsize \mbox{max}},j}(n_{m},n_{j};t)
           }
\\
& & \cdot \left[e^{\textstyle - \bar{q}^{2} 
(2 \bar{\ell}_s n_{m} - k - j - \bar{\ell}_s n_{j})} - 
e^{\textstyle - \bar{q}^{2} (k - j - \bar{\ell}_s n_{j})} \right]~.
\end{eqnarray*}
We now construct an approximate expression for 
${\cal P}^{(T)}_{{\scriptsize \mbox{max}},j}$, based on random walk theory.
We first present the essential steps of our approach and 
discuss the approximations involved thereafter. 
Some details of the calculations are given in Appendix A.

We introduce the auxiliary variable $n_{0} = - n (0,t)$ and write      
\begin{eqnarray}
\lefteqn{
{\cal P}^{(T)}_{{\scriptsize \mbox{max}},j} (n_{m},n_{j};t)}
\nonumber\\
 &=& \sum^{+ \infty}_{n_{0}= - \infty}\:
\overline{\Theta (j_{>} - j_{<}) \;
\delta_{n_{m},n_{\mbox{{\scriptsize max}}}(0,t)}\; \delta_{n_{j},n(j,t)}\; 
\delta_{n_{0},-n (0,t)}}
\nonumber \\
 &=& \sum^{+ \infty}_{n_{0}= - \infty}\:
{\cal P}^{(T)}_{\mbox{{\scriptsize max}},0,j} (n_{m},n_{0},n_{j};t)~.
\end{eqnarray}
Recall that $n (0,t)$ is the change in the occupation of reservoir $0$
within time $t$. We then factorize according to 
\begin{eqnarray}
\lefteqn{
{\cal P}^{(T)}_{\mbox{{\scriptsize max}},0,j} (n_{m},n_{0},n_{j};t)}
\\
&& \rightarrow 
\frac{{\cal P}^{(T)}_{\mbox{{\scriptsize max}},0} (n_{m},n_{0};t)\; 
{\cal P}_{2}(n_{0},n_{j};0,j,t)}{{\cal P}_{1}(n_{0};0,t)}
\nonumber
\end{eqnarray}
Here ${\cal P}^{(T)}_{\mbox{{\scriptsize max}},0} (n_{m},n_{0};t)$ 
is the simultaneous distribution of 
$n_{\mbox{{\scriptsize max}}}(0,t)$ and $-n (0,t)$, with the constraint 
$j_{>} \geq j_{<}$ taken into account. 
${\cal P}_{2}(n_{0},n_{j};0,j,t)$ is the simultaneous distribution 
of $- n (0,t)$ and $n(j,t)$, so that 
${\cal P}_{2}(n_{0},n_{j};0,j,t)/{\cal P}_{1}(n_{0};0,t)$ 
is the conditional probability to find $n_{j} = n(j,t)$, 
once $n_{0} = - n (0,t)$ is given. A rigorous expression for 
${\cal P}_{2}$ was given in \cite{Z7}, Sect.\ III. 
Here we again use the Gaussian approximation (\cite{Z7}, 
appendix C, Eq.\ (C.7)). Using also the Gaussian approximation 
(3.11) for ${\cal P}_{1}(n_{0};0,t)$ we find
\begin{eqnarray}
\lefteqn{
\frac{{\cal P}_{2}(n_{0},n_{j};0,j,t)}{{\cal P}_{1}(n_{0};0,t)} 
= \left(2 \pi \overline{n^{2}(j,t)}\right)^{-1/2} 
\left(1 - a^{2}\right)^{-1/2}}
\nonumber \\
&&~~~~~~~\cdot \exp \left[ - \frac{1}{2(1-a^{2})} 
\left(a^{2} z_{0}^{2} + z_{j}^{2} - 2 a z_{0} z_{j}\right)\right]~,
\end{eqnarray}
where 
\begin{eqnarray}
z_{j} &=& \frac{n_{j}}{\sqrt{\overline{n^{2}(j,t)}}}, ~\: \:   
z_{0} = \frac{n_{0}}{\sqrt{\overline{n^{2}(0,t)}}}
\\
&&a = a (j,t) = \frac{\tilde{A}_{3} (j,t)}
{\left(A_{1}(0,t)\; A_{1}(j,t)\right)^{1/2}}
\end{eqnarray}
with [cf.\ Ref.\ \cite{Z7}, Eq.\ (A.12)]
\begin{eqnarray}
\lefteqn{
\tilde{A}_{3} (j,t) = \frac{p t}{N} + \frac{N}{3} - \frac{1}{2} 
+ \frac{1}{6N} + \frac{j^{2}}{2N} - \left(1 - \frac{1}{2N}\right) j}
\nonumber \\
&&~~~~- \frac{1}{2N}\:\sum^{N-1}_{\kappa = 1}\:
\frac{\cos \left(\frac{\pi \kappa}{2N}\right) 
      \cos \left(\frac{\pi \kappa}{N} \left(j + \frac{1}{2}\right) \right)}
     {\sin^{2} \left(\frac{\pi \kappa}{2N}\right)}\; \alpha^{t}_{\kappa}~.
\end{eqnarray}
$A_{1}$ and $\alpha_{\kappa}$ are given  in Eqs.\ (3.5), (3.6), respectively.

This introduces the parameter function $a = a(j,t)$. It measures 
the coupling of the motion of bead $j$ to the motion of chain end $0$. 
If it vanishes, the conditional probability (4.15) reduces to 
${\cal P}_{1}(n_{j};j,t)$. This happens for $p t \ll j^{2}$ 
(see Ref.\ \cite{Z7}, Eq.\ (A.16)). The maximal value of $a$ is $1$, 
which is approached for $j \rightarrow 0$ and all $t$, or for $t \gg T_{2}$ 
and all $j$. In the latter limit, Eq.\ (4.15) yields $z_{j} = z_{0}$, 
and the motion of all segments is rigidly coupled to the motion of the 
end-segment. In this limit, we thus recover the basic assumption 
of the primitive chain model.

To find an acceptable functional form for 
${\cal P}^{(T)}_{\mbox{{\scriptsize max}},0} (n_{m},n_{0};t)$, 
we replace the correlated stochastic process $n (0,s)$ by a random walk 
$n'(s)$ on the integer numbers, with hopping rate $p'$.
We have to consider walks that start at $n'(0) = 0$, end at 
$n'(t) = n_{0}$, and attain the maximal value $n_{m} \geq n_{0}$ 
for some $s \in [0,t]$. To take care of the constraint
\begin{equation}
\frac{1}{\bar{\ell}_{s}} (j_{>} - j_{<}) 
= N' - n_{\mbox{{\scriptsize max}}}(N,t) 
- n_{\mbox{{\scriptsize max}}}(0,t) \geq 0~,
\end{equation}
we restrict the walk $n'(t)$ to the interval $[n_{m} - N' + 1,n_{m}]$, 
where $N'$ is the greatest integer less than $N/\bar{\ell}_s$, 
and we use absorbing boundary conditions. This amounts to the assumption 
that a particle entering the chain from the reservoir at chain end $N$ 
is transfered immediately to the reservoir at chain end $0$.
This assumption is in the spirit of the primitive chain model.

With these simplifications, 
${\cal P}^{(T)}_{\mbox{{\scriptsize max}},0}$ can be calculated 
as sketched in Appendix A. Our result reads
\begin{eqnarray}
\lefteqn{
{\cal P}^{(T)}_{\mbox{{\scriptsize max}},0} (n_{m},n_{0};t) }
\nonumber\\
&=& \Theta \left(n_{m}\!-\!\frac{|n_{0}|\!+\!n_{0}}{2}\right) \;
\Theta \left(N'\!-\!n_{m}\!-\!\frac{|n_{0}|\!-\!n_{0}}{2}\right) \;
\frac{1}{\sqrt{\pi} p' t}
\nonumber \\
& &
 \cdot\sum^{+\infty}_{\nu = - \infty}\:
\left\{ (\nu + 1) \;
\left(\frac{\nu N'}{\sqrt{p' t}} + \frac{n_{m}}{\sqrt{p' t}} 
      - \frac{n_{0}}{2 \sqrt{p' t}}\right)\right.\;
\nonumber\\ 
&&\qquad\qquad\qquad\cdot
\exp \left[ - \left(\frac{\nu N'}{\sqrt{p' t}} + \frac{n_{m}}{\sqrt{p' t}} 
           - \frac{n_{0}}{2 \sqrt{p' t}}\right)^{2} \right]
\nonumber \\
&&\qquad\qquad\:
-\;\nu\;\left(\frac{\nu N'}{\sqrt{p't}}+\frac{n_0}{2\sqrt{p't}}\right)\; 
\nonumber\\
&&\qquad\qquad\qquad\cdot\left.
\exp \left[-\left(\frac{\nu N'}{\sqrt{p't}}+\frac{n_0}{2\sqrt{p't}}\right)^2 
            \right] \right\}~.
\end{eqnarray}
It is valid for $p' t \gg 1$ and $N' \gg 1$, which is the region 
of interest here. We now note that for $p' t \gg 1$, the relations
\begin{eqnarray}
\overline{n'^{2}(t)} &=& 2 p' t
\nonumber \\
\bar{n}_{m} = \max_{s \in [0,t]}\: n'(s) &=& 2 \sqrt{\frac{p' t}{\pi}}
\end{eqnarray}
hold, and we use these relations to eliminate the factors $\sqrt{p' t}$:
\begin{eqnarray*}
\frac{n_{0}}{\sqrt{p' t}} 
&=& \sqrt{2}\:\frac{n_{0}}{\sqrt{\overline{n_{0}^{2}(t)}}} = \sqrt{2}\:z_{0}
\\
\frac{n_{m}}{\sqrt{p' t}} 
&=& \sqrt{2}\:\sqrt{\frac{2}{\pi}}\:\frac{n_{m}}{\bar{n}_{m}} = \sqrt{2}\:y
\\
\frac{N'}{\sqrt{p' t}} 
&=& \sqrt{2}\:\sqrt{\frac{2}{\pi}}\:\frac{N}{\bar{\ell}_s \bar{n}_{m}} 
= \sqrt{2}\:\hat{N} \: \: \:.
\end{eqnarray*}
In the last line, we used $\bar{n}_{m}$, since $N'$ is introduced 
via the constraint (4.19). Furthermore we write the prefactor in Eq.\ (4.20) as
\begin{eqnarray*}
\frac{1}{\sqrt{\pi} \;p' t} = \frac{2}{\sqrt{\pi}}\:
\left(\overline{n'^{2}(t)}\right)^{-1/2} 
\left(\sqrt{\frac{\pi}{2}}\:\bar{n}_{m}\right)^{-1} \: \: \:,
\end{eqnarray*}
and we again treat $z_0$ and $y$ as continuous variables, 
which for $p' t \gg 1$ is a valid approximation consistent 
with our derivation. With these substitutions, Eq.\ (4.20) reads 
\begin{eqnarray}
\lefteqn{
dn_0\;dn_m\;{\cal P}^{(T)}_{\mbox{{\scriptsize max}},0} (n_{m},n_{0};t) }
\nonumber\\
&=& \Theta \left(y\!-\!\frac{|z_{0}|\!+\!z_{0}}{2}\right) \;
\Theta \left(\hat{N}\!-\!y\!-\!\frac{|z_{0}|\!-\!z_{0}}{2}\right)\;
\sqrt{\frac{2}{\pi}}\;dz_{0}\;dy
\nonumber \\
& &
\cdot \sum^{+\infty}_{\nu = - \infty}\!\!\!
\Bigg\{(\nu\!+\!1)\; (2 \nu \hat{N}\!+ 2 y\!-\!z_{0}) \; 
\exp \left[ - \frac{(2 \nu \hat{N}\!+\!2 y\!-\!z_{0})^{2}}{2}\right] 
\nonumber \\
& & \qquad\qquad- \;\nu \;(2 \nu \hat{N}\!+\!z_{0})\:\exp 
\left[ - \frac{(2 \nu \hat{N}\!+\!z_{0})^{2}}{2}\right]\Bigg\}~.
\end{eqnarray}
We use this result which has been derived for a random walk, 
also for the correlated process $n(0,s)$ by reinterpreting 
the variables. $z_0$ is given by Eq.\ (4.16), and
\begin{equation}
y = \sqrt{\frac{2}{\pi}}\:\frac{n_{\mbox{{\scriptsize max}}}(0,t)}
{\;\overline{n_{\mbox{\scriptsize max}}(0)}\;}
\end{equation}
\begin{equation}
\hat{N} = N/c \: \: \:,
\end{equation}
with
\begin{equation}
c = c (t) = \sqrt{\frac{\pi}{2}}\:\bar{\ell}_s\:
\overline{n_{\mbox{{\scriptsize max}}}(0,t)} \: \: \:.
\end{equation} 
For $\overline{n_{\mbox{{\scriptsize max}}}(0,t)}$, the approximation 
(3.16) is used. Up to the factor $\sqrt{\frac{\pi}{2}}$, the parameter 
$c$ gives the distance up to which the tube has been destroyed on average. 
We now use expressions (4.22), (4.15), (4.14) to evaluate Eq.\ (4.13), 
where the sum over $n_0$ has to be replaced by the integral over $z_0$. 
Some exercise in Gaussian integrals yields our final result: 
\end{multicols}
\begin{equation}
d n_{j} \;d n_{m}\; 
{\cal P}^{(T)}_{\mbox{{\scriptsize max}},j} (n_{m},n_{j};t) 
= \frac{dz_{j}\:dy}{\sqrt{2 \pi}}\;\Theta(y)\; \Theta (\hat{N}-y) 
\sum^{+\infty}_{\nu = - \infty}
{\cal P}^{(\nu)}_{\mbox{{\scriptsize max}},j} (y,z_{j},a)
\end{equation}      
\begin{eqnarray}
\lefteqn{
{\cal P}^{(\nu)}_{\mbox{{\scriptsize max}},j} (y,z,a)= }
\nonumber\\
&\displaystyle (\nu + 1)\:
\exp \left[ - \frac{\textstyle (2a\nu\hat{N}\!+\!2ay\!-\!z)^2}{2}\right]
& \left\{ a (2 a \nu \hat{N}\!+\!2 a y\!-\!z) 
\left[ \mbox{erfc}\left(
\frac{a z\!-\!y\!+\!a_2 y\!+\!a_2\nu\hat{N}}{\sqrt{a_2}}\right)
- \mbox{erfc}
\left(\frac{a z\!-\!y\!+\!a_2 y\!+\!a_2 \nu\hat{N}\!+\!\hat{N}}
{\sqrt{a_2}}\right) 
\right]\right.
\nonumber \\
& & \quad+\; \sqrt{\frac{a_2}{\pi}} \;
\left.\left[ \exp \left( - \frac{\textstyle 
(a z\!-\!y\!+\!a_2 y\!+\!a_2\nu\hat{N})^2}{a_2}\right)
- \exp \left( - \frac{ \textstyle
(a z\!-\!y\!+\!a_2 y\!+\!a_2 \nu\hat{N}\!+\!\hat{N})^2}{a_2}
\right) 
\right] \right\}
\nonumber \\
&\displaystyle
-\;\nu\:\exp \left[ - \frac{\textstyle (2a\nu\hat{N}\!-\!z)^2}{2}\right] 
& \Bigg\{ a (2 a \nu \hat{N}\!-\!z)\;
\left[\mbox{erfc}\left(\frac{a z\!-\!y\!+\!a_2\nu\hat{N}}{\sqrt{a_2}}\right) 
- \mbox{erfc}\left(\frac{a z\!-\!y\! 
+\!a_2\nu \hat{N}\!+\!\hat{N}}{\sqrt{a_2}}\right)\right]
\nonumber \\
& & \quad+\; \sqrt{\frac{a_2}{\pi}}\; 
\left.\left[ \exp \left( - \frac{ \textstyle
(a z\!-\!y\!+\!a_2\nu\hat{N})^2}{a_2}\right)
- \exp \left( - \frac{ \textstyle
(a z\!-\!y\!+\!a_2\nu\hat{N}\!+\!\hat{N})^2}{a_2}\right)
\right] \right\}~,
\nonumber\\
&\displaystyle \mbox{ with the notation }&a_2=2(1-a^2)~.
\end{eqnarray}
\begin{multicols}{2}

Clearly, the approximations inherent in our construction of 
${\cal P}^{(T)}_{\mbox{{\scriptsize max}},j}$ need some justification. 
Steps like the replacement of discrete by continuous variables 
are well justified, since we need the result only in a time and 
chain-length regime where a continuous chain model is valid. 
The problematic steps are the factorization (4.14) of 
${\cal P}^{(T)} _{\mbox{{\scriptsize max}},0,j}$ and the calculation 
of the functional form of ${\cal P}^{(T)}_{\mbox{{\scriptsize max}},0}$ 
by random walk theory. 

Technically, the factorization (4.14) serves to reduce the problem 
to the treatment of the single stochastic process $n (0,s)$. 
It clearly is justified for large times, $t \gg T_{2}$, 
where $n (j,t)$ is firmly bound to $n (0,t)$, and where
${\cal P}^{(T)}_{\mbox{{\scriptsize max}},j}$ becomes equivalent 
to ${\cal P}^{(T)}_{\mbox{{\scriptsize max}},0}$. For shorter times 
$t \alt T_{2}$, it assumes that $- n(0,t)$ is a good measure of 
$n_{\mbox{{\scriptsize max}}}(0,t)$, which is certainly incorrect, 
in particular for $t \ll T_{2}$. However, for $t \ll T_{2}$, end effects 
influence only a small part of the chain. As can be seen from Eqs.\ 
(3.16), (3.7) and has been explicitly worked out in Ref.\ \cite{Z7}, 
Eq.\ (5.28), $\overline{n_{\mbox{{\scriptsize max}}}(0,t)}$ 
for $t \ll T_{2}$ behaves as $\overline{n_{\mbox{{\scriptsize max}}}(0,t)} 
\sim (p t)^{1/4} \sim N^{1/2} (t/T_{2})^{1/4} \ll N$. Since the coherent 
scattering function sums over all segments, it for such times is dominated 
by the motion of interior segments not influenced by end-effects 
and governed by the distribution function ${\cal P}_{1} (n;j,t)$. 
It is easily verified that in the appropriate limit 
$\hat{N} = N/c(t) \rightarrow \infty$, the distribution function (4.26), 
when integrated over $n_{m}$, reduces to the Gaussian approximation 
for ${\cal P}_{1} (n;j,t)$. (Note that in this limit only a part 
of the $\nu = 0$ contribution to the sum in Eq.\ (4.26) survives.)

We now turn to our construction of 
${\cal P}^{(T)} _{\mbox{{\scriptsize max}},0} (n_{m},n_{0})$. 
Our treatment of the constraint (4.19) should be adequate, 
since this constraint is relevant only for times of the order 
of the reptation time, $t \approx T_{3}$, where the internal degrees 
of freedom of the chain are irrelevant. Furthermore, 
${\cal P}^{(T)} _{\mbox{{\scriptsize max}},0}$ by construction 
obeys the constraint $n_{0} \leq n_{m}$, and integrating over 
$n_{0}$, we find a distribution with the desired first moment 
$\overline{n_{m}} = \overline{n_{\mbox{{\scriptsize max}}}(0,t)}$. 
Integrating over $n_{m}$, we find the correct (Gaussian) distribution 
of $n_{0}$. With these three important features guaranteed, 
we may hope that we have found a good approximation for the distribution 
function of the full correlated process.

To summarize, our construction interpolates among two limits where 
the full dynamics reduces to that of a single stochastic variable. 
For $t \ll T_{2}$, the motion of individual segments governed by 
$n (j,t)$, is essential. For $t \gg T_{2}$, the parameter function $a$ 
tends to 1 and the internal motion 
becomes irrelevant. Furthermore, the correlations of the stochastic process 
$n (0,s)$ have died out. We thus are concerned with a single uncorrelated 
process $n (0,s)$, as in the primitive chain model. 
Smoothly interpolating among these limits, we may hope to have found 
a good approximation also in the crossover region $t \approx T_{2}$. 
Indeed, as shown in Figs.~10 and 11 of Ref.\ \cite{Z8}, a simplified 
version of our theory almost quantitatively fits with Monte Carlo data 
for the motion of individual segments. Furthermore, as will be illustrated 
in Sect.\ VIII and in more detail in Ref.\ \cite{Z13}, 
our theory quantitatively accounts for data for the coherent scattering 
function $S_{c} (q,t;M,N)$. The agreement is equally good 
for the total chain $(M=N)$ where tube length fluctuations are 
very important, and for an interior piece $(M < N)$ where tube length 
fluctuations are irrelevant.      
    
\subsection{Distribution function for ${\cal S}_{2}(q,t;j,k,N)$}

To evaluate ${\cal S}_{2}(q,t;j,k,N)$ (Eq.\ (4.7)), 
we need the distribution function
\begin{eqnarray}
{\cal P}^{(T)}_{j}(n_{j};t) &=& 
\overline{\left[ \Theta (j_{>} - j_{<}) - 1\right]\; \delta_{n_{j},n(j,t)}}
\nonumber \\
&=& \sum^{\infty}_{n_{m}=0}\:
{\cal P}^{(T)}_{\mbox{{\scriptsize max}},j} (n_{m},n_{j};t) 
- {\cal P}_{1} (n_{j};j,t)~.
\end{eqnarray}
The first part can be determined by integrating 
${\cal P}^{(T)}_{\mbox{{\scriptsize max}},j}$ (Eq.\ (4.26)) over $n_{m}$. 
Eqs.\ (4.26)--(4.28) thus yield
\begin{equation}
d n_{j}\; {\cal P}^{(T)}_{j}(n_{j};t) 
= \frac{d z_{j}}{\sqrt{2 \pi}}\:
\sum^{+ \infty}_{\nu = - \infty}\:{\cal P}^{(\nu)}_{j}(z_{j},a)
\end{equation}  
\begin{equation}
{\cal P}^{(\nu)}_{j}(z_{j},a) 
= \int_{0}^{\hat{N}}d y\:
{\cal P}^{(\nu)}_{\mbox{{\scriptsize max}},j} (y,z_{j},a) 
- \delta_{\nu,0}\:\exp ( - z^{2}_{j}/2) ~.
\end{equation} 
The $y$-integral can be evaluated analytically and we quote the result, 
which is useful for the numerical evaluation of ${\cal S}_{2}$, in Appendix B.
    
\subsection{Result for the tube conserving contribution 
to the coherent structure function $S_{c}^{(T)} (q,t;M,N)$}

We consider the scattering from the $M+1$ central beads (Eq.\ (3.18)) 
and write
\begin{eqnarray}
S_{c}^{(T)}(q,t;M,N) &=& {\cal S}_{1} (q,t;M,N) + {\cal S}_{2} (q,t;M,N) 
\nonumber\\
&&+ {\cal S}_{3} (q,t;M,N)~,
\end{eqnarray}
where the ${\cal S}_{i} (q,t;M,N)$, $i = 1,2,3$, are the contributions 
${\cal S}_{i} (q,t;j,k,N)$ (Eq.\ (4.5)), summed over $j$ and $k$. 
The superscript $^{(T)}$ again recalls the constraint (4.1): 
$j_{>} - j_{<} \geq 0$.

Due to this constraint the relation
\begin{equation}
S_{c} (q,t;M,N)  = S_{c}^{(T)}(q,t;M,N)
\end{equation}
in general holds only for $t \ll T_{3}$. However, for large $q$ such that 
$q^{2} R_{g}^{2} \gg 1$, contributions in which the tube has been 
destroyed, contribute negligibly to $S_{c} (q,t;M,N)$, so that 
Eq.\ (4.32) in this limit holds for all times. 

Consider now the first contribution. 
\begin{eqnarray*}
{\cal S}_{1} (q,t;M,N) = \int^{\frac{N+M}{2}}_{\frac{N-M}{2}}dj\;dk\;
{\cal S}_{1} (q,t;j,k,N) 
\end{eqnarray*}
Using Eqs.\ (4.10) and (4.11), we can carry out the integral over $k$ to find  
\begin{eqnarray}
\lefteqn{{\cal S}_{1} (q,t;M,N)= \frac{1}{2\bar{q}^{2}}\:
\int^{\frac{N+M}{2}}_{\frac{N-M}{2}}dj}
\nonumber\\
&&\quad\cdot \Big[ e^{2 \Delta_{1} Q + Q^{2}}\:
\mbox{erfc}(Q\!+\!\Delta_{1}) - e^{- 2 \Delta_{1} Q + Q^{2}}\:
\mbox{erfc}(Q\!-\!\Delta_{1})
\nonumber\\
& &\quad\quad-\; e^{2 \Delta_{2} Q + Q^{2}}\:
\mbox{erfc}(Q\!+\!\Delta_{2}) + e^{- 2 \Delta_{2} Q + Q^{2}}\:
\mbox{erfc}(Q\!-\!\Delta_{2})
\nonumber\\ 
& & \quad\quad+\;2\:\mbox{erfc}\Delta_{2}-2\:\mbox{erfc}\Delta_{1}\Big]~,
\end{eqnarray}
where 
\begin{eqnarray}
Q &=& \bar{q}^{2} \bar{\ell}_{s} \sqrt{\rho_{0} A_{1}(j,t)}
\nonumber \\
\Delta_{1} &=& \frac{1}{2 \bar{\ell}_{s}}\; \left(\frac{N+M}{2} - j\right)\; 
(\rho_{0}  A_{1}(j,t))^{-1/2}
\nonumber \\
\Delta_{2} &=& \frac{1}{2 \bar{\ell}_{s}} \;\left(\frac{N-M}{2} - j\right)\; 
(\rho_{0}  A_{1}(j,t))^{-1/2}~.
\end{eqnarray}
The remaining integration in general must be done numerically, 
due to the $j$-dependence of $A_{1}(j,t)$. 

The integral over $k$ can be carried out also in ${\cal S}_{2},{\cal S}_{3}$. 
We introduce the notation
\begin{eqnarray}
&&\hat{q} = \bar{q}^{2} c; \: \: \: \: \hat{j} = j/c;\: \: \: \: \hat{M} = M/c
\\
&&b = \frac{\bar{\ell}_s}{c}\:
\overline{\left(n^2\left({\scriptstyle\frac{N-M}{2}}+j,t\right)\right)}^{1/2}~,
\end{eqnarray}
where the parameter $c$ has been defined in Eq.\ (4.25). 
With due regard of the $\Theta$-functions, 
a straightforward calculation yields
\end{multicols}
\begin{equation}
{\cal S}_{2} (q,t;M,N) = \frac{c^{2}}{\hat{q}}\:
\int_{0}^{\hat{M}}\:d \hat{j}\:
\left({\cal S}_{2}^{(1)} + {\cal S}_{2}^{(2)}\right)
\end{equation}
\begin{eqnarray}
{\cal S}_{2}^{(1)} &=& \frac{1}{b}\:\left(1 - e^{- \hat{q} \hat{M}}\right)\:
\int_{0}^{\infty}\!\!\!d z\:e^{- \hat{q} z}\:\frac{1}{\sqrt{2\pi}}
\sum_{\nu} \left[ 
{\cal P}_{j}^{(\nu)} \left(- \frac{z + \hat{j}}{b}, a\right)
+ {\cal P}_{j}^{(\nu)} \left(\frac{z - \hat{j}+\hat{N}}{b},a\right)\right]\\
{\cal S}_{2}^{(2)} &=& \frac{1}{b}\: \int_{0}^{\hat{M}}\!\!\!d z\:
\left[ 2 - e^{- \hat{q} z} - e^{- \hat{q}(\hat{M} - z)}\right] 
\frac{1}{\sqrt{2\pi}}
\sum_{\nu}\:{\cal P}_{j}^{(\nu)} \left(\frac{z-\hat{j}}{b}, a\right) 
\end{eqnarray}
\begin{equation}
{\cal S}_{3} (q,t;M,N) = \frac{2 c^{2}}{\hat{q}}\:\int_{0}^{\hat{M}}\:d  
\hat{j}\:\left({\cal S}_{3}^{(1)} + {\cal S}_{3}^{(2)} + {\cal S}_{3}^{(3)} 
+ {\cal S}_{3}^{(4)}\right)
\end{equation} 
\begin{eqnarray}
{\cal S}_{3}^{(1)} &=& \frac{1}{b}\:
\int_{0}^{(\hat{N} + \hat{M})/2}\!\!\!\!\!\!\!\!\!d y\:y\: 
\int_{0}^{1}\!\!\!d z\:
\left[ 2 e^{- \hat{q} y (1-z)} - 2 + e^{- \hat{q} y z} 
        - e^{- \hat{q} y (2-z)}\right]\;
\frac{1}{\sqrt{2\pi}}\:\sum_{\nu}\:
{\cal P}_{\mbox{{\scriptsize max}},j}^{(\nu)} 
\left( y + \frac{\hat{N} - \hat{M}}{2}, \frac{y}{b}\:z 
        - \frac{\hat{j}}{b}, a\right)
\\
{\cal S}_{3}^{(2)} &=& - \frac{1}{b}\:
\int_{0}^{(\hat{N} + \hat{M})/2}\!\!\!\!\!\!\!\!\!d y\: 
\int_{0}^{\infty}\!\!\!d z\: e^{- \hat{q} z} \;
\left(1 - e^{- \hat{q} y}\right)^2
\frac{1}{\sqrt{2\pi}}\:\sum_{\nu}\:
{\cal P}_{\mbox{{\scriptsize max}},j}^{(\nu)} 
\left( y + \frac{\hat{N} - \hat{M}}{2}, - \frac{z + \hat{j}}{b}, a\right)\\
{\cal S}_{3}^{(3)} &=& - \frac{1}{b}\:
\int_{0}^{(\hat{N} - \hat{M})/2}\!\!\!\!\!\!\!\!\!d y\:y\: 
\int_{0}^{1}\!\!\!d z\:
\left[ 2 e^{- \hat{q} y (1-z)} - 2 + e^{- \hat{q} y z} 
         - e^{- \hat{q} y (2-z)}\right]
\nonumber\\
&&\qquad\qquad\qquad\qquad\qquad\qquad\qquad\qquad\qquad
\cdot\frac{1}{\sqrt{2\pi}}\:\sum_{\nu}\:
{\cal P}_{\mbox{{\scriptsize max}},j}^{(\nu)} 
\left( y + \frac{\hat{N} + \hat{M}}{2}, \frac{y}{b}\:z 
           + \frac{\hat{M} - \hat{j}}{b}, a\right)
\\
{\cal S}_{3}^{(4)} &=& \frac{1}{b}\:
\int_{0}^{(\hat{N} - \hat{M})/2}\!\!\!\!\!\!\!\!\!d y\: 
\int_{0}^{\infty}\!\!\!d z\:
e^{- \hat{q} z} \;\left(1 - e^{- \hat{q} y}\right)^{2}
\frac{1}{\sqrt{2\pi}}\:\sum_{\nu}\:
{\cal P}_{\mbox{{\scriptsize max}},j}^{(\nu)} 
\left( y + \frac{\hat{N} + \hat{M}}{2}, - \frac{z}{b} 
         + \frac{\hat{M} - \hat{j}}{b}, a\right)~.
\end{eqnarray}
\begin{multicols}{2}
The prefactor of 2 in Eq.\ (4.40) accounts for the last two 
contributions in Eq.\ (4.8). We note that these results depend on time 
via the parameters $a$, $c$, and $b$. From its definition (4.36), 
the parameter function $b = b(j,t)$ measures the motion of an 
arbitrary segment relative to the motion of the end segment. 
It is weakly dependent on $j$ and tends to $1$ for $t \gg T_{2}$.

In ${\cal S}_{2}$ and ${\cal S}_{3}$, one more integration could 
be done analytically which, however, only blows up the number 
of terms without leading to any simplification. Due to the dependence 
on the segment index $j$ implicit in $a = a (j,t)$ and  $b = b (j,t)$, 
an analytical evaluation of all integrals is possible only in the limit 
$t \gg T_{2}$ where $a \rightarrow 1$ and  $b\rightarrow 1$. 
In general, we have to resort to numerical evaluation. 
In this context, we may note that the summations over $\nu$ for 
$t \alt T_{3}$ converge rapidly, so that in the range where 
$S_{c}^{(T)} (t \neq 0)/S_{c}(0)$ exceeds $10^{-3}$, 
we never need to go beyond $|\nu| \leq 4$. 


\section{Dynamics within the initial tube}

In a time region where end effects are unimportant, the results of 
the previous section can be simplified. In precise terms, 
the neglect of end effects amounts to considering a subchain 
of length $M$, in the center of an infinitely long chain. 
We here concentrate on this particular limit and compare our results 
to those derived for a Rouse chain in a coiled tube.

\subsection{Results of the reptation model}

In the limit $N \rightarrow \infty$, with $\tilde{j} = j - \frac{N}{2}$ 
and $\tilde{k} = k - \frac{N}{2}$ fixed, only the contribution 
${\cal S}_{1}$ to $S^{(T)}(q,t;j,k,N)$ (Eq.\ (4.5)) survives. 
Furthermore $A_{1}(j,t)$ (3.5) simplifies to [see Ref.\ \cite {Z7}, 
Eq.\ (4.11)]
\begin{eqnarray}
A_{1}(j,t) &=& \frac{\hat{t}^{1/2}}{\pi}\:\int_{0}^{4 \hat{t}}\:
\frac{d z}{\sqrt{z}}\:\sqrt{1 - \frac{z}{4 \hat{t}}}\: e^{-z}
\nonumber \\
&\stackrel{\hat{t} \gg 1}{\rightarrow}& \sqrt{\frac{\hat{t}}{\pi}}\: \: \:, 
\end{eqnarray}
independent of $j$. (Recall the definition $\hat{t} = p t$.) 
${\cal S}_{1}(q,t;j,k,N)$ (Eq.\ (4.10)) takes the form 
\begin{eqnarray}
\lefteqn{
{\cal S}_{1}(q,t;\tilde{j},\tilde{k})}
\\
&& = \frac{1}{2}\:e^{\hat{Q}^{2}} \Big[ e^{2 \hat{\Delta} \hat{Q}}\:
\mbox{erfc}\: (\hat{Q} + \hat{\Delta})
+ e^{- 2 \hat{\Delta} \hat{Q}}\:\mbox{erfc}\: (\hat{Q} - \hat{\Delta})\Big]~,
\nonumber
\end{eqnarray}
where now
\begin{eqnarray}
\hat{Q} &=& \bar{q}^{2} \sqrt{\bar{\ell}_{s}^{2} \rho_{0}}\:
\left(\frac{\hat{t}}{\pi}\right)^{1/4}
\nonumber \\
\hat{\Delta} &=& \frac{\tilde{k} - \tilde{j}}{2 \sqrt{\bar{\ell}_{s}^{2} 
\rho_{0}}}\:\left(\frac{\hat{t}}{\pi}\right)^{- 1/4} \: \: \:.
\end{eqnarray}
Integrating over $\tilde{j}$ and $\tilde{k}$, we find for the 
normalized dynamic structure function
\begin{eqnarray}
&&\bar S_c(q,t;M,\infty)=
\frac{S_{c}(q,t;M,\infty)}{M^{2}\; D (\bar{q}^{2} M)}
\nonumber\\ 
&&=1 - \frac{1}{\bar{q}^{2} M\; D (\bar{q}^{2} M)} 
\Bigg\{ 2 \:\mbox{erfc} 
\left( \left(\frac{\hat{T}_{4}}{\hat{t}}\right)^{1/4}\right)
\nonumber\\
&&\qquad\qquad\qquad\qquad\qquad
+ \frac{2}{\sqrt{\pi}} \left(\frac{\hat{t}}{\hat{T}_{4}}\right)^{1/2} 
\left(1 - e ^{- (\hat{T}_{4}/\hat{t})^{1/2}}\right) \Bigg\} 
\nonumber\\
&&\quad+ \;\frac{1}{(\bar{q}^{2} M)^{2}\; D (\bar{q}^{2} M)} 
\Bigg\{ 2 - 2 e^{- \bar{q}^{2} M}
\nonumber\\
&&\qquad\qquad+\; e^{\hat{Q}^{2}} 
\Bigg[ e^{\bar{q}^{2} M}\:
\mbox{erfc}\left(\hat{Q} 
+\left( \frac{\hat{T}_{4}}{\hat{t}}\right)^{1/4}\right) 
\\
&&
\qquad\qquad\quad+\; e^{- \bar{q}^{2} M}\:
\mbox{erfc} \left(\hat{Q} 
- \left(\frac{\hat{T}_{4}}{\hat{t}}\right)^{1/4}\right)
- 2 \:\mbox{erfc}(\hat{Q}) \Bigg] \Bigg\}.
\nonumber
\end{eqnarray}
We recall that $D(x) = \frac{2}{x^{2}} (e^{-x} - 1 + x)$ is the 
Debye function, and note that $\bar{q}^{2} M$ can also be written as
\begin{equation}
\bar{q}^{2} M = q^{2} R_{g}^{2} (M) \: \: \:,
\end{equation}
where $R_{g} (M)$ is the radius of gyration of the subchain. 
In Eq.\ (5.4), we introduced a new time scale
\begin{equation}
\hat{T}_{4} = \frac{\pi}{16} 
\left(\bar{\ell}_{s}^{2} \rho_{0}\right)^{-2} M^{4} \: \: \:,
\end{equation}  
which is of the order of the time the subchain needs to leave 
its original part of the tube. This interpretation is obvious 
from the results on segment motion quoted at the end of Sect.~II: 
$\left\langle (r_{j} (\hat{T}_{4}) - r_{j}(0))^{2} \right\rangle 
\sim \hat{T}_{4}^{1/4} \sim M$ for $\hat{T}_{4} \ll \hat{T}_{2}$. 
Also the variable $\hat{Q}$ can be expressed in terms of a time scale:
\begin{equation}
\hat{Q} = \left( \frac{\hat{t}}{\hat{T}_{q}}\right)^{1/4}
\end{equation}
\begin{equation}
\hat{T}_{q} = \pi \left(\bar{\ell}_s^{2} \rho_{0}\right)^{-2} 
\left(\bar{q}^{2}\right)^{-4} = 
\frac{16 \hat{T}_{4}}{(q^{2} R_{g}^{2}(M))^{4}} \: \: \:.
\end{equation}
It needs the time $\hat{T}_{q}$ before the distance diffused by the 
subchain, can be resolved by scattering of wave vector $q$. 
For comparison with previous work, we concentrate on 
$ q^{2} R_{g}^{2} (M) \gg 1$, so that $\hat{T}_{q} \ll \hat{T}_{4}$ 
and $D(x) \approx 2/x$. We then find the following limiting behavior 
in the various time regimes 
\begin{eqnarray}
&&\frac{S_{c}(q,t;M,\infty)}{M^{2}} - D \Big(q^{2} R_{g}^{2}(M)\Big) 
\\
&&~~~~~\approx \left\{
\begin{array}{ll}
- 2 \frac{1 - e^{- q^{2} R_{g}^{2}(M)}}{(q^{2} R_{g}^{2}(M))^{2}}\:
\left(\frac{\hat{t}}{\hat{T}_{q}}\right)^{1/2}~~~
& \mbox{for }\hat{t} \ll \hat{T}_{q}\\
- \frac{2}{\sqrt{\pi} q^{2} R_{g}^{2}(M)} 
\left( \frac{\hat{t}}{\hat{T}_{4}}\right)^{1/4}~~~ 
& \mbox{for }\hat{T}_{q} \ll \hat{t} \ll \hat{T}_{4}
\end{array}
\right.
\nonumber
\end{eqnarray}
\begin{equation}
\frac{S_{c}(q,t;M,\infty)}{M^{2}} \approx 
\frac{6}{\sqrt{\pi} q^{2} R_{g}^{2}(M)} 
\left(\frac{\hat{T}_{4}}{\hat{t}}\right)^{1/4}~~~\mbox{for }
\hat{T}_{4} \ll \hat{t} \: \: \:.
\end{equation}
Note that our results depend only on macroscopic parameters 
$q^{2} R_{g}^{2}(M)$ and $\hat{t}/\hat{T}_{4}$, which absorb 
any reference to the microscopic structure.

\subsection{Comparison with de Gennes' results for Rouse motion in a tube}

In Sect.\ II.B, we recalled de Gennes' results \cite{Z11}, 
derived for one dimensional Rouse-type motion in a coiled tube. 
In the derivation, the relation $q^{2} R_{g}^{2} \gg 1 \gg q^{2} \ell^{2}$ 
was assumed, with $\ell$ being the average segment size of a Gaussian chain. 
Of interest here is the `local' term $\bar{S}^{(\ell)}(q,t)$ (Eq.\ 2.12). 
A glance to the derivation shows that it implicitly exploits the limit 
considered here: a subchain (of length $M$) in an infinitely long chain. 
Combining Eqs.\ (2.10) to (2.14) we thus find 
\begin{equation} 
\frac{S_{dG}(q,t;M,\infty)}{M^{2} \;D (\bar{q}^{2} M)} 
= 1 - N_{e}\; \bar{q}^{2} \left[ 1 - e^{\textstyle t/T_{q}}\;
\mbox{erfc} \left( \sqrt{t/T_{q}} \right)\right]
\end{equation}
\begin{equation}
T_{q} = \frac{\mbox{const}}{(\bar{q}^{2})^{2}}~.
\end{equation}
Clearly this expression differs strongly from our result (5.4). 
It leads to very different asymptotics: 
\begin{eqnarray}
&&\frac{S_{dG}(q,t;M,\infty)}{M^{2}} - D \Big(q^{2} R_{g}^{2}(M)\Big) 
\\
&&= \frac{2 N_{e}}{M} \cdot  \left\{ 
\begin{array}{ll}
- \frac{2}{\sqrt{\pi}} \left(\frac{t}{T_{q}}\right)^{1/2}~~~
& \mbox{for }t \ll T_{q} \\
- 1 + \left(\frac{T_{q}}{\pi t}\right)^{1/2}~~~& \mbox{for } t \gg T_{q}~.
\end{array}
\right.
\nonumber
\end{eqnarray}
Furthermore the scaling with $q^2$, $M$, and $t$ is quite different.

\subsection{Closer inspection of Rouse motion in a tube}

The derivation of Eq.\ (5.11) in Ref.\ \cite{Z11}
involves some approximations, 
which greatly simplify the analysis but are not really necessary. 
For a general test of the validity of the model, we therefore
have repeated the analysis without these simplifications. 
The analysis is sketched  in Appendix C. For the further discussion, 
we here quote the result for the scattering from a given pair of 
beads (Eqs.\ (C.16) - (C.18)). 
\begin{eqnarray}
&&S_{RT}(q,t;\tilde{j},\tilde{k}) 
\\
&&= \frac{1}{2}\:e^{\tilde{Q}^{2}} \left\{ e^{2 \hat{\Delta} \tilde{Q}}\:
\mbox{erfc}\:(\tilde{Q} + \tilde{\Delta}) + e^{- 2 \hat{\Delta} \tilde{Q}}
\:\mbox{erfc}\:(\tilde{Q} - \tilde{\Delta})\right\}
\nonumber
\end{eqnarray}
\begin{eqnarray}
\tilde{Q} &=& q^{2} \ell^{2} \sqrt{\frac{N_{e}}{6}} 
\left[ | \tilde{j} - \tilde{k}| + \sqrt{2 \frac{\gamma_{0}}{\ell^{2}}\:t}\: 
\;g \left(\frac{| \tilde{j} - \tilde{k}|}
{\sqrt{2 \frac{\gamma_{0}}{\ell^{2}}\:t}}\right)\right]^{1/2}
\nonumber \\
\tilde{\Delta} &=& \sqrt{\frac{6}{N_{e}}} \frac{|\tilde{j} - \tilde{k}|}{2} 
\left[ | \tilde{j} - \tilde{k}| + \sqrt{2 \frac{\gamma_{0}}{\ell^{2}}\:t}\: 
\;g \left(\frac{| \tilde{j} - \tilde{k}|}
{\sqrt{2 \frac{\gamma_{0}}{\ell^{2}}\:t}}\right)\right]^{- 1/2}    
\nonumber\\
\end{eqnarray} 
\begin{equation}
g (z) = \frac{1}{\sqrt{\pi}}\:e^{- z^{2}} - z \:\mbox{erfc}\: z
\end{equation}
Here $\gamma_{0}$ is the segment mobility of the one-dimensional Rouse model. 

Clearly the structure of $S_{RT}$ (Eq.\ (5.14)) is identical to that 
of our result ${\cal S}_{1}$ (Eq.\ (5.2)). The difference is in the 
quantities $\tilde{Q}$, $\tilde{\Delta}$ (Eq.\ (5.15)), compared to 
$\hat{Q}$, $\hat{\Delta}$ (Eq.\ (5.3)). We note, however, that the relation
\begin{eqnarray*}
2 \tilde{Q} \tilde{\Delta} = q^{2} \ell^{2} |\tilde{j} - \tilde{k}| 
= q^{2} \frac{\ell_{0}^{2}}{6}\:|\tilde{j} - \tilde{k}| 
= 2 \hat{Q} \hat{\Delta}
\end{eqnarray*}
holds. Recall that the mean segment size $\ell$ of a Gaussian chain, 
which is asymptotically equivalent to a chain with fixed segment length 
$\ell_{0}$, obeys $\ell^{2} = \ell_{0}^{2}/6$ (in three dimensions).

To analyze the difference among the two models, we first consider 
the static limit, $t = 0$. ${\cal S}_{1}$ reduces to (recall the 
definition $\bar{q}^{2} = q^{2} \ell_{0}^{2}/6 \equiv q^{2} \ell^{2}$)
\begin{equation}
{\cal S}_{1} (q,0;\tilde{j},\tilde{k}) = 
e^{\textstyle - q^{2} \ell^{2} |\tilde{j} - \tilde{k}|}~,
\end{equation}
which is the exact result. The result for $S_{RT}$ can be written as 
\begin{eqnarray}
\lefteqn{S_{RT} (q,0;\tilde{j},\tilde{k})}
\nonumber\\ 
&\qquad&= \exp \left[ - q^{2} \ell^{2}|\tilde{j}\!-\!\tilde{k}| 
\left( 1 - q^{2} \ell^{2} \frac{N_{e}}{6}\right)\right]\;
\nonumber\\
&&\qquad\cdot\left\{ 1 - \frac{1}{2}\;\mbox{erfc}
\left[ \sqrt{\frac{6}{N_{e}} |\tilde{j}\!-\!\tilde{k}|} 
\left(\frac{1}{2}  - q^{2} \ell^{2} \frac{N_{e}}{6}\right)\right] \right\}
\nonumber \\
&&\quad+\;\frac{1}{2} \exp \left[q^{2} \ell^{2}|\tilde{j}\!-\!\tilde{k}| 
\left( 1 + q^{2} \ell^{2} \frac{N_{e}}{6}\right)\right]\;
\nonumber\\
&&\qquad\cdot
\mbox{erfc}\left[ \sqrt{\frac{6}{N_{e}} |\tilde{j}\!-\!\tilde{k}|} 
\left(\frac{1}{2}  + q^{2} \ell^{2} \frac{N_{e}}{6}\right)\right]\:.
\end{eqnarray}
Even if we ignore the terms $q^{2} \ell^{2} N_{e}/6 \ll 1$, taking them 
to be irrelevant micro-structure effects, this result does {\em not} 
reduce to the exact form (5.17).
To recover this form, we rather consistently have to take the limit 
$q^{2} \ell^{2} \rightarrow 0$ with $q^2\ell^{2}(\tilde{j}-\tilde{k})$ 
fixed, i.e., $|\tilde{j} - \tilde{k}|\equiv|j-k| \rightarrow \infty$.
This just demonstrates that a one-dimensional Gaussian chain, folded 
into the three dimensional random walk configuration of the tube, 
does not yield the exact distribution of a three dimensional Gaussian 
chain. In other words, the model of a Rouse chain in a tube violates 
the equilibrium initial conditions by micro-structure terms on scale 
$|\tilde{j} - \tilde{k}| \sim N_{e}$.

For the dynamics, this discussion implies that the model gives a wrong 
estimate for the number of wiggles in the initial configuration. 
To eliminate this effect of unphysical initial conditions, we have 
to take the same limit $q^{2} \ell^{2} \rightarrow 0$ with 
$q^{2} \ell^{2}(\tilde{j} - \tilde{k})$ fixed, also in the full 
time dependent expression (5.14). To facilitate the discussion, 
we rewrite Eqs.\ (5.15) in a form which exhibits the fixed
combination of variables $q^{2} \ell^{2} |\tilde{j} - \tilde{k}|$.
\begin{eqnarray}
\tilde{Q} &=& q \ell \sqrt{\frac{N_{e}}{6}} 
\left[q^{2} \ell^{2} |\tilde{j} - \tilde{k}| 
+ q^{2} \ell^{2} \sqrt{2 \frac{\gamma_{0}}{\ell^{2}}\:t}\; 
g \left(\frac{q^{2} \ell^{2} |\tilde{j} - \tilde{k}|}
{q^{2} \ell^{2} \sqrt{2 \frac{\gamma_{0}}{\ell^{2}}\:t}}\right)\right]^{1/2}
\nonumber \\
\tilde{\Delta} &=& \frac{1}{2 q \ell} \sqrt{\frac{6}{N_{e}}} q^{2} 
\ell^{2} |\tilde{j} - \tilde{k}|
\\
&&\qquad\cdot \left[q^{2} \ell^{2} | \tilde{j} - \tilde{k}| 
+ q^{2} \ell^{2} \sqrt{2 \frac{\gamma_{0}}{\ell^{2}}\:t}\;
g \left(\frac{q^{2} \ell^{2} |\tilde{j} - \tilde{k}|}
{q^{2} \ell^{2} \sqrt{2 \frac{\gamma_{0}}{\ell^{2}}\:t}}\right)\right]^{- 1/2}
\nonumber    
\end{eqnarray} 
The function $g(z)$ obeys the relations
\begin{eqnarray*}
g(0) &=& \frac{1}{\sqrt{\pi}}
\\
g(z) &\sim& e^{-z^{2}}, \: \: \: \: z \rightarrow \infty \: \: \:.
\end{eqnarray*}
Now the limiting result for $S_{RT}$ sensitively depends on the way 
we scale the time. We first consider times such that $q^2\sqrt{t}$
stays finite upon taking the limit $q^2\ell^2\to0$:
\begin{eqnarray*}
q^{2} \ell^{2} \sqrt{2 \frac{\gamma_{0}}{\ell^{2}}\:t} \sim q^{\alpha}, 
\: \: \: \alpha \geq 0 \: \: \:.
\end{eqnarray*}
We then find
\begin{eqnarray*}
\tilde{Q} \rightarrow 0, \: \: \: \tilde{\Delta} \rightarrow \infty
\end{eqnarray*}
and recover the static limit (5.17). Indeed, for such times the
scattering cannot resolve the internal motion:
\begin{eqnarray*}
q^2\;\left\langle~\overline{({\bf r}_j(t)-{\bf r}_j(0))}~\right\rangle
\sim q^2\;t^{1/4} \sim q^{1+\alpha/2}\to0~.
\end{eqnarray*}
Effects of internal dynamics can be seen only for times such that
$q^2\sqrt{t}$ diverges:
\begin{eqnarray*}
q^{2} \ell^{2} \sqrt{2 \frac{\gamma_{0}}{\ell^{2}}\:t} \sim q^{- \alpha}, 
\: \: \: \alpha > 0 \: \: \:.
\end{eqnarray*}
Then the contribution proportional to $g$ dominates 
the square brackets in Eqs.\ (5.19), and furthermore the argument 
of $g$ tends to zero. Thus 
\begin{eqnarray}
\tilde{Q} &\rightarrow& q^{2} \ell^{2} \sqrt{\frac{N_{e}}{6}} 
\left(\frac{2}{\pi} \frac{\gamma_{0}}{\ell^{2}}\:t\right)^{1/4}
\nonumber \\
\tilde{\Delta} &\rightarrow& \frac{1}{2} 
\sqrt{\frac{6}{N_{e}}} |\tilde{j} - \tilde{k}| \left(\frac{2}{\pi}\:
\frac{\gamma_{0}}{\ell^{2}}\:t\right)^{-1/4} \: \: \:,
\end{eqnarray}
which is the same functional dependence on $t$ and 
$|\tilde{j} - \tilde{k}|$ as that of $\hat{Q}$ or $\hat{\Delta}$ 
(Eq.\ (5.3)). The limiting expression for 
$S_{RT} (q,t;\tilde{j},\tilde{k})$ becomes identical to our result 
${\cal S}_{1} (q,t;\tilde{j},\tilde{k})$ if we identify 
\begin{equation}
q^{2} \ell^{2} = \bar{q}^{2} = q^{2}\:\frac{\ell_{0}^{2}}{6}
\end{equation}
\begin{equation}
\frac{N_{e}^{2}}{18} \frac{\gamma_{0}}{\ell^{2}}\:t = 
\left(\bar{\ell}_{s}^{2} \rho_{0}\right)^{2} \hat{t} \: \: \:.
\end{equation}
In summary, we have found that the model of a Rouse chain in a tube
is equivalent to the reptation model only in the limit $q^2\ell^2\to0$
with $q^2\ell^2(j-k)$ fixed. Outside this limit, it exhibits an unphysical 
relaxation of nonequilibrium initial conditions.

\subsection{Relation among the microscopic parameters of the different models}

As a byproduct of our analysis, we with Eq.\ (5.22) have found
a relation among the microscopic parameters of our model and those used
in more standard Rouse type modeling of chain dynamics. Analyzing in
Sect.\ VII the relation of our model with the primitive chain model,
we will find as an additional result:
\begin{equation}
\frac{N_e}{6}\;\frac{\gamma_0}{\ell^2}\;t=\bar\ell_s^2\;\rho_0\;\hat t~.
\end{equation}
Combining Eqs.\ (5.22) and (5.23), we find
\begin{eqnarray}
N_{e} &=& 3 \;\bar{\ell}_{s}^{2}\; \rho_{0} \: \: \:,
\\
\frac{\gamma_{0}}{2 \ell^{2}}\:t  &=& \hat{t} \: \: \:.
\end{eqnarray}
We now can give a quantitative definition of the equilibration time $T_2$,
which we identify with the Rouse time of a free chain of $N$ segments:
\begin{eqnarray*}
T_2=\frac{2}{\pi^2}\;(N+1)^2\;\frac{\ell^2}{\gamma_0}~.
\end{eqnarray*}
With Eq.\ (5.25), we find
\begin{equation}
\hat T_2=p\;T_2=\frac{(N+1)^2}{\pi^2}~.
\end{equation}

\subsection{Implications for the coherent structure function}

The artifact of the model of a Rouse chain in a tube concerns 
only small parts of the chain of the order of the tube diameter 
and thus might be thought to be negligible. In the static structure 
function, the error sums up to a term of order $N$, small compared 
to $S_{c}(q,0;N) = N^{2} D (\bar{q}^{2} N)$. Since, however, 
for $t \alt T_{2}$ the time dependence of $S_{c}$ is weak, even such 
a small effect is relevant. Indeed, it can greatly change the picture. 
To illustrate this, we in Fig.~3 compare our result (5.4) for 
the normalized coherent structure function $S_{c}(q,t;M,\infty)$ 
to the result found by integrating $S_{RT} (q,t;\tilde{j},\tilde{k})$ 
(Eq.\ (5.14)) over $- \frac{M}{2} \leq \tilde{j} \leq \frac{M}{2}$
and $- \frac{M}{2} \leq \tilde{k} \leq \frac{M}{2}$. 
To relate the models, we used relations (5.21), (5.24), and (5.25). 
To include also de Gennes' approximate form (5.11), we used the large value 
$q^{2} R_{g}^{2} (M) = 50$. We note that de Gennes enforced the correct 
$t = 0$ behavior by artificially subtracting his result for $S_{RT}$. 
We furthermore note that repeating his calculation in our notation, 
we found $t/T_{q} = \bar{q}^{4} \hat{t}$. The remaining parameter 
$\bar{\ell}_{s}^{2} \rho_{0} = 1.23$ has been taken from our previous 
work \cite{Z8}.

Fig.~3 shows that the effect of the artificial initial conditions 
can be quite large and dies out only slowly. It extends up to the Rouse
time of the subchain considered. This result is generic. For longer chains, 
the amplitude of the effect decreases for the normalized structure 
function, as expected for a micro-structure effect, but the range 
stays of order $T_2(M)$. This is obvious since a non-equilibrium 
initial condition generically will relax only on time scale $T_{2}$. 
As a side issue, we note that de Gennes' approximation (5.11)
agrees quite well with the shifted form of $S_{RT}$. 
                                
To close this section, a general remark on micro-structure corrections 
for the dynamics may be appropriate. Our result shows no such corrections, 
which would give rise to an additional $1/M$-dependence in Eq.\ (5.4), 
which is not in the form of the scaling variables $\bar{q}^{2} M$ and
$\hat{t}/\hat{T}_{4}$. 
Thus our model succeeded in singling out the universal aspects of 
reptation dynamics. This, however, does not imply that (non-universal) 
terms yielding some additional $1/M$ dependence could
not show up for a microscopically realistic model, which takes the details 
of the microscopic motion into account. But we want to stress that 
any model first of all has to yield the correct static structure function. 
Otherwise some unphysical relaxation will influence the dynamics. 
Such results can safely be trusted only in a range where 
$S_{c}(q,0;N) - S_{c}(q,t;N)$ exceeds the error in $S_{c}(q,0;N)$.

We finally note that here we have been concerned exclusively with 
the reptation aspect of the dynamics, modeled as {\em one} dimensional 
Rouse motion in a tube. This is to be clearly distinguished from 
{\em three} dimensional Rouse motion among fixed entanglement points, 
as treated by Des Cloizeaux \cite{Z14}, for instance. The latter model 
is concerned with motion in melts for `microscopic' times: $t \alt T_{0}$.  


\section{Analysis of complete tube destruction}

In this section, we derive an integral equation extending
$S_{c} (q,t;M,N)$ to arbitrarily large times  
(subsection A). Basically it is an equation for 
$S (q,t;j,k,N)$, which incorporates $S^{(T)}(q,t;j,0,N)$ as inhomogeneity. 
To calculate $S_{c}$, we need to sum this inhomogeneity over $j$. 
We construct this function in subsection B, following the approach 
of Section IV. The kernel of the integral equation involves some 
distribution function which is calculated in Subsect.\ C with the help 
of the mean hopping rate approximation. Quantities like the probability 
density of tube destruction at time $t$, which can be derived from 
this distribution function, are discussed in Subsection D. 
The numerical evaluation of our results for $S_{c} (q,t;M,N)$ 
is deferred to Sect.\ VIII, after we have shown that our theory 
in the appropriate limit yields the results of the primitive chain model.   

\subsection{Derivation of an integral equation for the structure function }

Up to now, we only considered stochastic processes for which some part 
of the initial tube still exists at time $t$. To get rid of this 
constraint, we have to deal with situations as shown in Fig.~4: 
at time $t_{0}$, $0 < t_{0} < t$, the chain leaves the original tube, 
which means that the remainder of the original tube is the single point 
${\bf r}_{j_{0}}(0)$. This point is occupied by a chain end. The rest 
of the chain has found a completely new configuration.

Let ${\cal P}^{*}(j_{0},t_{0} \mid 0)$ or 
${\cal P}^{*}(j_{0},t_{0} \mid N)$ be the probability that the tube 
is finally destroyed at time $t_{0}$, the last point 
${\bf r}_{j_{0}}(0)$ being occupied by chain end $0$ or $N$, 
respectively. We assume that ${\cal P}^{*}$ does not depend 
on the initial configuration, which should be satisfied except 
for rare extreme cases. 

We then can write the full time dependent scattering (3.17) 
from a pair of beads $j,k$ as 
\begin{eqnarray}
\lefteqn{S (q,t;j,k,N) = S^{(T)} (q,t;j,k,N)}
\\
&&+\sum^{t-1}_{t_{0}=1} \sum^{N}_{j_{0} = 0} 
\Big[ {\cal P}^{*}(j_{0},t_{0} \mid 0)\: \: S (q,t,j,k \mid j_{0},t_{0},0) 
\nonumber\\
&&\qquad \qquad\quad+ {\cal P}^{*}(j_{0},t_{0} \mid N)\: \:  
S (q,t,j,k \mid j_{0},t_{0}, N) 
\Big]~,
\nonumber
\end{eqnarray}
where $S (q,t,j,k \mid j_{0},t_{0},m)$ with $m = 0,N$ denotes the scattering 
with tube destruction specified by $j_{0}$, $t_{0}$ and $m$. We now factorize 
$S (q,t,j,k \mid j_{0},t_{0},m)$ according to 
\begin{eqnarray}
\lefteqn{S (q,t,j,k \mid j_{0},t_{0},m)}
\nonumber\\
&=& \left.\left\langle ~\overline{
e^{\textstyle i {\bf q} ({\bf r}_{j}(t) - {\bf r}_{k}(0))}}
                ~\right\rangle \right|_{j_{0},t_{0},m}
\nonumber \\
&=& \left.\left\langle ~\overline{
e^{\textstyle i {\bf q} ({\bf r}_{j}(t) - {\bf r}_{m}(t_{0}))}\;
e^{\textstyle i {\bf q} ({\bf r}_{j_{0}}(0) - {\bf r}_{k}(0))}} 
~\right\rangle \right|_{j_{0},t_{0},m}
\nonumber \\
&\approx& \left\langle ~\overline{
e^{\textstyle i {\bf q} ({\bf r}_{j}(t) - {\bf r}_{m}(t_{0}))}} 
~\right\rangle\; \left\langle \;
e^{\textstyle i {\bf q} ({\bf r}_{j_{0}}(0) - {\bf r}_{k}(0))} 
\right\rangle
\nonumber \\
&&m = 0,N~, 
\end{eqnarray}
where the second factor in the last line is a purely static average. 
We have exploited $r_{m}(t_{0}) = r_{j_{0}}(0)$. This factorization 
should be well justified, since the chain at time $t_{0}$ has attained 
a completely new internal configuration. Now the first factor in the 
last line of Eq.\ (6.2) equals $S (q,t-t_{0};j,m,N)$, whereas 
the second factor is the static structure function 
$\exp [ - \bar{q}^{2} \mid j_{0} - k \mid]$. 
Combining Eqs.\ (6.1), (6.2) we thus find 
\begin{eqnarray*}
\lefteqn{
S (q,t;j,k,N) = S^{(T)} (q,t;j,k,N)}
\nonumber \\ 
& & + \sum^{t-1}_{t_{0}=1} \sum_{j_{0}} \Big[
{\cal P}^{*}(j_{0},t_{0} \mid 0) \;e^{\textstyle- \bar{q}^{2} |j_{0} - k|} 
\;S (q,t-t_{0};j,0,N)
\nonumber \\ 
& &\qquad\qquad
+ {\cal P}^{*}(j_{0},t_{0} \mid N)\; e^{\textstyle- \bar{q}^{2} |j_{0} - k|} 
\;S (q,t-t_{0};j,N,N) \Big] \: \: \:. 
\end{eqnarray*}
Reflection symmetry along the chain implies 
\begin{eqnarray*}
{\cal P}^{*}(j_{0},t_{0} \mid N) &=& {\cal P}^{*} (N-j_{0},t_{0} \mid 0)
\\
S (q,t;j,N,N) &=& S (q,t;N-j,0,N) \: \: \:,
\end{eqnarray*}
so that our result takes the form 
\begin{eqnarray}
\lefteqn{
S (q,t;j,k,N) = S^{(T)} (q,t;j,k,N)}
\\ 
& & + \sum^{t-1}_{t_{0}=1} \sum_{j_{0}} {\cal P}^{*}(j_{0},t_{0} \mid 0)\; 
\Big[ e^{\textstyle - \bar{q}^{2} |j_{0} - k|}\; S (q,t-t_{0};j,0,N)
\nonumber \\ 
& & \qquad\qquad
+ \;e^{\textstyle - \bar{q}^{2} |N - j_{0} - k|}\; 
S (q,t-t_{0};N-j,0,N)\Big]~.\nonumber
\end{eqnarray}
We now sum $j$ and $k$ over the central piece of the chain to find 
\begin{eqnarray}
\lefteqn{S_{c}(q,t;M,N) = S_{c}^{(T)} (q,t;M,N)}
\nonumber\\
&&+ 2 \sum^{t-1}_{t_{0}=1} \left\{ \sum_{j_{0} = 0}^{N} 
\sum^{(N+M)/2}_{k=(N-M)/2} {\cal P}^{*}(j_{0},t_{0} \mid 0)\: \: 
e^{\textstyle -\bar{q}^2 |j_0 - k|} \right\}
\nonumber\\
&&\qquad\qquad\cdot S_{E}(q,t-t_{0};M,N) \: \: \:,
\end{eqnarray}
where
\begin{equation}
S_{E}(q,t;M,N) = \sum^{(N+M)/2}_{j = (N-M)/2}\:S (q,t;j,0,n)\: \: \:.
\end{equation}
For $S_{E}$, Eq.\ (6.3) yields
\begin{eqnarray}
\lefteqn{
S_{E}(q,t;M,N) = S_{E}^{(T)}(q,t;M,N)}
\nonumber\\
& & + \sum^{t-1}_{t_{0} = 1} \left\{ \sum^{N}_{j_{0} = 0} 
{\cal P}^{*} (j_{0},t_{0} \mid 0) \left(e^{\textstyle- \bar{q}^{2} j_{0}} 
+ e^{\textstyle- \bar{q}^{2} (N - j_{0})}\right)\right\}
\nonumber\\ 
&&\qquad\qquad\cdot S_{E} (q, t-t_{0};M,N)~.
\end{eqnarray}
With time (and segment index) taken continuous, this is the basic 
integral equation of our approach. We note that it is of Volterra-type 
and therefore has a unique solution. 

\end{multicols}

\subsection{Expression for $S_{E}^{(T)}(q,t;M,N)$} 

The tube conserving contribution to $S_{E} (q,t;M,N)$ is easily found 
from the results of Section IV. Following Eq.\ (4.31), we write
\begin{equation}
S_{E}^{(T)}(q,t;M,N) = {\cal S}_{E,1} (q,t;M,N) + {\cal S}_{E,2} (q,t;M,N) 
+ {\cal S}_{E,3} (q,t;M,N) \: \: \:.
\end{equation} 
\begin{eqnarray}
{\cal S}_{E,1} (q,t;M,N) &=& \int^{\frac{N+M}{2}}_{\frac{N-M}{2}}\:d j\: 
{\cal S}_{1} (q,t;j,0,N) 
\\
{\cal S}_{E,2} (q,t;M,N) 
&=& c\:\int^{(\hat{N} + \hat{M})/2}_{(\hat{N} - \hat{M})/2} d \hat{j}\: 
\int_{0}^{\infty}\: d z\:\frac{e^{- \hat{q} z}}{b}\:
\frac{1}{\sqrt{2\pi}} 
\:\sum_{\nu} \left[ {\cal P}_{j}^{(\nu)} \left(\frac{z - \hat{j}}{b},a\right)
+ {\cal P}_{j}^{(\nu)} \left(\frac{- z - \hat{j}}{b},a\right)\right]
\\
{\cal S}_{E,3} (q,t;M,N) 
&=& c\:\int^{(\hat{N} + \hat{M})/2}_{(\hat{N} - \hat{M})/2} d \hat{j}\: 
\frac{1}{b} \int_{0}^{\hat{N}}\:d y \;
\Bigg\{\int_{0}^{\infty}\!\!\! d z 
\left[ e^{\textstyle-\hat{q}(2y+z)}-e^{\textstyle- \hat{q} z} \right]
\;\frac{1}{\sqrt{2\pi}} \:\sum_{\nu} 
{\cal P}_{\mbox{{\scriptsize max}},j}^{(\nu)}
\left(y,-\frac{z+\hat{j}}{b},a\right)
\\
&&\qquad\qquad\qquad\qquad\qquad\qquad
+\; \int_{0}^{1}\!\!\!d z\;y \left[e^{\textstyle-\hat{q} y (2-z)} 
-e^{\textstyle-\hat{q}yz} \right]
\;\frac{1}{\sqrt{2\pi}} \:\sum_{\nu} 
{\cal P}_{\mbox{{\scriptsize max}},j}^{(\nu)}
\left(y,\frac{yz-\hat{j}}{b},a\right)\Bigg\} \: \: \:.
\nonumber
\end{eqnarray}         
The notation is the same as in Sect.\ IV (see, in particular, 
Eqs.\ (4.27), (4.35), (4.36)).

\subsection{Expression for ${\cal P}^{*} (j_{0},t_{0} \mid 0)$}

To construct an expression for ${\cal P}^{*} (j_{0},t_{0} \mid 0)$, 
we again use random walk theory, closely following the derivation 
of ${\cal P}_{\mbox{{\scriptsize max}},0}^{(T)}$ in Sect.\ IV.C. 
The calculation is sketched in Appendix A. It yields the result
\begin{eqnarray}
d t_{0} \;{\cal P}^{*} (j_{0},t_{0} \mid 0) 
&=& \frac{d t_{0}}{t_{0} - 1} \left(\pi p' (t_{0} - 1)\right)^{-1/2}
\;\sum^{+ \infty}_{\nu = - \infty}\:\nu 
\;\left[1-\frac{2\;(\nu N+j_0/2)^{2}}{\bar{\ell}_s^2 p'(t_0-1)} \right] 
\;\exp \left[-\frac{(\nu N+j_0/2)^2}{\bar{\ell}_s^2 p' (t_0-1)}\right]~.
\end{eqnarray} 
We again express $t_{0} - 1 \approx t_{0}$ in terms of the maximal 
excursion $\bar{n}_{m}$ (Eq.\ (4.21)) and identify $\bar{n}_{m}$ 
with $\overline{n_{\mbox{{\scriptsize max}}}(0,t_{0})}$. This yields 
the replacement
\begin{eqnarray*}
\bar{\ell}_{s} \sqrt{p' (t_{0} - 1)} \rightarrow \frac{\sqrt{\pi}}{2}\:
\bar{\ell}_{s} \overline{n_{\mbox{{\scriptsize max}}}(0,t_{0})} 
= \frac{c}{\sqrt{2}} \: \: \:,
\end{eqnarray*}
resulting in 
\begin{eqnarray*}
\frac{N}{\bar{\ell}_{s} \sqrt{p'(t_{0}-1)}} 
\rightarrow \sqrt{2}\:\hat{N} = \sqrt{2} \frac{N}{c}\qquad&,&\qquad 
\frac{j_{0}}{\bar{\ell}_{s} \sqrt{p'(t_{0}-1)}} \rightarrow 
\sqrt{2}\:\hat{j}_{0} = \sqrt{2} \frac{j_{0}}{c}~,
\\           
\frac{d t_{0}}{t_{0} - 1} \rightarrow 2 \:\frac{dc}{c} \qquad&,&\qquad
\frac{d j_{0}}{\bar{\ell}_{s} \sqrt{\pi p'(t_{0}-1)}} \rightarrow 
\sqrt{\frac{2}{\pi}}\:d \hat{j}_{0}~.
\end{eqnarray*}
With these replacements, we find
\begin{equation}
d j_{0}\; d t_{0}\; {\cal P}^{*} (j_{0},t_{0} \mid 0) 
\rightarrow 2 \sqrt{\frac{2}{\pi}}\:\frac{d c}{c}\:d \hat{j}_{0} 
\sum^{+ \infty}_{\nu = - \infty}\:\nu    
\left[ 1 - 4 ( \nu \hat{N} + \hat{j}_{0}/2)^{2}\right] 
\;\exp \left[ - 2 (\nu \hat{N} + \hat{j}_{0}/2)^{2}\right]~.
\end{equation}
To construct the kernels for the integral equations (6.4) and (6.6), 
we basically need
\begin{equation}
dt_0 \int_{0}^{X} d j_{0}\; {\cal P}^{*} (j_{0},t_{0} \mid 0) 
\;e^{\textstyle- \bar{q}^{2} j_{0}} 
= 2 \;\frac{d c}{c} \;\tilde{{\cal P}}^{*} (\hat{q},\hat{X},\hat{N}) \: \: \:,
\end{equation}
where $\hat{X} = X/c$, and where 
$\tilde{{\cal P}}^{*} (\hat{q},\hat{X},\hat{N})$ is given by
\begin{eqnarray}
\lefteqn{
\tilde{{\cal P}}^{*} (\hat{q},\hat{X},\hat{N}) 
= \sqrt{\frac{2}{\pi}}\:\sum^{+ \infty}_{\nu = - \infty}\:\nu\:
\int_{0}^{\hat{X}} d \hat{j}_{0} 
\left[ 1 - 4 \left(\nu \hat{N} + \hat{j}_{0}/2\right)^{2} \right] 
\;\exp \left[-2\left(\nu\hat{N}+\hat{j}_0/2\right)^2-\hat{q}\hat{j}_0\right]} 
\nonumber \\
&=& \sum^{+ \infty}_{\nu = - \infty}\nu \;\Bigg\{ \hat{q}^{2} 
\;e^{2\nu \hat{q} \hat{N} + \hat{q}/2}\; 
\left[\mbox{erfc}\left(\frac{2\nu\hat{N}+\hat{X}+\hat{q}}{\sqrt{2}}\right) - 
\mbox{erfc} \left(\frac{2\nu\hat{N}+\hat{q}}{\sqrt{2}}\right)\right]   
\nonumber \\
& & \qquad\qquad
- \sqrt{\frac{2}{\pi}} \;2 \nu \hat{N} \;e^{-2\nu^{2}\hat{N}^2} 
+ \sqrt{\frac{2}{\pi}}\;(2 \nu \hat{N} + \hat{X} - \hat{q})\; 
\exp \left[-\frac{(2\nu\hat{N} + \hat{X})^2}{2} - \hat{q} \hat{X}\right]
\Bigg\}~.
\end{eqnarray}
In terms of $\tilde{{\cal P}}^{*} (\hat{q},\hat{X},\hat{N})$, 
the kernel of Eq.\ (6.6) takes the form
\begin{eqnarray}
\lefteqn{
2\; \frac{d c}{c}\;{\cal K}_{E} (\hat{q},\hat{N}) 
= dt_0\int_{0}^{N} d j_{0}\; {\cal P}^{*} (j_{0},t_{0} \mid 0)
\left( e^{- \bar{q}^{2} j_{0}} + e^{- \bar{q}^{2}(N - j_{0})}\right)}
\nonumber \\        
& & = 2 \;\frac{d c}{c} \;\left[\tilde{{\cal P}}^{*} (\hat{q},\hat{N},\hat{N}) 
+e^{-\hat{q}\hat{N}}\;\tilde{{\cal P}}^{*}(-\hat{q},\hat{N},\hat{N})\right]~. 
\end{eqnarray}
The kernel of Eq.\ (6.4) can be written as
\begin{eqnarray}
\lefteqn{
2\:\frac{d c}{\hat{q}}\; {\cal K}_{c} (\hat{q},\hat{N},\hat{M}) =  
dt_0\int_{0}^{N} d j_{0}\int^{(N + M)/2}_{(N - M)/2}d k \;
{\cal P}^{*} (j_{0},t_{0} \mid 0)\; e^{\textstyle- \bar{q}^{2} |j_{0} - k|}}
\nonumber \\
&= & 2\;\frac{d c}{\hat{q}} \;\bigg[ 
2\;\tilde{{\cal P}}^*(0,{\scriptstyle\frac{\hat{N}+\hat{M}}{2}},\hat{N}) 
-2\;\tilde{{\cal P}}^*(0,{\scriptstyle\frac{\hat{N} - \hat{M}}{2}},\hat{N})
\\
& &\qquad\quad -\; e^{\hat{q} (\hat{N} - \hat{M})/2} 
\;\left(\tilde{{\cal P}}^*(\hat{q},\hat{N},\hat{N})
- \tilde{\cal P}^*(\hat{q},{\scriptstyle\frac{\hat{N}-\hat{M}}{2}},\hat{N})
\right) + e^{- \hat{q}(\hat{N} - \hat{M})/2} 
\;\tilde{{\cal P}}^{*} 
(- \hat{q},{\scriptstyle\frac{\hat{N} - \hat{M}}{2}},\hat{N})
\nonumber \\
& & \qquad\quad+\; e^{\hat{q}(\hat{N} + \hat{M})/2} 
\;\left(\tilde{{\cal P}}^{*} (\hat{q},\hat{N},\hat{N}) 
- \tilde{{\cal P}}^{*} 
(\hat{q},{\scriptstyle\frac{\hat{N} + \hat{M}}{2}},\hat{N})\right) 
- e^{- \hat{q}(\hat{N} + \hat{M})/2} \;\tilde{{\cal P}}^{*} 
(- \hat{q},{\scriptstyle\frac{\hat{N} + \hat{M}}{2}},\hat{N})\bigg]~.
\nonumber
\end{eqnarray}

\begin{multicols}{2}

\subsection{Discussion of the probability density of tube destruction}

Comparing the present results with those of Sect.\ IV.C, we can verify 
the internal consistency of our random walk approximation. 
From ${\cal P}_{\mbox{{\scriptsize max}},j}^{(T)}(n_{m},n_{j},t)$ 
(Eq.\ (3.20)), we can derive the probability that a part of the initial 
tube still exists at time $t$:
\begin{equation}
{\cal P}^{(T)}(t) = \sum^{- \infty}_{n_{j} 
= - \infty}\:\sum^{N}_{n_{m} = 0}\:
{\cal P}_{\mbox{{\scriptsize max}},j}^{(T)} (n_{m},n_{j},t) \: \: \:.
\end{equation}
A straightforward calculation starting from Eqs.\ (4.26), (4.27) yields 
\begin{eqnarray}
{\cal P}^{(T)}(t) &=& 1 + 4 \sum^{\infty}_{\nu = 1}\:(-1)^{\nu} \nu \:
\mbox{erfc}\:\left(\frac{\nu}{\sqrt{2}}\:\hat{N}\right)
\nonumber \\ 
&=& \tilde{{\cal P}}^{(T)} \left(\frac{N}{c}\right)~.
\end{eqnarray} 
The probability that the tube is destroyed within time interval 
$d t_{0}$, corresponding to
\begin{eqnarray*}
d c = c (t_{0} + d t_{0}) - c (t_{0}) \: \: \:,
\end{eqnarray*}
can be calculated as $- d c\:\frac{\partial}{\partial c}\:{\cal P}^{(T)}$.
\begin{eqnarray}
&&- d t_{0} \;\frac{\partial}{\partial t_{0}} \;{\cal P}^{(T)} 
= - d c\:\frac{\partial}{\partial c}\:\tilde{{\cal P}}^{(T)} 
\left(\frac{N}{c}\right) 
\nonumber\\
&&= - 4 \;\frac{d c}{c}\;\sqrt{\frac{2}{\pi}}\:\hat{N}\:
\sum^{\infty}_{\nu = 1} (-1)^{\nu} \nu^{2}\; 
e^{- \frac{1}{2} \nu^{2} \hat{N}^{2}}
\end{eqnarray}
On the other hand, we can calculate this probability also as
\begin{equation}
2 \;d t_{0} \sum^{N}_{j_{0} = 0}\:{\cal P}^{*} (j_{0},t_{0} \mid 0) 
= 4 \;\frac{d c}{c}\:\tilde{{\cal P}}^{*} (0, \hat{N}, \hat{N}) \: \: \:,
\end{equation}
where the factor of $2$ takes the two chain ends into account. 
It is easily verified that these two expressions are identical. 
Thus the following relation holds
\begin{equation}
- \;\frac{\partial}{\partial t}\:{\cal P}^{(T)} (t) 
= 2 \sum^{N}_{j_{0} = 0} {\cal P}^{*} (j_{0},t_{0} \mid 0)~.
\end{equation}
This identity guarantees the validity of the normalization
\begin{equation}
S_{E}(q = 0,t;M,N) \equiv \sum^{\frac{M-N}{2}}_{j = \frac{M-N}{2}}\:1 = M + 1
\end{equation}
for all times. From the definition of $S_{E}^{(T)}$, we have
\begin{eqnarray}
S_{E}^{(T)} (0,t;M,N) &=& \sum^{\frac{N+M}{2}}_{j = \frac{N-M}{2}}\:
\overline{\Theta (j_{>} - j_{<})}
\nonumber \\
&=& (M + 1)\;{\cal P}^{(T)} (t) \: \: \:.
\end{eqnarray}
Substituting Eqs.\ (6.21) and (6.23) into the integral equation (6.6), we find
\begin{eqnarray}
&&S_{E} (0,t;M,N) = (M + 1)\;{\cal P}^{(T)} (t) 
\\
&&~~~- \int_{0}^{t} d t_{0}\;
\left(\frac{\partial}{\partial t_{0}}\:{\cal P}^{(T)} (t_{0})\right)
\;  S_{E} (0, t-t_{0}; M,N)~.
\nonumber
\end{eqnarray}
Partial integration together with
\begin{equation}
S_{E} (0,0;M,N) = M + 1~,~~~ {\cal P}^{(T)} (0) = 1
\end{equation}
yields
\begin{equation}
0 \equiv \int_{0}^{t} d t_{0}\;{\cal P}^{(T)} (t_{0}) \;
\frac{\partial}{\partial t_{0}}\; S_{E} (0,t-t_{0}; M,N) \: \: \:,
\end{equation}
with only the trivial solution 
\begin{equation}
\frac{\partial}{\partial t_{0}}\:S_{E} (0,t-t_{0}; M,N) \equiv 0 \: \: \:.
\end{equation}
Together with Eq.\ (6.25), this proves Eq.\ (6.22). 
The corresponding analysis can be applied to Eq.\ (6.4), 
yielding the correct normalization
\begin{equation}
S_{c} (0,t;M,N) \equiv (M + 1)^{2}~.
\end{equation}
      
To get an impression of the time dependence of complete tube destruction, 
we in Fig.~5~a show $\hat{N}^{2} \tilde{{\cal P}}^{*} (0,\hat{N},\hat{N}) 
\sim - \frac{\partial}{\partial t} {\cal P}^{(T)}(t)$ 
(cf.\ Eqs.\ (6.20), (6.21)) as a function of $2/\hat{N}^{2} = 2 (c/N)^{2} 
= \pi (\bar{\ell}_{s}\:\overline{n_{\mbox{{\scriptsize max}}}(0,t)}/N)^{2}$. 
This choice of the variable is motivated by the relation $\hat{N}^{-2} 
\sim t/T_{3}$, cf.\ Eq.\ (7.3). As we see, noticeable tube destruction 
starts at $2/\hat{N}^{2} \approx 0.1$ and is essentially completed at 
$2/\hat{N}^{2} \approx 3.5$. The variation of $\hat{N}^{2} 
\tilde{{\cal P}}^{*}$ as shown here, dominates the time dependence 
of the kernels (6.15), (6.16). It allows us to solve the integral equation 
(6.6) for finite time $t$ by a finite number of iterations, the result 
being exact within the numerical accuracy of our calculation. 

To close this section, we evaluate the probability that the initial tube 
finally is destroyed at the position of segment $j_{0}$, with chain end 
$0$ being the last part residing in the initial tube (Eq.\ (6.12)). 
Fig.~5~b shows the dependence on $j_{0}/N$ for several values of 
$\hat{N}$. As expected, for shorter times $2/\hat{N}^{2} < 1$, 
chain end $0$ leaves the tube close to the other end $(j_{0}/N \approx 1)$. 
With increasing time the most probable point of final destruction slowly shifts 
to the center of the tube, but for times where the rate of the tube 
destruction is maximal, (corresponding to the maximum in Fig.~5~a), 
the shape of ${\cal P}^{*} (j_{0},t_{0} \mid 0)$ is still quite un-symmetric. 

Obviously, the distribution functions considered here are closely 
related to the r.h.s.\ of Eq.\ (2.9), which is determined by the part of the 
original tube that is still occupied at time $t$ (see Refs.\ \cite{Z1,Z11}). 
In this context, it is interesting to note that Des Cloizeaux \cite{Z15} 
modified the expression (2.9) by replacing $\tau = t/\tau_{d}$ 
in the exponent by some more complicated time dependence, 
meant to take the local motion near an entanglement point into account. 
This modification is quite similar to our introduction
of the quantity $\hat{N}^{2}$ replacing $t/\tau_{d}$. 
We note, however, that $\hat{N} = N/c$ via the crossover behavior 
of $c = c (t,N)$ takes end effects like tube length fluctuations
into account rather than internal motion.     


\section{The limit of large time and the primitive chain model}

\subsection{Special cases}

If the time is large compared to the equilibration time $T_{2}$ 
of the chain, our results simplify since the parameters $a$ and $b$ 
can be replaced by their limiting values
\begin{equation}
a = 1 = b\qquad\mbox{for } t \gg T_{2}~.
\end{equation}
This implies that all segments experience the same curvilinear shift, 
which is the basic assumption of the primitive chain model. 
Furthermore $A_{1}(j,t) \rightarrow \hat{t}/N$ for $t \gg T_{2}$ 
(cf.\ Eq.\ (3.5)), and the parameter $c$ (Eq.\ (4.25)) reduces to
\begin{eqnarray}
c &=&\sqrt{\frac{\pi}{2}}\:\bar{\ell}_{s} 
\overline{n_{\mbox{\scriptsize max}}(0,t)} \\
&&\rightarrow 
\left(\frac{\bar{\ell}_{s}^{2} \rho_{0}}{2}\right)^{1/2} 
\sum^{t}_{s=1}\:\frac{1}{s}\:A_{1}^{1/2}(0,s)
=\left(2 \frac{\bar{\ell}_s^2\rho_0}{N}\right)^{1/2}\hat{t}^{1/2}~.
\nonumber
\end{eqnarray}
Thus
\begin{equation}
\hat{N}^{-1} = \frac{c}{N} \rightarrow 
\left(\frac{\hat{t}}{2 \hat{T}_{3}}\right)^{1/2}
\end{equation}
becomes a direct measure of $\hat{t}/\hat{T}_{3}$, 
where for brevity we introduced 
\begin{equation}
p T_{3} = \hat{T}_{3} = \frac{N^{3}}{4 \bar{\ell}_{s}^{2} \rho_{0}}
\end{equation}
as measure of the reptation time. With relations (7.1), all the integrals 
in Eqs.\ (4.37) - (4.44) can be evaluated analytically, resulting 
in a fairly lengthy expression for $S_{c}^{(T)}(q,t;M,N)$ as a sum 
of terms involving error functions and Gaussians. We here quote 
the result in those limits, in which $S_{c}^{(T)}$ becomes identical 
to the full scattering function $S_{c}$, which is the case for either 
short time: $t/T_{3} \ll 1$ or large wave vectors: $q^{2} R_{g}^{2} \gg 1$.

\subsubsection{Limit $t/T_{3} \rightarrow 0$ with fixed $Q = q^{2} R_{g}^{2}$}

We find
\begin{equation}
\frac{S_{c} (q,t;N,N)}{N^{2}} = D (Q) - \frac{t}{2 T_{3}} (1 - e^{-Q}) 
+ O \left(\frac{t}{T_{3}}\right)^{3/2}~.
\end{equation}
Recall that $D(Q)$ denotes the Debye function. Of course, this limit 
can be attained only for an extremely long chain, since relation (7.1) 
implies $T_{2}/T_{3} \rightarrow 0$, i.e., $N \rightarrow \infty$. 
The result (7.5) shows that for such a chain relaxation becomes 
observable only for $t \gg T_{2}$. Furthermore, with increasing $Q$, 
the time variation of $S_{c}$ becomes rapidly insensitive to the 
scattering vector. 

\subsubsection{Limit $Q = q^{2} R_{g}^{2} \rightarrow \infty$ with
fixed $\bar{q}^{2}$ and $t/T_{3}$} 

In this limit, our result reads 
\begin{eqnarray}
\lefteqn{\frac{S_{c}(q,t;N,N)}{N^{2} D(Q)} = 
1 - \sqrt{\frac{2}{\pi}}\:\frac{2}{\hat{N}} }
\\
&&+ 4 \sum^{\infty}_{\nu = 1}\:(- 1)^{\nu} 
\left[ \nu \:\mbox{erfc}\:
\frac{\nu \hat{N}}{\sqrt{2}} - \sqrt{\frac{2}{\pi}}\:
\frac{1}{\hat{N}} e^{- \frac{\nu^{2}}{2} \hat{N}^{2}}\right]~,
\nonumber
\end{eqnarray}
which is the Poisson transform of
\begin{eqnarray}
\lefteqn{\frac{S_{c}(q,t;N,N)}{N^{2} D (Q)} =}
\\
&& \frac{8}{\pi^{2}}\:\sum^{\infty}_{p=1}\:(2 p - 1)^{-2} 
\exp \left[ - (2 p - 1)^{2} \frac{\pi^{2}}{2 \hat{N}^{2}}\right]~.
\nonumber
\end{eqnarray}
We thus recover the result of Refs.\ \cite{Z11,Z12}, Eq.\ (2.9), 
provided we identify 
\begin{eqnarray*}
\frac{t}{\tau_{d}} = \frac{\pi^{2}}{2 \hat{N}^{2}} = 
\frac{\pi^{2}}{4} \frac{\hat{t}}{\hat{T}_{3}} \: \: \:,
\end{eqnarray*}
leading to
\begin{equation}
p \tau_{d} = \frac{N^{3}}{\pi^{2} \bar{\ell}_{s}^{2} \rho_{0}} \: \: \:.
\end{equation}
From Ref.\ \cite{Z2}, Eq.\ (6.19), we can take the relation of $\tau_{d}$ 
to the parameters of the underlying Rouse model, which in our notation reads 
\begin{eqnarray*}
\tau_{d} = \frac{N^{3} \ell_{0}^{2}}{\pi^{2} \gamma_{0} N_{e}} \: \: \:.
\end{eqnarray*}
(Replacement $\zeta \rightarrow 1/\gamma_{0}$, $b \rightarrow \ell_{0}$, 
$a^{2} \rightarrow \ell_{0}^{2} N_{e}$, $k_{B}T = 1$ in Ref.\ \cite{Z2}, 
Eq.\ (6.19).) ~Thus 
\begin{eqnarray*}
\frac{p \ell_{0}^{2}}{\gamma_{0} N_{e}} = 
\left(\bar{\ell}_{s}^{2} \rho_{0}\right)^{-1} \: \: \:,
\end{eqnarray*}
and Eq.\ (5.23) results.

\subsection{Proof of asymptotic equivalence to the primitive chain model}

Having recovered the results of Doi and Edwards for $q^{2} R_{g}^{2} \gg 1$,
$t \gg T_{2}$, we clearly may ask whether for $t \gg T_{2}$, the two 
approaches yield identical results irrespective of $q^{2} R_{g}^{2}$. 
This is not obvious since formally the approaches are quite different. 
Doi and Edwards \cite{Z12,Z2} start from a diffusion equation for 
$S(q,t;j,k,N)$. With the relation among model parameters 
established in Sect.\ V.D, this equation takes the form
\begin{equation}
\left[ \frac{N}{p \bar{\ell}_{s}^{2} \rho_{0}} \frac{\partial}{\partial t} 
- \frac{\partial^{2}}{\partial j^{2}}\right] S (q,t;j,k,N) = 0~.
\end{equation}
This is amended by the initial condition
\begin{equation}
S (q,0;j,k,N) = \exp (- \bar{q}^{2} |j-k|)
\end{equation}
and boundary conditions
\begin{eqnarray}
\lim_{j \rightarrow 0}\:\frac{\partial}{\partial j}\:S (q,t;j,k,N) 
&=& \bar{q}^{2} S (q,t;0,k,N)
\nonumber \\
\lim_{j \rightarrow N}\:\frac{\partial}{\partial j}\:S (q,t;j,k,N) 
&=& - \bar{q}^{2} S (q,t;N,k,N) \: \: \:.
\end{eqnarray}  
   
On the other hand, according to our theory, $S (q,t;j,k,N)$ obeys 
Eq.\ (6.3), written in the continuous chain model as
\begin{eqnarray}
\lefteqn{S (q,t;j,k,N) = S^{(T)} (q,t;j,k,N)}
\\
&&+ \int^{t}_{0}\!\!d t_{0}\int^{N}_{0}\!\!\!d j_{0}\;
{\cal P}^{*}(j_{0},t_{0} \mid 0) \;
\Big\{e^{- \bar{q}^{2} |j_{0} - k|}\; S (q,t - t_{0};j,0,N)
\nonumber \\
&&\qquad\qquad\qquad
+\; e^{- \bar{q}^{2} |N - j_{0} - k|} \;S (q,t - t_{0};N-j,0,N)\Big\}~.
\nonumber
\end{eqnarray}
The inhomogeneity takes the form
\begin{eqnarray}
\lefteqn{S^{(T)} (q,t;j,k,N) }
\\
&&= \int_{0}^{N} dy \int^{y}_{y-N} d z\:\sum_{\nu}\:
\hat{{\cal P}}_{\nu}(y,z) \;F (y,z;j,k)
\nonumber
\end{eqnarray}
where 
\begin{eqnarray}
\lefteqn{
\hat{{\cal P}}_{\nu}(y,z) 
= \frac{1}{\sqrt{2\pi}\;c^{2}}\;\lim_{a \to 1}\;
{\cal P}_{\mbox{{\scriptsize max}},j}^{(\nu)} 
\left(\frac{y}{c},\frac{z}{c}, a\right)}
\nonumber \\
&=& \sqrt{\frac{2}{\pi}}\:\frac{1}{c^{3}} \Big[ (\nu + 1)\;(2 \nu N + 2 y - z)
\;e^{- \frac{1}{2 c^{2}} (2 \nu N + 2 y - z)^{2}}
\nonumber \\
&&\qquad\quad-\;\nu\;(2\nu N-z)\;e^{-\frac{1}{2c^2}(2\nu N-z)^2} \Big]~.
\end{eqnarray}
The rescaling of the variables $y$ and $z$ serves to isolate 
the time dependence which now is contained in $c$ 
only (cf.\ Eq.\ (7.2)). The function $F$ collects all contributions 
contained in ${\cal S}_{1}$, ${\cal S}_{2}$, ${\cal S}_{3}$ (Eq.\ (4.5)), 
and is found to be
\begin{eqnarray}
F (y,z;j,k) &=& e^{- \bar{q}^{2}|k - j - z|}
\\
&&+ \Theta  (k - j - z)\; \Theta (y-k) 
\nonumber\\
&&\qquad\cdot\left[ e^{- \bar{q}^{2}(2y - z - k - j)} 
                   - e^{- \bar{q}^{2}(k - j - z)}\right]
\nonumber \\
&&+ \Theta  (j + z - k) \;\Theta (y-j-z) 
\nonumber\\
&&\qquad\cdot\left[ e^{- \bar{q}^{2}(2y - z - k - j)} - 
e^{- \bar{q}^{2}(j - k + z)}\right]
\nonumber \\
&&+ \Theta (k - j + z)\; \Theta (j - z + y - N) 
\nonumber\\
&&\qquad\cdot\left[ e^{- \bar{q}^{2}(k + j - z + 2 y - 2 N)} 
- e^{- \bar{q}^{2}(k - j + z)}\right]
\nonumber \\
&&+ \Theta (j - k - z) \;\Theta (k - N + y) 
\nonumber\\
&&\qquad\cdot\left[ e^{- \bar{q}^{2}(k + j - z + 2 y - 2 N)} 
- e^{- \bar{q}^{2}(j - k - z)}\right]    ~.
\nonumber
\end{eqnarray} 
Here the last two contributions arise from the last two terms 
in Eq.\ (4.8), which a priori involve the distribution function 
\begin{eqnarray*}
\overline{\Theta (j_{>} - j_{<}) 
\;\delta_{n_{m},n_{\mbox{{\scriptsize max}}}(N,t)} 
\;\delta_{n_{j},n(j,t)}} \: \: \:.
\end{eqnarray*}
Interchange of the chain ends transforms this distribution to 
${\cal P}_{\mbox{{\scriptsize max}},j}^{(T)}(n_{m},-n_{j};t)$ 
(Eq.\ (3.20)) and implies that we have to take $j_{>} = N-y$
and $j(t) = j-z$ in the corresponding contributions to $F (y,z;j,k)$.

By construction, our form of $S (q,t;j,k,N)$ obeys the initial condition 
(7.10). To derive Eq.\ (7.9), we apply the operator
\begin{equation}
{\cal D} = \frac{N}{p \bar{\ell}_{s}^{2} \rho_{0}}\:
\frac{\partial}{\partial t} - \frac{\partial^{2}}{\partial j^{2}}
\end{equation}
to the integral equation (7.12) to find 
\begin{eqnarray}
\lefteqn{{\cal D} S(q,t;j,k,N) = {\cal D} S^{(T)}(q,t;j,k,N)}
\\
&+& \frac{N}{p \bar{\ell}_{s}^{2} \rho_{0}}\:\int_{0}^{N}\:d j_{0}\:
{\cal P}^{*}(j_{0},t \mid 0)
\nonumber\\
&&\quad  \cdot\Big\{ e^{- \bar{q}^{2}|j_{0} - k| - \bar{q}^{2} j}
+ e^{- \bar{q}^{2}|N - j_{0} - k| - \bar{q}^{2}(N-j)} \Big\}
\nonumber \\
&+& \int^{t}_{0}d t_{0}\:\int^{N}_{0}d j_{0} \;
{\cal P}^{*}(j_{0},t_{0} \mid 0) 
\nonumber\\
&&\quad\cdot\Big\{ e^{- \bar{q}^2|j_{0} - k|}\;{\cal D} S (q,t - t_{0};j,0,N)
\nonumber \\
&&\qquad+\; e^{- \bar{q}^{2} | N - j_{0} - k |}\;{\cal D} 
S (q,t - t_{0};N-j,0,N)\Big\}     ~.
\nonumber
\end{eqnarray} 
This is an integral equation for ${\cal D} S$ which has only 
the trivial solution
\begin{equation}
{\cal D} S (q,t;j,k,N) \equiv 0 \: \: \:,
\end{equation}
provided that the inhomogeneity vanishes. We first consider 
the contribution ${\cal D} S^{(T)}$ and note that in view of Eq.\ (7.2), 
${\cal D}$ can be written as
\begin{eqnarray*}
{\cal D} = \frac{1}{c} \frac{\partial}{\partial c} 
- \frac{\partial^{2}}{\partial j^{2}} \: \: \:,
\end{eqnarray*}
resulting in  
\begin{eqnarray}
{\cal D} S^{(T)} &=& \sum_{\nu} \int^{N}_{0}\!\!\!d y
\int^{y}_{y-N}\!\!\!d z \;
\Bigg\{\left( \frac{1}{c}\; \frac{\partial}{\partial c}\; 
\hat{{\cal P}}_{\nu}(y,z)\right) F (y,z;j,k)
\nonumber \\
&&\qquad\qquad\qquad\qquad-\;\hat{{\cal P}}_{\nu}(y,z) \;
\frac{\partial^{2}}{\partial j^{2}} F (y,z;j,k)\Bigg\}~.
\nonumber\\
\end{eqnarray}
$F (y,z;j,k)$ (Eq.\ (7.15)) is a sum of terms which depend on $j$ and $z$ 
exclusively via the combinations $j + z$ or $j-z$, respectively. 
Thus $\frac{\partial^{2}}{\partial j^{2}}$ is equivalent to 
$\frac{\partial^{2}}{\partial z^{2}}$, and partial integration yields 
\end{multicols}
\begin{eqnarray*}
{\cal D} S^{(T)} 
&=& \sum_{\nu} \int^{N}_{0} dy\int^{y}_{y-N} dz \;F (y,z;j,k)\; 
\left( \frac{1}{c}\; \frac{\partial}{\partial c} 
- \frac{\partial^{2}}{\partial z^{2}}\right) \hat{{\cal P}}_{\nu}(y,z)     
\\
&&- \;\sum_{\nu}\int^{N}_{0}d y \;\Bigg\{ \hat{{\cal P}}_{\nu}(y,y) 
\;\left.\frac{\partial}{\partial z}\right|_{y} F (y,z;j,k) 
- \hat{{\cal P}}_\nu(y,y-N)\;\left.\frac{\partial}{\partial z}\right|_{y-N}
F (y,z;j,k) 
\\
&&\qquad\qquad\qquad
-\;F(y,y;j,k)\;\left.\frac{\partial}{\partial z}\right|_y
\hat{{\cal P}}_{\nu}(y,z) + F(y,y-N;j,k) \;
\left.\frac{\partial}{\partial z}\right|_{y-N}\hat{\cal P}_\nu(y,z)\Bigg\}~.     
\end{eqnarray*}
\begin{multicols}{2} 
It is easily verified that in this expression the first term vanishes 
identically, and after some calculation exploiting relations like
\begin{eqnarray*}
f (z,\nu) 
&=& (2 \nu N - z)\; \exp \left[ - \frac{(2 \nu N - z)^{2}}{2 c^{2}} \right] 
\\
&\equiv& f (z + 2 N, \nu + 1)~,
\end{eqnarray*}
we find
\begin{equation}
\sqrt{2\pi}\;c^{3} \;{\cal D} S^{(T)} = e^{- \bar{q}^{2} j}\; {\cal C}(k) 
+ e^{- \bar{q}^{2}(N-j)}\; {\cal C}(N - k)
\end{equation}
\begin{eqnarray}
{\cal C}(k) &=& 2 N \sum_{\nu} 
\Big[ 4 \nu^{2}\; f_{0} (0,\nu)\; e^{- \bar{q}^{2} k} 
\nonumber\\
&&\qquad\qquad- (2\nu - 1)^{2}\; f_0(N,\nu)\;e^{- \bar{q}^2(N-k)}\Big]
\nonumber \\
&&- 4 \bar{q}^{2} c^{2} \sum_{\nu}\;\nu \left[f_{0} (N,\nu)\; 
e^{- \bar{q}^{2} (N-k)} - 2 f_{0} (k,\nu)\right]
\nonumber \\
&&- 4 \bar{q}^{4} c^{2} \int_{0}^{N}dy\sum_{\nu}\nu\; 
f_{0} (y,\nu) \; e^{- \bar{q}^{2}|k-y|} \: \: \:,
\end{eqnarray}
where 
\begin{equation}
f_{0}(y,\nu) = \exp \left[ - \frac{(2 \nu N - y)^{2}}{2 c^{2}} \right] 
\: \: \:.
\end{equation}

We now turn to the second part of the inhomogeneity in Eq.\ (7.17), 
and we use Eq.\ (6.12) together with $d c/ d t = p \bar{\ell}_{s}^{2} 
\rho_{0}/(N c)$ (cf.\ Eq.\ (7.2)), to write 
\begin{eqnarray*}
\lefteqn{\sqrt{2\pi}\;c^{3}\; \frac{N}{p \bar{\ell}_{s}^{2} \rho_{0}}\; 
{\cal P}^{*}(j_{0},t \mid 0) =}
\\
&&4 \;\sum_{\nu}\nu \;\left( 1 - \frac{4}{c^{2}} 
\left(\nu N + \frac{j_{0}}{2}\right)^{2}\right)
\;\exp\left[-\frac{2}{c^2}\left(\nu N+\frac{j_0}{2}\right)^2\right]~.
\end{eqnarray*}
Thus
\begin{eqnarray}
&&\sqrt{2\pi}\;c^{3}\; \frac{N}{p \bar{\ell}_{s}^{2} \rho_{0}} 
\int^{N}_{0}\!\!dj_{0}\;{\cal P}^{*}(j_{0},t \mid 0) 
\nonumber\\
&&\qquad\cdot
\Big\{ e^{- \bar{q}^{2}|j_{0}-k| - \bar{q}^{2} j} 
+ e^{- \bar{q}^{2}|N-j_{0}-k| - \bar{q}^{2} (N-j)} \Big\}
\nonumber \\
& & = e^{- \bar{q}^{2}j} \;\tilde{{\cal C}}(k) + e^{- \bar{q}^{2}(N-j)} 
\;\tilde{{\cal C}}(N-k) \: \: \:,
\end{eqnarray}
with 
\begin{eqnarray}
\tilde{{\cal C}}(k) &=& 4\; \sum_{\nu}\nu\int_{0}^{N}dj_{0} \;
\left[ 1 - \frac{(2 \nu N + j_{0})^{2}}{c^{2}} \right] 
\nonumber \\
&&\cdot \exp \left[ - \frac{(2 \nu N + j_{0})^2}{2 c^2}  - 
\bar{q}^{2} |j_{0} - k|\right] \: \: \:.
\end{eqnarray}
A short calculation shows that indeed
\begin{equation}
\tilde{{\cal C}}(k) = - {\cal C} (k) \: \: \:.
\end{equation}
Thus the inhomogeneity in Eq.\ (7.17) vanishes and $S (q,t;j,k,N)$ 
obeys the diffusion equation (7.18). 

Checking the boundary conditions (7.11) is an even simpler task. 
Direct calculation yields
\begin{eqnarray}
\lefteqn{\left.\frac{\partial}{\partial j}\right|_{0}S^{(T)}(q,t;j,k,N)}
\nonumber\\
&=& \int_{0}^{N}dy\int^{y}_{y-N}d z\sum_{\nu}
\hat{{\cal P}}_{\nu}(y,z)\; 
\left.\frac{\partial}{\partial j}\right|_{0}\:F (y,z;j,k)
\nonumber \\
&\equiv& \bar{q}^{2} S^{(T)}(q,t;0,k,N) 
\end{eqnarray}
\begin{equation}
\frac{\partial}{\partial j}\:S^{(T)}(q,t;j,k,N) \equiv 
- \bar{q}^{2} S^{(T)}(q,t;N,k,N) \: \: \:,
\end{equation}
and differentiating the integral equation (7.12), we find
\begin{eqnarray}
\lefteqn{
S'(q,t;j,k,N) \equiv \frac{\partial}{\partial j}\:S (q,t;j,k,N)}
\nonumber \\
& & = \frac{\partial}{\partial j}\:S^{(T)}(q,t;j,k,N) + 
\int_{0}^{t}d t_{0}\int_{0}^{N}d j_{0}\;{\cal P}^{*}(j_{0},t_{0} \mid 0 )
\nonumber \\
&&\qquad\qquad\quad \cdot 
\Big\{ e^{- \bar{q}^{2}|j_{0}-k|}\;  S'(q,t-t_{0};j,0,N)
\\
&&\qquad\qquad\quad~~~ 
- e^{- \bar{q}^{2}|N - j_{0}-k|}\; S'(q,t-t_{0}, N-j,0,N)\Big\}~.
\nonumber
\end{eqnarray}
Now writing
\begin{eqnarray}
S'(q,t;0,k,N) &=& \bar{q}^{2} \hat{S}_{1} (q,t,k,N)~,
\nonumber \\
S'(q,t;N,k,N) &=& - \bar{q}^{2} \hat{S}_{2} (q,t,k,N)~,
\end{eqnarray}
we find from Eqs.\ (7.26) - (7.28)
\begin{eqnarray}
\lefteqn{\hat{S}_{1} (q,t,k,N) = S^{(T)}(q,t;0,k,N)}
\nonumber \\
&+& \int_{0}^{t}\!\!\!d t_{0}\int_{0}^{N}\!\!\!d j_{0}\;
{\cal P}^{*}(j_{0},t_{0} \mid 0 )\; 
\Big\{ e^{- \bar{q}^{2}|j_{0}-k|} \;\hat{S}_{1} (q,t-t_{0};0,N)
\nonumber \\
&&\qquad\qquad+  e^{- \bar{q}^{2}|N - j_{0}-k|}\; 
\hat{S}_{2} (q,t-t_{0},0,N)\Big\}
\nonumber \\
\lefteqn{\hat{S}_{2} (q,t,k,N) = S^{(T)}(q,t;N,k,N)}
\nonumber \\
&+& \int_{0}^{t}\!\!\!d t_{0}\int_{0}^{N}\!\!\!d j_{0}\;
{\cal P}^{*}(j_{0},t_{0} \mid 0 )\; 
\Big\{ e^{- \bar{q}^{2}|j_{0}-k|} \;\hat{S}_{2} (q,t-t_{0};0,N)
\nonumber \\
&&\qquad\qquad + e^{- \bar{q}^{2}|N - j_{0}-k|}\; 
\hat{S}_{1} (q,t-t_{0},0,N)\Big\}~.
\end{eqnarray}    
This is exactly the system of equations obeyed by $S(q,t;0,k,N)$ 
and $S(q,t;N,k,N)$ (cf.\ Eq.\ (6.3)). The uniqueness of the solution 
together with Eq.\ (7.29) thus guarantees that the boundary conditions 
(7.11) are obeyed.

We thus have shown that in the limit $N \rightarrow \infty$ with 
$ t/T_{3}$ and $q^{2} R_{g}^{2}$ fixed, our theory reproduces 
the results of Doi and Edwards.  



\section{Numerical evaluation and comparison to Monte Carlo data}

\subsection{Technicalities of solving the integral equations}

As shown by relations like Eqs.\ (6.19), (6.20), the natural measure 
of time in our theory is the parameter $c  = c (t)$. It measures 
the motion of the chain ends, i.e., the time dependence of tube 
destruction, and is defined by Eqs.\ (4.25), (3.16). In evaluating 
the theory, we therefore replace time by the variable
\begin{equation}
x = {\cal X} (t) = \frac{c(t,N)}{N} = \frac{1}{\hat{N}} \: \: \:.
\end{equation}
Using Eq.\ (6.15), we write the integral equation (6.6) in the form
\begin{eqnarray}
\lefteqn{\bar{S}_{E}(x) = \bar{S}_{E}^{(T)}(x)}  \\
&&+ 2 \int_{0}^{x}
\frac{d x'}{x'}\;{\cal K}_{E}\left(\bar{q}^{2} N x',\frac{1}{x'}\right) 
\;\bar{S}_{E} ({\cal X}(\tau (x) - \tau (x')))~,
\nonumber
\end{eqnarray}
where $t = \tau(x)$ is the inverse function to $x = {\cal X}(t)$.  
$\bar{S}_{E}(x)$ and $\bar{S}_{E}^{(T)}(x)$ denote the scattering functions 
$S_{E}$ and $S_{E}^{(T)}$, normalized with the static coherent 
structure function, e.g.
\begin{equation}
\bar{S}_{E}(x) = \frac{S_{E}(q,t;M,N)}{S_{c}(q,t=0;M,M)} \: \: \:.
\end{equation}
With corresponding notation, Eqs.\ (6.4) and (6.16) yield for 
the normalized coherent structure function
\begin{eqnarray}
\lefteqn{\bar{S}_{c}(x) = \bar{S}_{c}^{(T)}(x)} \\ 
&&+ \frac{4}{\bar{q}^{2}}\int^{x}_{0}\frac{d x'}{x'}\; 
{\cal K}_{c} \left(\bar{q}^{2} N x', \frac{1}{x'},\frac{M}{x' N}\right) 
\;\bar{S}_{E} ({\cal X}(\tau (x) - \tau (x')))~.\nonumber
\end{eqnarray}
In Eqs.\ (8.2), (8.4), we then transform from variables $x'$ 
to $\hat{x} = {\cal X}(\tau(x) - \tau(x'))$ to find equations 
of standard Volterra form, which are solved by discretizing 
$x$, $\hat{x}$ and iteration. We note that both $\bar{S}_{E}^{(T)}(x)$ 
and $\bar{S}_{c}^{(T)}(x)$ for $x > 2$ are negligibly small, 
less than $10^{-7}$, to be compared to the normalization 
$\bar{S}_{c}(0) = 1$. Also the kernels ${\cal K}_{E}$, ${\cal K}_{c}$ 
exceed $10^{-7}$ only in the interval $0.1 < x' < 2.2$. This allows 
for an accurate evaluation, simply using computer algebra. 
The numerical uncertainty of our final results is less than 0.5 \%. 
In all our analysis, we used the same parameter values as in 
Ref.\ \cite{Z8}. Specifically, $\bar{\ell}_{s}^{2} \rho_{0} = 1.23$,
$p = 1/5$, and $\bar{\ell}_{s} = 2.364$. (We recall that the precise
values of $p$ and $\bar\ell_s$ in fact are irrelevant for our numerical results.)

\subsection{Typical results}

We first want to illustrate the magnitude of the different contributions 
to the normalized structure function $\bar{S}_{c}$. Fig.~6 shows results 
for two very different values of wave number: 
$q^{2} R_{g}^{2} \approx 0.27$ in Fig.~6~a, and 
$q^{2} R_{g}^{2} \approx 53$ in Fig.~6~b. The thick lines give 
the full results for $\bar{S}_{c}$, including end effects and 
tube destruction. Long dashes represent $\bar{S}_{c}^{(T)}$, 
i.e., the contribution without complete tube destruction. 
Short dashes represent the (normalized) contribution 
$\bar{\cal S}_{1}$ (Eq.\ (4.33)), which omits all end-effects
and treats the chain as embedded in an infinitely long tube. 
The arrows point to the internal equilibration time $\hat{T}_{2}$, 
defined by Eq.\ (5.26): 
\begin{eqnarray*}
\hat{T}_{2} = \frac{(N+1)^2}{\pi^2}~.
\end{eqnarray*}
Finally, the heavy slashes in the time axes 
give the reptation time defined as the first moment of the time 
dependent probability density of complete tube destruction:
\begin{equation}
\hat{T}_{3} = p \int_{0}^{\infty}dt_{0}\;t_{0}\; 
\left( - \frac{\partial}{\partial t_{0}}\right) {\cal P}^{(T)} (t_{0}) \: \: \:,
\end{equation}
Here we use Eq.\ (6.19) for $\partial {\cal P}^{(T)}/\partial (t_{0})$. 
For long chains, the thus defined reptation time tends to the value 
given in Eq.\ (7.4). 

Fig.~6~a shows the typical behavior of $\bar{S}_{c}(q,t,N)$ 
for wave numbers which are too small to resolve the internal structure 
of the tube. $\bar{S}_{c}^{(T)}$ stays close to 1 up to times of order 
$\hat{T}_{3}$ and then rapidly drops to zero. The contribution of 
processes with complete tube destruction (dot-dashed line in Fig.~6~a) 
is very important, and end effects become visible only at a time 
where also tube destruction plays a role. Note that according to 
Eq.\ (6.4), the total structure function $\bar{S}_{c}$ is a sum of 
two independently calculated terms: $\bar{S}_{c}^{(T)}$ (long dashed)
and the contribution of complete tube destruction (dot-dashed). 
These terms add up to a smooth curve (thick solid), an observation 
which demonstrates the consistency of our approach on the quantitative 
level. It is only for very short chains, $N \alt 30$, that these two 
contributions do not quite match. We trace this back to our approximate 
calculation of the kernel ${\cal K}_{c}$. For such short chains, 
tube length fluctuations and internal relaxation presumably play 
a role also for the kernels.

The limit of large wave vectors is illustrated with Fig.~6~b. Here 
configurations where the original tube has been destroyed, 
essentially do not contribute to the scattering. Indeed, in Fig.~6~b, 
the curves for $\bar{S}_{c}$ and $\bar{S}_{c}^{(T)}$ fall right on top 
of each other. However, end effects like tube length fluctuations 
have a strong influence, as shown by the deviations among the full line 
and the dashed line representing $\bar{\cal S}_{1}$. They lead to a gradual 
decrease of $\bar{S}_{c}$, starting long before complete tube destruction 
becomes effective.        

In the previous section, we have shown that our theory asymptotically 
reduces to the primitive chain model of Doi and Edwards. To test the range 
of validity of the asymptotic result, we have evaluated our theory 
for the fairly large value $N = 637$ of the chain length 
(corresponding to a Monte Carlo chain of 640 beads, cf.\ Ref.\ \cite{Z8}, 
Sect.\ II.C). Fig.~7 shows the results for the normalized coherent 
structure function $\bar{S}_{c}$ as function of $\log_{10} \hat{t}$ 
for a set of wave vectors: $q^{2} R_{g}^{2} =$ 0.1, 1.0, and 10. 
The dashed lines give the asymptotic result (2.5), (2.6), where we used 
Eq.\ (7.8) for $\tau_{d}$. Obviously, the time scales do not quite match: 
even for this long chain, the reptation time does not yet follow 
the $N^{3}$-law. A shift of $\log_{10} \hat{t}$ by $- 0.1$, 
equivalent to a decrease of the time scale by 20 \%, for small $q$, 
such that $q^{2} R_{g}^{2} \alt 1$, brings the asymptotic results close to 
the results of our full model. For $q^{2} R_{g}^{2} = 10$, however, 
even with such a shift, there remains a definite difference: 
the result of the full theory initially decreases faster and approaches 
the shifted asymptotic curve only for $\hat{t} \agt 10^{2} \hat{T}_{2}$. 
This is an effect of internal relaxation and tube length fluctuations. 
The absence of a visible mismatch in the shape of the curves for 
smaller $q$-values just implies that with such small values again 
the structure of the tube cannot be resolved. 

To examine more closely the influence of internal relaxation and 
tube length fluctuation, we in Figs.~8 and 9 show results for the 
scattering from internal pieces of a chain. Fig.~8 shows results 
for $q = 0.5$ and a subchain of about 80 beads, corresponding to 
$q^{2} R_{g}^{2} \approx 3.29$. The subchain is embedded as central 
piece in chains of different lengths, precisely: $~(N = 77$, total chain$)~$, 
$~(N = 157, M = 79)~$, $~(N = 317, M = 79)~$, $~(N = \infty, M = 79)~$. 
This figure illustrates the suppression of end-effects with increasing $N$. 
The asymptotic result $N = \infty$ is due to internal relaxation only. 
For finite $N$, the curves start to deviate from the asymptotic form 
as soon as wiggles created at a chain end have a nonnegligible 
probability to reach the central piece. The characteristic time 
for this process scales with $(N - M)^{2}$. Fig.~9 shows results 
for $q = 0.5$, $N = 317$, and central pieces of lengths $M = 317$ 
to $M = 39$. Due to tube length fluctuations, the normalized scattering 
function of the total chain initially decreases faster than the result 
for $M = 159$. Tube destruction on the average reaches the sub-chains 
at times between $\hat{t} \approx 10^{5.8}~(M = 159)$ or 
$\hat{t} \approx 10^{6.4}~(M = 39)$, so that for large time regimes, 
the results for the sub-chains are not affected by tube length 
fluctuations. The decrease of the normalized scattering intensity 
with decreasing length of the subchain rather is due to the fact 
that a shorter subchain leaves its initial position in the tube earlier, 
i.e., it is due to internal relaxation.  

In Figs.~8 and 9, we included results from a simulation of 
the Evans-Edwards model \cite{Z16}. This model takes the chain 
configuration as a random walk on a cubic lattice and allows only 
moves of `hairpin'-configurations ${\bf r}_{j+1} - {\bf r}_{j} 
= {\bf r}_{j-1} - {\bf r}_{j}$ as internal motion. An illustration 
for a two-dimensional system is shown in Fig.~1. We used the same 
implementation of the model as in our previous work \cite{Z8}, 
to which we refer for details. In comparing theory and simulations 
therefore all parameters are fixed by our previous analysis of segment 
motion. Since, however, the new simulations lead to better statistics, 
we allow for some readjustment of the relation among $\hat{t}$ and 
the Monte Carlo time scale: $\hat{t} = 6.8 \cdot 10^{-2}\;t_{MC}$ 
instead of $\hat{t} = 6.09 \cdot 10^{-2}\;t_{MC}$ taken previously. 
This yields a shift of $-0.048$ of the logarithmic time scale and 
leaves our previous results essentially unaffected.

As shown in Fig.~8, our theory in all details reproduces 
the time variations of the data, but the data systematically lie 
somewhat below the theoretical curves. This is not due to our 
approximations which essentially only concern the treatment of 
tube length fluctuations. Considering for instance the data for 
$N = 317$, we note that the deviations from the theory are strongest 
for $\hat{t} \alt 10^{5}$, where tube destruction and tube length 
fluctuations are irrelevant and our theory for the internal part $M=79$ 
is an exact evaluation of the reptation model. Furthermore, the deviations 
are fairly independent of the lengths of the end pieces $(N-M)/2$. 
This suggests that we see some (non-universal) relaxation of the 
micro-structure. Clearly particle hopping, which is the elementary 
dynamics of the reptation model, is no faithful representation 
of the Monte Carlo hairpin dynamics on the microscopic level. 
The wave-vector $|{\bf q}| = 0.5$ is large enough to resolve such details. 
Since the dynamic effects of micro-structure should saturate at larger 
times, this suggests that we should scale down the theoretical curves 
by some factor $B_{R} < 1$. This was done in Fig.~9, where $B_{R}$ 
ranges from 0.981 to 0.990, depending on $M$. For $\hat{t} \agt 10^{3}$, 
theory and data agree excellently. The same level 
of agreement can be reached for the data of Fig.~8. We therefore believe 
that our theory adequately describes the universal part of the coherent 
scattering function, including tube length fluctuations and (universal) 
internal relaxation.

A more extensive presentation of simulation results, comparing with 
the present and previous theories, will be published elsewhere. 
Here we only note that we have taken data for values of $|{\bf q}|$ 
ranging from 1.0 to 0.1, and the $q$-dependence found for the initial 
deviation among theory and simulations strongly supports 
the interpretation as micro-structure effects.


\section{Summary and conclusions}

In this work, we have exploited the pure reptation model to calculate 
the coherent structure function $S_{c}(q,t;M,N)$ of a flexible chain 
moving through an array of impenetrable topological obstacles. 
Our analysis is rigorous for a subchain (length $M$) in the interior 
of an infinitely long chain $(N \rightarrow \infty)$. This allows for 
a detailed comparison with an approach where the interior motion of the 
chain is modeled as one dimensional Rouse motion along a coiled tube. 
We found that the latter model starts from unphysical non-equilibrium 
initial conditions, which relax only on the scale of the Rouse
time $T_{2}(M)$ of the subchain. This relaxation completely distorts 
the time dependence of $S_{c}(q,t;M,N = \infty)$ for times 
$t \alt T_{2} (M)$. Only for times $t\gg T_2(M)$, this model is equivalent 
to the reptation model. If applied to the total chain,
'local relaxation' calculated as Rouse motion in a tube
therefore is unphysical.
A realistic system may show some relaxation which is specific 
to the microscopic dynamics, and which is not contained in the pure 
reptation model. However, our analysis 
sheds strong doubts on an interpretation of such non-universal effects 
within the framework of the model of a Rouse chain in a tube. 

To evaluate the total structure function for all times, we have derived 
integral equations which split $S_{c}$ into a contribution $S_{c}^{(T)}$ 
of configurations where some part of the initial tube still exists, 
and the remainder. The kernel and in particular the inhomogeneities 
(like $S_{c}^{(T)}$) of these equations cannot be calculated rigorously. 
They involve distribution functions coupling the motion of a given segment 
to tube renewal, which is a non-Markovian process with memory time 
of the order of the Rouse time $T_{2} (N)$. To calculate 
the functional form of these distributions, we used a random walk 
approximation. 
We thus at each instant of time replaced the correlated process 
by an uncorrelated process which as closely as possible reproduces 
the instantaneous distributions of the correlated process. 
This `mean hopping rate' approximation introduces functions 
$c = c(t)$, $a = a (j,t)$, and $b = b (j,t)$ which appear as parameters 
in the distribution functions and can be calculated from the microscopic 
hopping process of spared length in the reptation model. They also 
have a simple physical meaning: $c(t)$ measures the average extent 
of tube destruction, $a(j,t)$ describes the coupling of motion of 
segment $j$ to tube destruction, and $b (j,t)$ takes care of the 
inhomogeneity of the effective segment mobility along the chain 
which arises from the fact that the mobile units, i.e., the wiggles 
of spared length, can be created and destroyed only at the chain ends. 

In the limit of long chains and time $t \gg T_{2}$, the parameters $a$
and $b$ tend to 1 irrespective of $j$, and $c \sim t^{1/2}$. Our theory 
then reproduces the results of the `primitive chain' model. Our proof of 
this asymptotic result amounts to a derivation of the primitive 
chain model from microscopic reptation dynamics. Combined with the 
analysis of the internal motion for $t \ll T_{2}$, this result allows 
for a mapping of the microscopic parameters of our reptation model 
to the more commonly used Rouse-type parameters.

The parameter functions $a (j,t)$, $b (j,t)$, and $c(t)$ approach 
their asymptotic behavior only slowly, and it needs chain lengths 
of order $N/N_{e} \agt 300$ to find a time region where the primitive 
chain model is valid. In particular, for shorter chains the reptation 
time, if extracted by fitting the Doi-Edwards result for the primitive 
chain model to the large time behavior of $S_{c}(q,t,N)$ in the reptation 
model, does not obey the asymptotic power law $T_{3} \sim N^{3}$. 
As will be shown in Ref.\ \cite{Z13}, it rather exhibits the well 
known behavior $T_{3} \sim N^{z_{\rm eff}}$, with an effective 
exponent $z_{\rm eff} > 3$. The deviation of $a (j,t)$, $b (j,t)$, and
$c(t)$ from their asymptotic primitive chain behavior incorporates 
the effect of internal relaxation and tube length fluctuations. 
Our numerical evaluation of the full theory illustrates that these 
effects in general are quite important. In particular, we find a 
clear difference among the time variation of scattering from the 
total chain compared to scattering from internal pieces. 
The latter are less influenced by tube length fluctuations 
but are more strongly affected by internal relaxation. 
This leads to a peculiar behavior of $S_{c}(q,t;M,N)$ with varying 
length $M$ of the internal piece, as shown in Fig.~9. Quite generally, 
for the total chain $(M=N)$ it is the tube length fluctuations, 
that determine $S_{c}$ for times up to $t \approx 10\: T_{2}$.

All our quantitative numerical results are well supported by simulations 
of pure reptation, exploiting the lattice model of Evans and Edwards. 
In view of the unavoidable approximations inherent in the theory, 
the quantitative agreement is quite remarkable. It suggests that our 
mean hopping rate approximation adequately takes care of the 
coupling among internal relaxation, tube length fluctuations, 
and global creep. A more extensive comparison to Monte Carlo data 
including a numerical parametrization of our analytical results
will be published elsewhere \cite{Z13}.  


{\bf Acknowledgement:}
This work was supported by the Deutsche Forschungsgemeinschaft, SFB 237.


\end{multicols}

\newpage

\appendix

\section{Random walk model for distribution functions}

\subsection{The function
${\cal P}^{(T)}_{\mbox{{\scriptsize max}},0}(n_{m},n_{0};t)$}

As explained in Sect.\ IV.C, we consider a random walk $n'(s)$ 
on the integer numbers, with hopping rate $p'$. The walk starts 
at $n'(0) = 0$ and ends at $n'(t) = n_{0}$. It is restricted 
to the interval $[n_{m} - N'+1, n_{m}]$, with absorbing boundary 
conditions. To simplify the notation, we shift the interval by $N'-n_{m}$ to
\begin{equation}
{\cal I} = [1, N'] \: \: \:.
\end{equation}
with the starting point of the walk $n'_{0} = N'-n_{m}$, and
the endpoint $n'(t) = n_{0} + N'-n_{m}$. 
The hopping matrix of the walk takes the form
\begin{equation}
\hat{W} (N')_{j,j'} = (1 - 2 p') \delta_{j j'} 
+ p' (\delta_{j ,j'+1} + \delta_{j',j-1}),\: \: \: \:  
j, j' \in\: {\cal I} \: \: \:.
\end{equation}
${\cal P}^{(T)}_{\mbox{{\scriptsize max}},0}(n_{m},n_{0};t)$ gives 
the weight of the walk under the constraint that $n'(s) = n_{m}$ 
is attained for at least one $s \in [0,t]$. It is easily found as 
\begin{eqnarray}
\lefteqn{
{\cal P}^{(T)}_{\mbox{{\scriptsize max}},0}(n_{m},n_{0};t) 
= \Theta \left(n_{m} - \frac{|n|+n}{2}\right)\; 
\Theta \left(\frac{n-|n|}{2} - n_{m} + N' - 1\right)}
\nonumber \\
& & ~\cdot \Big\{ (\hat{W}^{t}(N'))_{1+n_{m}-n,1+n_{m}} - 
(1 - \delta_{n_{m},0})\; (\hat{W}^{t} (N' - 1))_{n_{m}-n,n_{m}}\Big\}~.
\end{eqnarray}
$\hat{W}^{t}(N')$ can be written as
\begin{eqnarray}
(\hat{W}^{t}(N'))_{j,j'} = \frac{2}{N'+1} \sum^{N'}_{\kappa = 1}
\sin\frac{\pi \kappa j}{N' + 1} \;\sin\frac{\pi \kappa j'}{N' + 1}\; 
\left(1 - 4 p' \sin^{2}\frac{\pi \kappa}{2 (N' + 1)}\right)^{t} \: \: \:.
\end{eqnarray}
For $t \gg 1$, the last factor can be replaced by 
$\exp \left( - p' t \frac{\pi^{2} \kappa^{2}}{(N' + 1)^{2}}\right)$, 
and a little calculation yields 
\begin{eqnarray}
\lefteqn{
{\cal P}^{(T)}_{\mbox{{\scriptsize max}},0}(n_{m},n,t) \approx 
\Theta \left(n_{m} - \frac{|n| + n}{2}\right)\; 
\Theta \left( \frac{n - |n|}{2} - n_{m} + N' - 1\right)}
\nonumber \\
& & ~~\cdot\Bigg\{ \frac{1}{N'+1} \sum^{\infty}_{\kappa = 1}
e^{\textstyle - p' t \;\frac{\pi^{2} \kappa^{2}}{(N' +1)^{2}}} 
\left[ \cos \frac{\pi \kappa\; n}{N' + 1} 
- \cos \frac{\pi \kappa\;(2 n_{m} + 2 - n)}{N' + 1} \right]
\nonumber \\
& & ~~\qquad-\; \frac{1}{N'} \sum^{\infty}_{\kappa = 1}
e^{\textstyle - p' t \;\frac{\pi^{2} \kappa^{2}}{N'^{2}}} 
\left[ \cos \frac{\pi \kappa\;n}{N'} 
- \cos \frac{\pi \kappa\;(2 n_{m} - n)}{N'} \right] \Bigg\}~.
\end{eqnarray}
In extending the sums to infinity, we neglect terms of order 
$\exp (- \pi^{2} p' t)$.

Now we note that ${\cal P}^{(T)}_{\mbox{{\scriptsize max}},0}$ 
rapidly decreases for $t > T_{3}$. As mentioned in Sect.\ III.B, 
the effective hopping rate for $t \sim T_{3} \gg T_{2}$ behaves as 
$p' \sim 1/N \sim 1/N'$, and as a consequence, the argument of the 
exponential in Eq.\ (A.5) takes the form $- \mbox{const}\: 
\kappa^{2} t/T_{3}$. Thus ${\cal P}^{(T)}_{\mbox{{\scriptsize max}},0}$ 
yields a relevant contribution only for smaller times: $t \alt T_{3}$, 
and for treating this time range, it is preferable to replace 
the summations in Eq.\ (A.5) by their Poisson transform. This yields 
\begin{eqnarray}
\lefteqn{
{\cal P}^{(T)}_{\mbox{{\scriptsize max}},0}(n_{m},n;t) 
= \Theta \left(n_{m} + \frac{|n| + n}{2}\right)\; 
\Theta \left( \frac{n - |n|}{2} - n_{m} + N' - 1\right) 
\;\frac{1}{\sqrt{4 \pi p' t}}}
\nonumber \\
& & \cdot \sum^{+\infty}_{\nu = - \infty} 
\Bigg\{\exp\left(-\frac{1}{p't}\left(\nu N' + \frac{n}{2}\right)^{2} \right) 
\left[ \exp \left(- \frac{2\nu}{p' t} \left(\nu N' + \frac{n}{2}\right) 
          - \frac{\nu^{2}}{p' t}\right) - 1\right]
\\
& &\qquad-\;\exp
\left(-\frac{1}{p't}\left(\nu N'+n_m-\frac{n}{2}\right)^2\right) 
\left[\exp\left(-\frac{2\nu+2}{p't}\left(\nu N'+n_m-\frac{n}{2}\right) 
             - \frac{(\nu + 1)^{2}}{p' t}\right) - 1 \right] \Bigg\} ~.
\nonumber
\end{eqnarray}
Now the Gaussian prefactors of the square brackets allow for an 
essential contribution only for $p' t \agt N'^{2}$, and in this region, 
the square brackets can be replaced by the linear approximation 
\begin{eqnarray*}
&&\exp \left[ - \frac{2\nu}{p' t} \left(\nu N' + \frac{n}{2}\right) 
- \frac{\nu^{2}}{p' t}\right] - 1 \\
&\approx& - \frac{2\nu}{p' t} \left(\nu N' + \frac{n}{2}\right)\;
\exp \left[ - \frac{2\nu + 2}{p' t} \left(\nu N' + n_{m} - \frac{n}{2}\right) 
- \frac{(\nu + 1)^{2}}{p' t}\right] - 1 \\
&\approx& - \frac{2(\nu +1)}{p't}\;\left(\nu N'+n_m-\frac{n}{2}\right)~.
\end{eqnarray*}
Eq.\ (4.20) from Sect.\ IV.C is the result, which is correct 
to leading order in $1/N'$. The neglect of $1/N'$-corrections 
is consistent with treating segment indices as continuous.
 
\subsection{The function ${\cal P}^{*} (j_{0},t_{0}  \mid  0)$}

By definition, ${\cal P}^{*} (j_{0},t_{0}  \mid  0)$ gives the probability 
that the initial tube is destroyed completely at time step $t_{0}$, 
with $j_{0}$ being the last point, occupied  by chain end $0$ 
(see Sect.\ VI.A). In the random walk model, this probability is given 
by the weight of a walk starting at $n'(0) = 0$ and ending at 
$n'(t_{0}) = j'_{0} = j_{0}/\ell_{s}$. The point $j'_{0}$ is reached 
at $t = t_{0}$ for the first time, but the point $j'_{0} - N' + 1$ 
is attained for some $s \in [0,t_{0} - 1]$. The walk is restricted 
to the interval $[j'_{0} - N' + 1, j'_{0}]$. Shifting the interval 
by $N' - j'_{0}$, we can express this weight as
\begin{equation}
{\cal P}^{*} (j_{0},t_{0}  \mid  0) 
= p' \left[ (\hat{W}^{t_{0}-1}(N' - 1))_{N'-1,N'-j'_{0}} 
- (\hat{W}^{t_{0}-1}(N' - 2))_{N'-2,N'-j'_{0}-1}\right]\: \: \:,
\end{equation}
where the prefactor $p'$ gives the probability of the last step, 
leading from $N'-1$ to $N'$ (in the shifted walk). Using the explicit 
expression (A.4) for $\hat{W}^{t}$ and exploiting $t = t_{0} - 1 \gg 1$, 
we find
\begin{eqnarray}
{\cal P}^{*} (j_{0},t_{0}  \mid  0) 
&=& \frac{p'}{N'} \sum^{\infty}_{\kappa=1}
\left(\cos \frac{\pi \kappa\;(j'_{0} - 1)}{N'}
- \cos \frac{\pi \kappa\;(j'_{0} + 1)}{N'}\right) 
\;e^{\textstyle - p'(t_{0}-1)\;\frac{\pi^{2}\kappa^{2}}{N^{'2}}}
\nonumber \\
&&- \;\frac{p'}{N'-1} \sum^{\infty}_{\kappa = 1}
\left(\cos \frac{\pi \kappa\;(j'_{0} - 1)}{N'-1} 
- \cos \frac{\pi \kappa\;(j'_{0} + 1)}{N'-1}\right) 
\;e^{\textstyle - p'(t_{0}-1)\;\frac{\pi^{2}\kappa^{2}}{(N'-1)^{2}}}
\end{eqnarray}    
correct up to exponentially small terms (cf.\ Eq.\ (A.5)). 
The Poisson transform yields 
\begin{eqnarray}
{\cal P}^{*} (j_{0},t_{0}  \mid  0) 
&=& \frac{p'}{\sqrt{4\pi p'(t_{0}-1)}}\:\sum^{+\infty}_{\nu = - \infty}\:
\exp\left[-\frac1{p'(t_0-1)}\;\left(\nu N'+\frac{j'_0}{2}\right)^2\right]
\nonumber \\
&&\cdot \Bigg\{\exp \left[\frac{1}{p'(t_{0}-1)}\; 
\left(\nu N' + \frac{j'_{0}}{2} - \frac{1}{4}\right)\right]
- \exp \left[\frac{-1}{p'(t_{0}-1)}\;
\left(\nu N' + \frac{j'_{0}}{2} + \frac{1}{4}\right)\right]
\nonumber\\
&& \quad- \;\exp \left[\frac{1}{p'(t_{0}-1)} \;
\left(2 \nu\;\left(\nu N' + \frac{j'_{0}}{2} - \frac{1}{2}\right)
+ \nu N' + \frac{j'_{0}}{2} - \frac{1}{4} - \nu^{2}\right)\right] 
\nonumber \\
&& \quad+\; \exp \left[\frac{1}{p'(t_{0}-1)}\; 
\left(2 \nu\;\left(\nu N' + \frac{j'_{0}}{2} + \frac{1}{2}\right) 
- \nu N' - \frac{j'_{0}}{2} - \frac{1}{4} - \nu^{2}\right)\right]
\Bigg\}  ~.
\end{eqnarray} 
As in Eq.\ (A.6), we can expand in the square brackets, keeping 
the first non-vanishing terms which here are of second order. 
Identifying now $N' = N/\bar{\ell}_{s}$ and 
$j'_{0} = j_{0}/\bar{\ell}_{s}$, Eq.\ (6.11) in Sect.\ VI.C results.


\section{Expression for ${\cal P}_{\lowercase{j}}^{(T)}({\lowercase{n_j;t}})$}

According to Eqs.\ (4.28) and (4.30), ${\cal P}_{j}^{(T)}(n_{j};t)$ 
can be determined by integrating 
${\cal P}^{(T)}_{\mbox{{\scriptsize max}},j}$ over $n_{m}$ 
(or $y$, equivalently). This integral can be carried through 
analytically to yield  
\begin{eqnarray}
\lefteqn{
{\cal P}^{(\nu)}_{j}(z,a) 
= \exp \left(-\frac{(z - 2 a \nu \hat{N})^{2}}{2}\right)}
\nonumber\\ 
&&~\cdot\Bigg\{ 2 (2-a^{2})\nu\;\mbox{erfc}
\left(\frac{a z + a_2 \nu\hat{N}}{\sqrt{a_2}}\right)
- \left((2-a^{2}) \nu + \frac{1}{2}\right)\;\mbox{erfc}
\left(\frac{a z + a_2\nu\hat{N} + \hat{N}}{\sqrt{a_2}}\right)
\nonumber \\
& &\qquad-\; \left((2-a^{2}) \nu - \frac{1}{2}\right)\;\mbox{erfc}
\left(\frac{a z + a_2\nu\hat{N}-\hat{N}}{\sqrt{a_2}}\right)
\nonumber \\
& &\qquad+\;\nu a(z-2\nu a\hat{N}) \Bigg[ 2 (a z + a_2\nu\hat{N}) 
\;\mbox{erfc}\left(\frac{a z + a_2\nu\hat{N}}
{\sqrt{a_2}}\right)
\nonumber \\
& &\qquad\qquad\qquad\qquad\qquad
-\; (a z + a_2\nu\hat{N}+\hat{N})\;\mbox{erfc}
\left(\frac{a z + a_2\nu\hat{N} + \hat{N}}{\sqrt{a_2}}\right)
\nonumber \\
& &\qquad\qquad\qquad\qquad\qquad
-\;(a z + a_2\nu\hat{N}-\hat{N}) \;\mbox{erfc}
\left(\frac{a z + a_2\nu\hat{N}-\hat{N}}{\sqrt{a_2}}\right) 
\Bigg]
\nonumber \\
& &\qquad-\; \sqrt{\frac{a_2}{\pi}}\;\nu a (z - 2 \nu a \hat{N}) 
\;\Bigg[ 2 \exp 
\left(- \frac{(a z + a_2\nu\hat{N})^{2}}{a_2}\right)
- \exp
\left(- \frac{(a z + a_2\nu\hat{N} + \hat{N})^{2}}{a_2}\right) 
\nonumber\\
& &\qquad\qquad\qquad\qquad\qquad\qquad -\; \exp
\left(- \frac{(a z + a_2\nu\hat{N}-\hat{N})^{2}}{a_2}\right) 
\Bigg] \Bigg\}
\end{eqnarray}
with $a_2=2(1-a^2)$.


\section{Modeling a Rouse chain in a coiled tube}

The model describes the internal dynamics of the reptating chain 
as that of a one-dimensional Rouse chain, stretched so as to span 
the contour length of the tube. The potential energy takes the form
\begin{equation}
\frac{{\cal V}}{k_{B}T} = \frac{1}{4\ell^{2}} 
\sum^{N}_{j=1}\:(x_{j} - x_{j-1})^{2} - \frac{h}{\ell}\:(x_{N} - x_{0})~,
\end{equation}
where the $x_{j}$, $j = 0,\ldots,N$ are the bead positions, 
$\ell$ measures the mean segment size, and $h/\ell$ is the stretching 
force acting on the end beads. The average extension of the chain 
is easily calculated as 
\begin{equation}
L = \left\langle x_{N} - x_{0}\right\rangle = 2 \ell\:h\:N \: \: \:.
\end{equation}

The dynamics of the chain is given by a Langevin equation.
\begin{equation}
\frac{d}{d t}\: x_{j} = - \gamma_{0} \;\frac{\partial}{\partial x_{j}} 
\;\frac{{\cal V}}{k_{B}T} + \xi_{j}
\end{equation}
The fluctuating force $\xi_{j}$ is Gaussian distributed:
\begin{equation}
{\cal P} [\xi_{j}(t)] = {\cal N}^{-1}\:\exp 
\left[-\frac1{4\gamma_0}\int^{+\infty}_{-\infty}\:dt\:\xi_{j}^2(t)\right]~.
\end{equation}

It is a standard exercise to calculate the dynamical structure functions. 
Indeed, for this system of coupled harmonic oscillators, the stretching 
force does not influence the dynamics but changes the static prefactor only. 
For the scattering from a pair $(j,k)$ of beads, one finds
\begin{equation}
S^{(1 d)}(p,t;j,k) = \left\langle ~
\overline{e^{\textstyle i p (x_{j}(t) - x_{k}(0))}}~\right\rangle 
= e^{\textstyle 2 i p \ell h (j-k)}\; e^{\textstyle - p^{2} D_{j k}(t)}~,
\end{equation}
where
\begin{eqnarray}
D_{j k}(t) &=& \frac{\gamma_{0}|t|}{N+1} + \ell^{2} |j - k|
\nonumber \\
&&+ \;\frac{2 \gamma_{0}}{N+1} \sum^{N}_{\kappa = 1}\:
\cos \left(\pi \kappa \frac{j + \frac{1}{2}}{N+1}\right)\; 
\cos \left(\pi \kappa \frac{k + \frac{1}{2}}{N+1}\right)\; 
\frac{1 - e^{- \omega_{\kappa}|t|}}{\omega_{\kappa}}\: \: \:,
\end{eqnarray}
\begin{equation}
\omega_{\kappa} = 2 \;\frac{\gamma_{0}}{\ell^{2}}\;\sin^{2}
\frac{\pi \kappa}{2(N+1)} 
\approx \frac{\gamma_{0}}{2 \ell^{2}}\;\pi^{2}\;
\frac{\kappa^{2}}{(N+1)^{2}} \: \: \:.
\end{equation}
Using the approximate form of $\omega_{\kappa}$, we neglect 
some exponentially small micro-structure effects.

In the analysis of Sect.\ V, we need this result for two segments 
deep inside a very long chain, for times small compared to the Rouse 
relaxation time of the total chain. Writing $j = N/2 + \tilde{j}$, 
$k = N/2 + \tilde{k}$ and taking the limit $N \rightarrow \infty$, 
with $\gamma_{0} t/\ell^{2} \gg 1$ fixed, one finds 
\begin{equation}
D_{j k}(t) = \ell^{2}\; |\tilde{j} - \tilde{k}|  + 
\ell\; \sqrt{2 \gamma_{0} t}\;
g\left(\frac{ |\tilde{j} - \tilde{k}|\:\ell}{\sqrt{2 \gamma_{0} t}}\right)~,
\end{equation}
where 
\begin{equation}
g(z) = \frac{1}{\sqrt{\pi}}\;e^{- z^{2}} - z\;\mbox{erfc} z \: \: \:.
\end{equation}

Now assume that the chain is embedded in a coiled tube, 
consisting of $N/N_{e}$ segments of fixed length $\ell_{T}$. 
The contour length of the tube equals the length of the stretched chain:
\begin{equation}
\ell_{T}\:\frac{N}{N_{e}} = L = 2 \ell\:h\:N~.
\end{equation}
The end-to-end distance of the tube $R_{e}^{2} = \ell_{T}^{2} 
\frac{N}{N_{e}}$ must match the end-to-end distance of the physical chain. 
The potential energy of the latter is given by the three-dimensional 
version of Eq.\ (C.1) in the absence of the stretching force, 
which results in $R_{e}^{2} = 6 \ell^{2} N$. Thus
\begin{equation}
\frac{\ell^{2}_{T}}{\ell^{2}} = 6 \:N_{e} \: \: \:,
\end{equation}
and Eq.\ (C.10) yields
\begin{equation}
2 h = \sqrt{\frac{6}{N_{e}}} \: \: \:.
\end{equation}

To calculate the scattering from segments $j$ and $k$ of the stretched 
one-dimensional Rouse chain embedded in the tube, we write 
\begin{eqnarray}
\lefteqn{
S^{(R)}(q,t;j,k) = \left\langle~ 
\overline{e^{\textstyle i {\bf q}({\bf r}_{j}(t) - {\bf r}_{k}(0))}}
~\right\rangle^{\rm Tube}}
\nonumber \\
& & = \int_{- \infty}^{+\infty}d x\;\left\langle~ 
\overline{e^{\textstyle i {\bf q}({\bf r}_{j}(t) - {\bf r}_{k}(0))}\; 
\delta (x_{j}(t) - x_{k}(0) - x)}~\right\rangle^{\rm Tube}
\nonumber \\
& & = \int_{- \infty}^{+\infty}d x\;\left.\left\langle~ 
e^{\textstyle i {\bf q}({\bf r}_{j}(t) - {\bf r}_{k}(0))}~\right\rangle 
\right|_{x_{j}(t) - x_{k}(0) = x}  
\left\langle~ \overline{ \delta (x_{j}(t) - x_{k}(0) - x)}~\right\rangle~. 
\end{eqnarray}  
Here the first factor is to be calculated under the constraint 
that the points ${\bf r}_{j}(t)$ and ${\bf r}_{k}(0)$ have distance $x$ 
measured along the tube. It is thus given by the static correlation 
function of a chain of $x/\ell_{T}$ segments of fixed length $\ell_{T}$:
\begin{equation}\left.
\left\langle e^{\textstyle i{\bf q}({\bf r}_j(t)-{\bf r}_k(0))}\right\rangle 
\right|_{x_{j}(t) - x_{k}(0) = x} = e^{\textstyle-\frac{q^2}{6}\ell_T|x|}~.
\end{equation}
The second factor in Eq.\ (C.13) is the (1-dimensional) Fourier transform 
of $S^{(1 d)}$ (Eq.\ (C.5))
\begin{eqnarray}
\left\langle~\overline{ \delta (x_{j}(t) - x_{k}(0) - x)}~\right\rangle 
&=& \int^{+\infty}_{- \infty}\frac{d p}{2\pi}\;
e^{\textstyle - i p x} \;S^{(1 d)}(p,t;j,k)
\nonumber \\
&=& (4 \pi D_{j k}(t))^{-1/2} 
\exp \left[ - \frac{(x - 2 h \ell (j-k))^{2}}{4 D_{jk}(t)}\right]~.
\end{eqnarray}
Substituting Eqs.\ (C.14) and (C.15) into Eq.\ (C.13), we can carry out 
the integral to find
\begin{equation}
S^{(R)}(q,t;j,k) = \frac{1}{2}\;e^{\textstyle Q^{2}} 
\Big\{ e^{\textstyle 2 \Delta Q}\:\mbox{erfc}\: (Q + \Delta) 
+ e^{\textstyle - 2 \Delta Q}\:\mbox{erfc}\: (Q-\Delta)\Big\}
\end{equation}
\begin{equation}
Q = q^{2} \ell \sqrt{\frac{N_{e}}{6} D_{jk}(t)}
\end{equation}
\begin{equation}
\Delta = \frac{\ell (j-k)}{2 \sqrt{\frac{N_{e}}{6} D_{jk}(t)}}~.
\end{equation}
If we take for $D_{jk}(t)$ the result (C.8), the variables $Q$ and
$\Delta$ reduce to $\tilde{Q}$ and $\tilde{\Delta}$ given in Eq.\ (5.15), 
and the result (C.16) becomes identical to the expression (5.14).

A final remark on De Gennes' approximation \cite{Z11} may be appropriate. 
The derivation starts from Eq.\ (C.5) with $D_{jk}$ taken from Eq.\ (C.8). 
Aiming directly at the coherent structure function, one integrates 
this expression over $j$ and $k$. If we ignore end effects, this yields
\begin{eqnarray*}
\frac{1}{N}\int dj \;dk\; S^{(1 d)}(p,t;j,k) 
&\approx& \int_{0}^{\infty}ds\;\exp 
\left[- p^{2} \ell^{2} s - p^{2} \ell \sqrt{2 \gamma_{0} t}\; 
g \left(\frac{s \ell}{\sqrt{2 \gamma_{0} t}}\right)\right]\\
&&\cdot\left(e^{\textstyle 2ip\ell hs} + e^{\textstyle-2ip\ell hs}\right)~,
\end{eqnarray*}
where $s=|j-k|$. To evaluate this integral analytically, one expands 
$\exp \left[ - p^{2} \ell \;\sqrt{2 \gamma_{0} t}\;
g \left(\frac{s \ell}{\sqrt{2 \gamma_{0} t}}\right)\right]$ 
up to first order. The remaining steps closely follow our derivation 
given above and result in the form (2.12) of the `local' contribution 
to the coherent structure function. It should be noted that the expansion 
is valid only for $p^{2} \ell \;\sqrt{2\gamma_{0} t} \ll 1$. The analysis 
supposes that $t$ is small compared to the Rouse relaxation time, 
and in this time regime, $\sqrt{2\gamma_{0} t}$ is of the order 
of the mean square distance moved by a segment along the tube. 
Thus the condition $p^{2} \ell\; \sqrt{2\gamma_{0} t} \approx p^{2} 
\left\langle ~\overline{(x_{j}(t) - x_{j}(0))^{2}}~\right\rangle \ll 1$ 
implies that the wave number $p$ cannot resolve the motion of a segment.       


\begin{multicols}{2}

\end{multicols}

\newpage

\begin{figure}
\label{fig1}
\begin{center}
\epsfig{figure=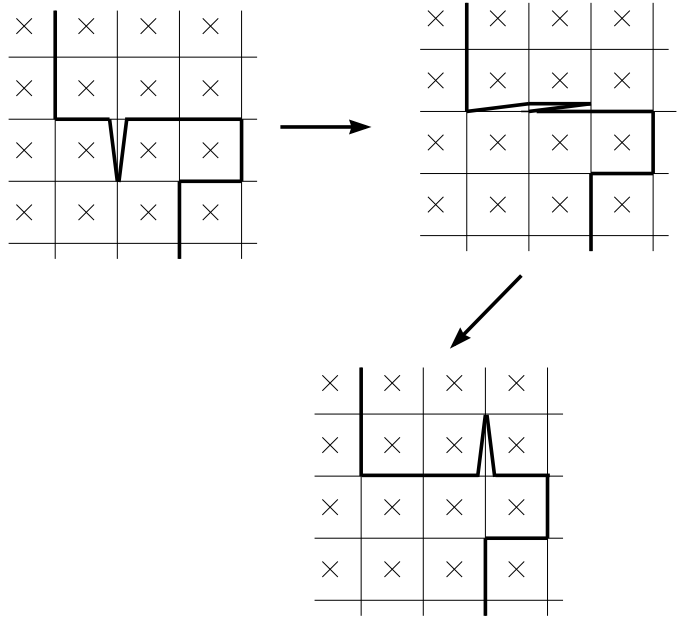,width=0.5\linewidth} 
\caption[]{A realization of reptational dynamics in the Evans-Edwards 
lattice model (two dimensional illustration). The crosses denote 
impenetrable obstacles which allow only for `hairpin' moves 
as shown by the sequence of pictures. The hairpins represent 
the wiggles of spared length.}
\end{center}
\end{figure}


\begin{figure}
\label{fig2}
\begin{center}\epsfig{figure=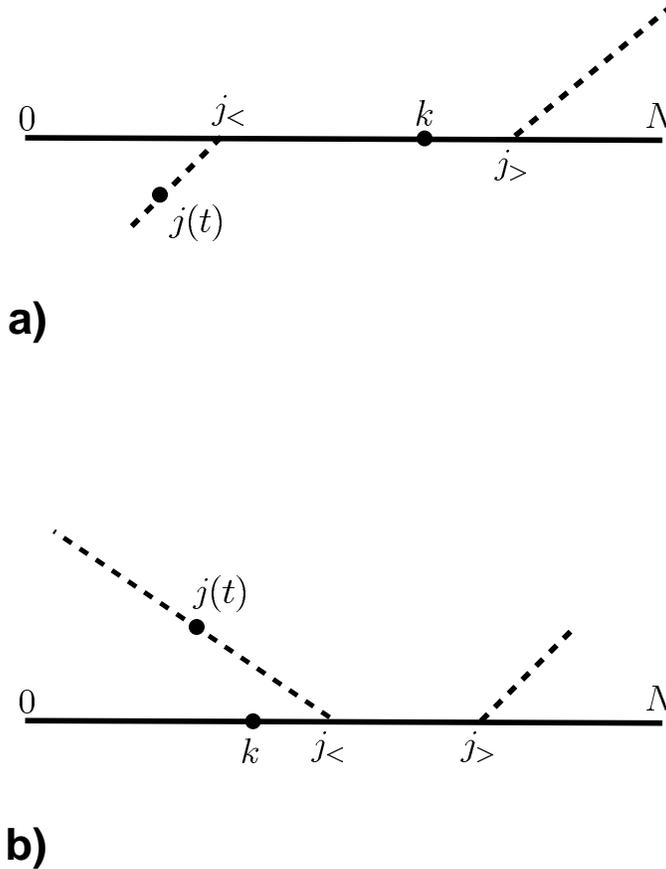,width=0.5\linewidth} 
\caption[]{
Schematic drawings of the cases discussed in the text. The full line 
represents the unfolded initial tube. Broken lines represent the new 
end pieces of the unfolded tube created up to time $t$.}
\end{center}
\end{figure}


\begin{figure}
\label{fig3}
\begin{center}
\epsfig{figure=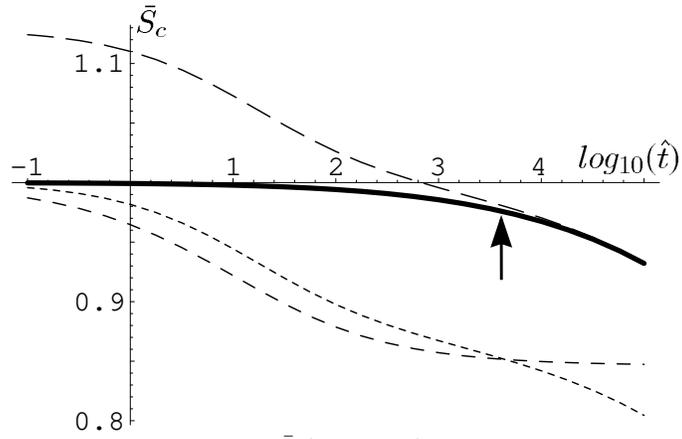,width=0.5\linewidth} 
\caption[]{
Coherent normalized structure function $\bar S_{c} (q,t;M,\infty)$ 
of the central piece of an infinitely long chain for 
$q^{2} R_{g}^{2}(M) = 50$, $M = 200$. Fat line: 
reptation result (Eq.\ (5.4)); long dashes and short dashes: 
Rouse chain in a coiled tube. For the latter curve, a constant has been
subtracted such that $\bar S_c(q,0;M,\infty)=1$.
Medium size dashes: De Gennes' approximation (Eq.\ (5.11)).
The arrow points to $\hat T_2(M=200)$.}
\end{center}
\end{figure}


\begin{figure}
\label{fig4}
\begin{center}
\epsfig{figure=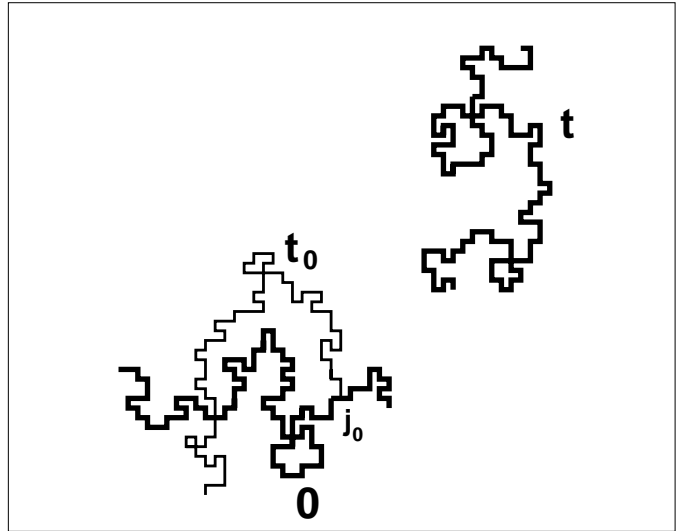,width=0.5\linewidth} 
\caption[]{
Initial $(0)$ and final $(t)$ configuration of the chain (fat lines), 
together with the configuration at time $t_{0}$ (thin line).
At time $t_0$, the chain leaves the last piece of the 
initial tube, with one chain end at the position of bead $j_{0}$
in the initial tube.}
\end{center}
\end{figure} 


\begin{figure}
\label{fig5}
\begin{center}
\epsfig{figure=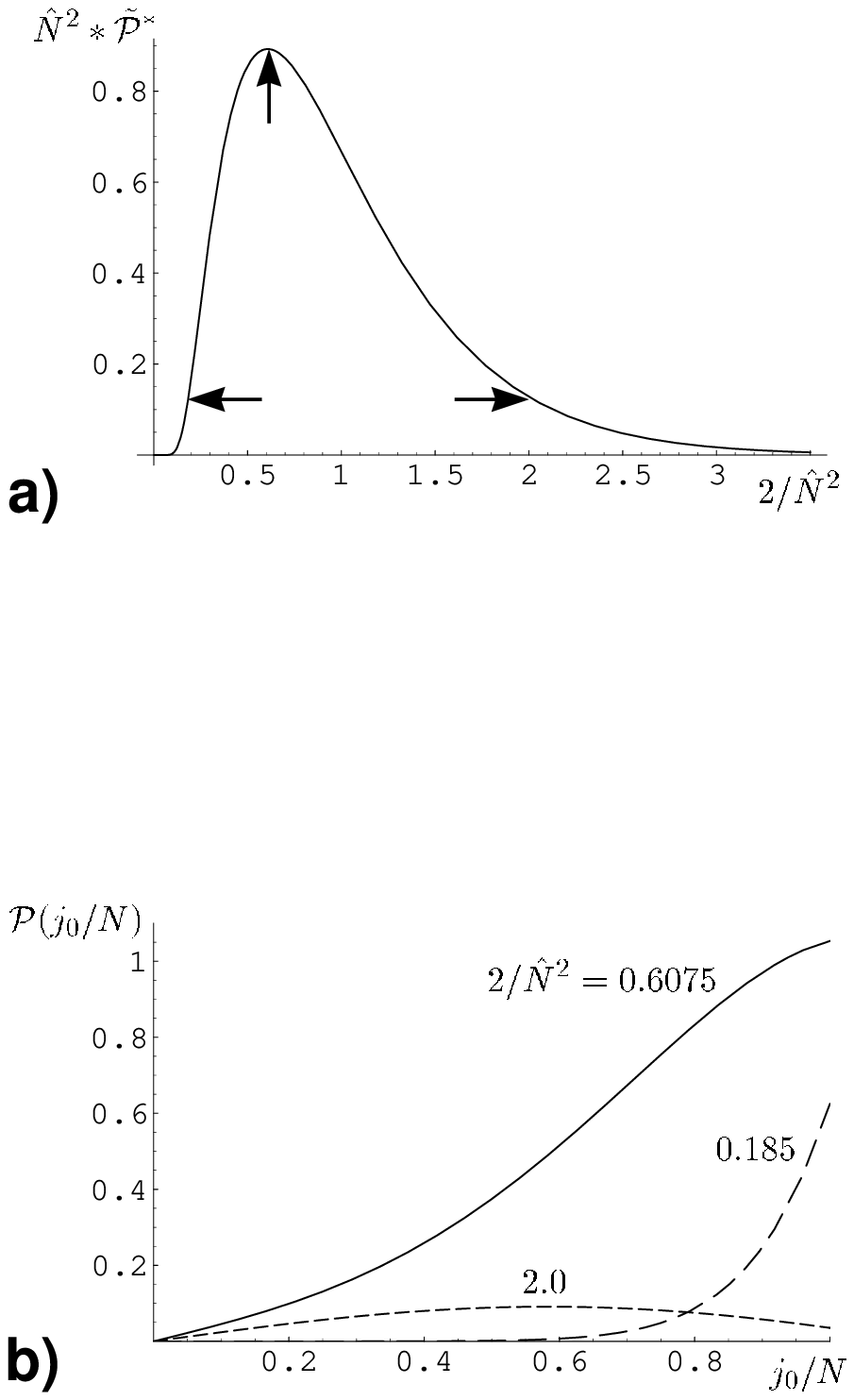,width=0.5\linewidth} 
\caption[]{
Distribution functions for complete tube destruction. 
a) Probability density of complete tube destruction 
as function of $2/\hat{N}^{2} \sim t/T_{3}$. Normalization: 
$\int_{0}^{\infty} d (2/\hat{N}^{2})\; \hat{N}^{2}\; 
\tilde{{\cal P}}^{*} (0,\hat{N},\hat{N}) = 1$.\\
b) Probability density of complete tube destruction 
as function of the position $j_{0}/N$ of the final segment 
of the original tube: $d (2/\hat{N}^{2})\; {\cal P} (j_{0}/N) = 
d t_{0} \; {\cal P}^{*} (j_{0},t_{0} \mid 0)$. 
The chain leaves the tube with end $0$. The values of 
$2/\hat{N}^{2}$ chosen are indicated by arrows in Fig.~5~a.}
\end{center}
\end{figure}


\begin{figure}
\label{fig6}
\begin{center}
\epsfig{figure=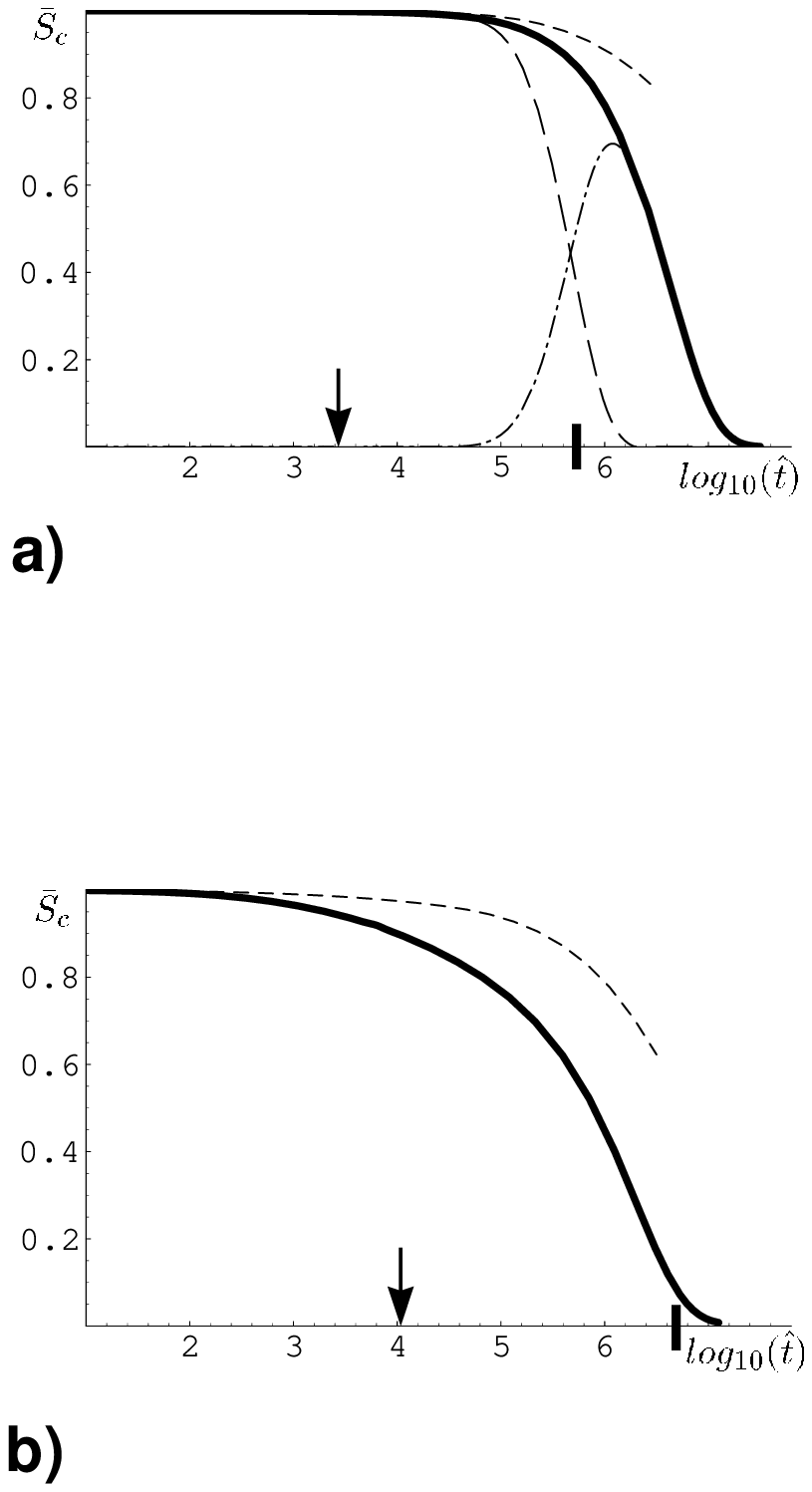,width=0.5\linewidth} 
\caption[]{
Normalized coherent scattering function in different wave number 
regions as function of $\log_{10}(\hat{t})$. 
a) $q^{2} = 0.01$, $N = M = 157$; b) $q^{2} = 1.0$, $N = M = 317$. 
The thick solid lines give the full functions $\bar{S}_{c}$. 
Long dashes represent $\bar{S}_{c}^{(T)}$, which in b) coincides 
with $\bar{S}_{c}$. Short dashes are the results neglecting 
all end effects. The dot-dashed line in a) is the contribution of 
complete tube destruction. See the text for further explanations.}
\end{center}
\end{figure}


\begin{figure}
\label{fig7}
\begin{center}
\epsfig{figure=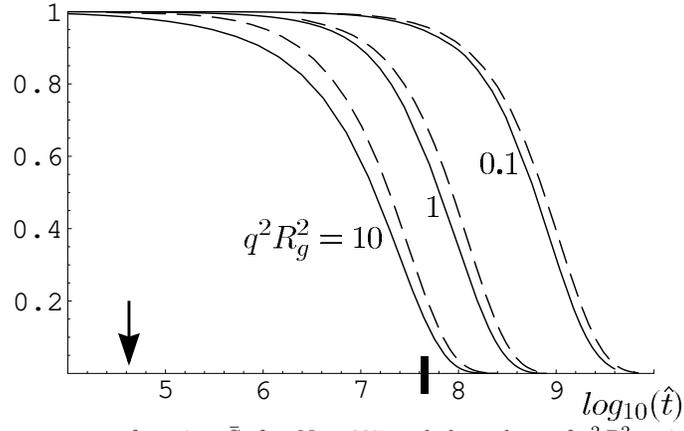,width=0.5\linewidth} 
\caption[]{
Normalized coherent structure function $\bar{S}_{c}$ for $N = 637$ 
and the values of $q^{2} R_{g}^{2}$ as indicated (full lines). 
The dashed lines denote $\bar{S}_{DE}$. Arrow and slash indicate 
$\hat{T}_{2}$ or $\hat{T}_{3}$, respectively.}
\end{center}
\end{figure}


\begin{figure}
\label{fig8}
\begin{center}
\epsfig{figure=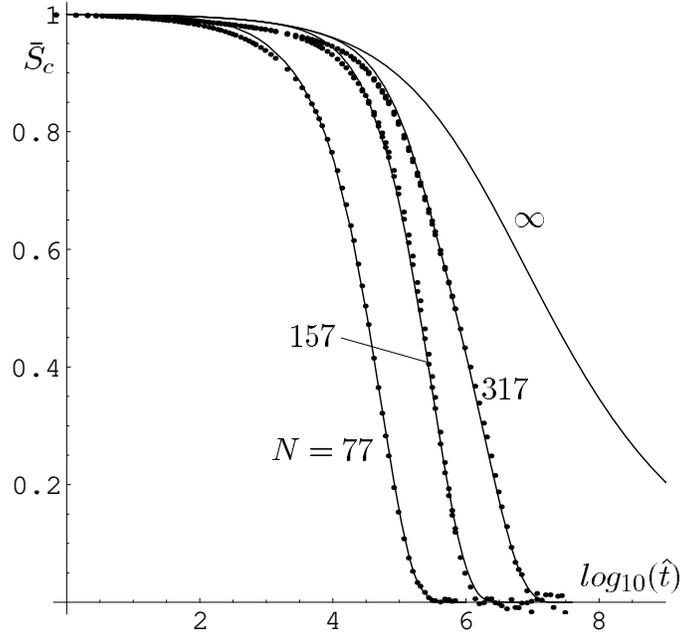,width=0.5\linewidth} 
\caption[]{
Normalized coherent structure function $\bar{S}_{c}$ of a subchain 
of about $M=80$ beads in chains of the total lengths $N$ 
as indicated in the figure. Wavenumber $q = 0.5$. Solid line: theory.
Data points result from a simulation of the Evans-Edwards model.}
\end{center}
\end{figure}


\begin{figure}
\label{fig9}
\begin{center}
\epsfig{figure=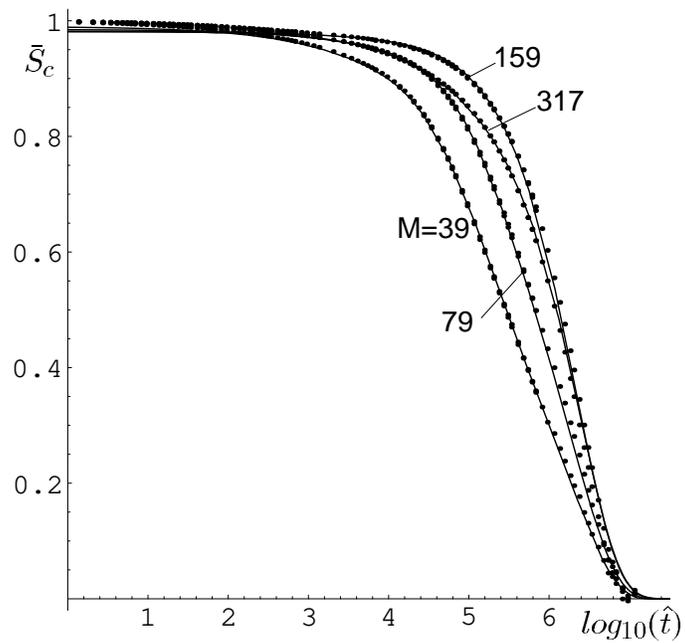,width=0.5\linewidth} 
\caption[]{
Results for $\bar{S}_{c}$, $q = 0.5$, $N = 317$. Theoretical results 
(solid lines) for central sub-chains of lengths $M = $39, 79, and 159 
are compared to simulations (dots). 
Results for the total chain $(M = 317)$ are also shown.}
\end{center}
\end{figure}





\end{document}